\documentclass[12pt]{iopart}
\usepackage{epsfig}    

\def\Journal#1#2#3#4{#4 {#1} {\bf #2} #3 }
\def\ANP {\it Advances in Nucl.\ Phys.\/}
\def\ANN {\it Ann.\ Phys.\ (N.Y.)\/}

\def\EPJA{{\it Eur.\ Phys.\ J\/} A}
\def\EPJC{{\it Eur.\ Phys.\ J\/} C}
\def\FBS{\it Few Body Syst.\/}
\def\FBSS{\it Few Body Syst.\ Supplement\/}
\def\HPA{\it Helv.\ Phys.\ Acta\/}
\def\JETPL{\it JETP Lett.\/}
\def\Nature{\it Nature\/}
\def\NCA{\it Nuovo Cimento\/}
\def\NCL{\it Nuovo Cimento Lett.\/}
\def\NCS{\it Nuovo Cimento Supp.\/}
\def\NIM{\it Nucl.\ Instrum.\ Methods\/}
\def\NIMA{{\it Nucl.\ Instrum.\ Methods\/} A}
\def\NPA{{\it Nucl.\ Phys.\/} A}
\def\NPB{{\it Nucl.\ Phys.\/} B}
\def\PLB{{\it Phys.\ Lett.\/}  B}
\def\PRP{\it Phys.\ Reports\/}
\def\PR{\it Phys. Rev.\/}
\def\PRL{\it Phys.\ Rev.\ Lett.\/}
\def\PRD{{\it Phys.\ Rev.\/} D}
\def\PRC{{\it Phys.\ Rev.\/} C}
\def\PZETF{\it Pis'ma Zh.\ Eksp.\ Teor.\ Fiz.\/}
\def\SJNP{\it Sov.\ J.\ Nucl.\ Phys.\/}
\def\YF{\it Yad.\ Fiz.\/}
\def\ZPA{{\it Z.\ Phys.\/} A}
\def\ZPC{{\it Z.\ Phys.\/} C}

\begin{document} 

\setlength{\topmargin}{-0.4 in}

{\small\vspace*{-0.2in}\rightline{\parbox{1.5in}
                {\leftline{JLAB-PHY-01-25}
                \leftline{WM-01-113}
}}}

\review{Electromagnetic structure of 
the deuteron}{}

\author{ R.~Gilman$^{\dag\ddag}$ and Franz~Gross$^{\ddag\S}$}

\address{$^\dag$\ Rutgers University, 126 Frelinghuysen Rd, Piscataway, NJ  08855 USA}
\address{$^\ddag$\ Jefferson Lab, 12000 Jefferson Ave, Newport News, VA  23606 USA}
\address{$^\S$\ College of William and Mary, Williamsburg, VA  23187 USA}

\eads{\mailto{gilman@jlab.org}, \mailto{gross@jlab.org}}
\begin{abstract}
Recent measurements of the deuteron electromagnetic
structure functions $A$, $B$, and $T_{20}$ extracted
from high energy elastic $ed$ scattering,  and the cross
sections and asymmetries extracted from
high energy  photodisintegration $\gamma+d\to n+p$, are reviewed
and compared to theory.  The theoretical calculations 
range from nonrelativistic and relativistic models using the
traditional meson and baryon degrees of freedom, to
effective field theories, to models based on the underlying
quark and gluon degrees of freedom of QCD,
including nonperturbative quark cluster models and
perturbative QCD.  We review what has been learned from these
experiments, and discuss why elastic
$ed$ scattering and photodisintegration seem to require very
different theoretical approaches, even though they are
closely related experimentally.

\end{abstract}
 
\vspace{-0.3in} 


\vspace{-0.3in} 
 
\tableofcontents   
 
\vspace{-0.2in} 

 
\article[Electromagnetic structure of 
the deuteron]{}{}


     
   
\section{Introduction}
  
The deuteron, the only $A = 2$ nucleus, provides the simplist
microscopic test of the {\it conventional nuclear
model\/}, a framework in which nuclei and nuclear interactions
are explained as baryons interacting through the exchange of
mesons. With improved nucleon-nucleon force models from the
1990s \cite{nnreview}, and advances in our understanding of
relativistic bound state techniques, more accurate calculations
of deuteron structure are possible.  

During the 1990s there have also been revolutionary 
improvements  in our experimental knowledge of deuteron
electromagnetic structure. The start of experiments at the
Thomas Jefferson National Accelerator Facility (JLab) has now
made available continuous high energy beams,  with high
currents and large polarization, along with new detector 
systems. Several experiments have now significantly extended
the energy and momentum transfer range of deuteron
electromagnetic studies,  including $A$ and $t_{20}$ for
elastic $ed$ scattering, and photodisintegration
cross sections and polarizations. Existing experimental
proposals promise to continue this trend. Other laboratories
have also made several important measurements, generally at
lower momentum transfer.

In this context, a review of the deuteron electromagnetic 
studies, examining the current status of the agreement between
experiments and theory, is appropriate.
We attempt to cover our current knowledge of 
the deuteron electromagnetic structure, focussing on the recent
JLab results, and prospects for the future.
We do not consider experiments that use the deuteron as a
neutron target, for  example,  or for studies of the (extended)
Gerasimov-Drell-Hearn sum rule,  deep inelastic scattering, or
baryon resonance production in nuclei. Our interest is on
experiments that probe the conventional picture of a nucleus as
composed of baryons and mesons,  and that probe how far models
with these effective degrees of freedom can be extended.
Table~\ref{dexptab} is a summary of some of the JLab exeriments that
fit this description, and that we will review in the sections 
below.

The high precision, large momentum transfer measurements may be
sensivitive to effects not incorporated in the conventional
nuclear model.  It seems self-evident that probes of short
distances, well below the size of the nucleon, should require
explicit consideration of the quark substructure of the
nucleons.
Our review suggests that evidence for the appearance of
these effects seems to depend on the nature of the reaction.
In elastic scattering, where only the $NN$ chanel is 
expicitly excited, a successful description is obtained 
using a relativistic description of the $NN$ channel together 
with a minor modification of the short-range structure
of the nucleon current (see Sec.~\ref{sec:elasticscat}).
In photodisintegration by 4 GeV photons, where hundreds of
$N^*N^*$ channels are explicitly excited, an efficient explanation
seems to require the explicit use of quark degrees of freedom
(see Sec.~\ref{photodis}).

\begin{table}
\caption{\label{dexptab} Some JLab deuterium experiments.}
\begin{indented}
\item[]\begin{tabular}{@{}lll}
\br
Experiment & Reaction / Observables & Status \\
\mr
        & {\it elastic scattering} & \\ 
91-026 & $A$ & paper published \cite{alexa99} \\
        & $B$ & analysis in progress \\
94-018 & $A$ & paper published \cite{abbott99} \\
        & $t_{20}$ & paper published \cite{abbott00} \\
\mr
        & {\it electrodisintegration} & \\
89-028 & recoil proton polarization & analysis in progress \\
94-004 & in-plane response functions & analysis in progress \\
94-102 & high momentum structure & awaiting beam time \\
00-103 & threshold $d(e,e^{\prime})pn$ & proposal \\
\mr
        & {\it photodisintegration} & \\
89-012 & cross sections      & paper published \cite{bochna98} \\
89-019 & $p_y$, $C_{x'}$, $C_{z'}$ & paper published \cite{wije01} \\
96-003 & cross sections      & paper published \cite{schulte01}  \\
93-017 & cross sections      & analysis in progress \\
99-008 & cross sections      & analysis in progress \\
00-007 & $p_y$, $C_{x'}$, $C_{z'}$ & awaiting beam time \\
00-107 & $p_y$, $C_{x'}$, $C_{z'}$ & awaiting beam time \\
\br
\end{tabular}
\end{indented}
\end{table}

This review begins with a survey of deuteron wave
functions, and then discusses the deuteron form factors,
threshold electrodisintegration, and high energy deuteron
photodisintegration.  We also call attention to recent reviews
by Gar\c{c}on and Van Orden \cite{GVO01}, and by Sick
\cite{Si01}.  These reviews contain a discussion of
the static properties of the deuteron and a survey of recent
models of the nucleon form factors, two topics we have decided
to omit from this work. Both also have an
extensive discussion of the deuteron form factors.


\section{Deuteron Wave Functions}

Calculations of deuteron form factors and photo and
electrodisintegration to the $NN$ final state require a
deuteron wave function, the final state
$NN$ scattering amplitude (if the transition is inelastic), and
the current operator, all of which should be consistently
determined from the underlying dynamics.  Deuteron wave
functions used in the {\it conventional nuclear model\/}
will be reviewed in this section.  
 
\subsection{Nonrelativistic wave functions}

The nonrelativistic $NN$ wave function of the deuteron can be
written in terms of two scalar wave functions.  In coordinate
space the full wave function is
\begin{eqnarray}
\fl{\Psi}_{abm}^{+}({\bf r})=\sum_\ell\sum_{m_s}
\frac{z_\ell(r)}{r}Y_{\ell\, m-m_s}(\hat{\bf r})\;\chi^{1m_s}_{ab}
\left<\ell\,1\,m-m_s\, m_s|1\, m\right>\nonumber\\
\lo=\frac{u(r)}{r}Y_{00}(\hat{\bf r})\,\chi^{1m}_{ab}+
\frac{w(r)}{r} \sum_{m_s}  Y_{2\,m-m_s}(\hat{\bf r}) \; 
\chi^{1m_s}_{ab}
\left<2\,1\,m-m_s\, m_s|1\, m\right> \label{rspacewf}
\end{eqnarray}
where $Y_{\ell m_\ell}$ are the spherical harmonics normalized to unity on the
unit sphere, $z_0=u$ and $z_2=w$ are the reduced $S$ and
$D$-state wave functions, and the $+$ distinguishes this from
other (relativistic) components of the wave function to be
described below.  The spin part of the wave function is 
\begin{equation}
\chi^{1m_s}_{ab}=\cases{\qquad\qquad
\left|+\right>_a\left|+\right>_b & $m_s=+$\cr 
\frac{1}{\sqrt{2}}\Biggl\{\left|+\right>_a\left|-\right>_b
+ \left|-\right>_a\left|+\right>_b\Biggr\} & $m_s=0$\cr 
\qquad\qquad
\left|-\right>_a\left|-\right>_b & $m_s=-$ \, .}
\label{spinwf}
\end{equation}
Introducing the familar compact notation for matrix operations 
on each of the two nucleon subspaces 1 and 2
\begin{equation}
A_{aa'}\,\left|+\right>_{a'}= A_1\,\left|+\right>_{1}\, ,
\end{equation}
where $A$ is any $2\times2$ operator, we can show that
\begin{eqnarray}\fl
Y_{00}\,\chi^{1m}_{ab}\;=\frac{1}{\sqrt{4\pi}}\,\sigma_1\cdot\sigma_2
\;\chi^{1m}_{_{12}} =
\frac{1}{\sqrt{4\pi}}\,\chi^{1m_s}_{_{12}} \nonumber\\ 
\fl\sum_{m_s} Y_{2\,m-m_s}(\hat{\bf r}) \;\chi^{1m_s}_{ab}
\left<2\,1\,m-m_s\, m_s|1\,m\right>
=\frac{1}{\sqrt{32\pi}}(3\,\sigma_1\cdot\hat{\bf
r}\,\sigma_2\cdot\hat{\bf r} - \sigma_1\cdot\sigma_2 )\;\chi^{1m}_{_{12}}\, .
\label{identity}
\end{eqnarray} 
These identities permit us to write the wave function 
(\ref{rspacewf}) in a convenient operator form \cite{BG79}:
\begin{eqnarray}
{\Psi}_{abm}^{+}({\bf r})=\frac{1}{\sqrt{4\pi}\;r}\left[u(r)\,
\sigma_1\cdot\sigma_2 + \frac{w(r)}{\sqrt{8}}\,
(3\,\sigma_1\cdot\hat{\bf r}\,\sigma_2\cdot\hat{\bf r} -
\sigma_1\cdot\sigma_2 )
\right]\;\chi^{1m}_{_{12}}
 \label{rspacewf2}
\end{eqnarray}

In momentum space the deuteron wave function becomes
\begin{eqnarray}
{\Psi}_{abm}^{+}({\bf p})&\equiv&\frac{1}{\sqrt{(2\pi)^3}}
\int d^3r\,e^{-i{\bf p\cdot r}}\;{\Psi}_{abm}^{+}({\bf
r})\nonumber\\
&=&\frac{1}{\sqrt{4\pi}}\left[u(p)\,
\sigma_1\cdot\sigma_2 - \frac{w(p)}{\sqrt{8}}\,
(3\,\sigma_1\cdot\hat{\bf p}\,\sigma_2\cdot\hat{\bf p} -
\sigma_1\cdot\sigma_2 )
\right]\;\chi^{1m}_{_{12}}
\, . \label{ft1}
\end{eqnarray}
We use the same notation for both coordinate and momentum space wave
functions.  If $u(p)=z_0(p)$ and $w(p)=z_2(p)$, then 
\begin{eqnarray}
z_\ell(p)&=& \sqrt{\frac{2}{\pi}} \int_0^\infty
rdr\,z_\ell(r)\,j_\ell(pr)\nonumber\\
\frac{z_\ell(r)}{r}&=& \sqrt{\frac{2}{\pi}} \int_0^\infty
p^2dp\,z_\ell(p)\,j_\ell(pr)\, .
\label{ft2}
\end{eqnarray}
Note the appearance of the factors $\sqrt{2/\pi}$, a feature of the
symmetric definition (\ref{ft1}).

\begin{figure}
\begin{center}
\mbox{
   \epsfxsize=3.5in
\epsffile{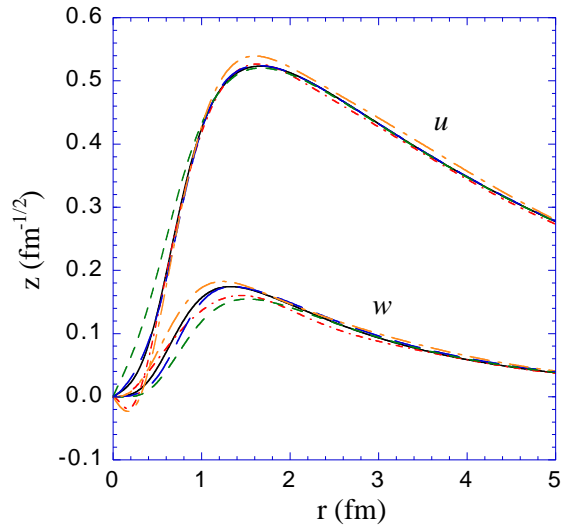} 
}
\end{center}
\vspace*{-0.2in}
\caption{Reduced coordinate space wave functions for five 
models discussed  in the text:  AV18 (solid), Paris (long
dashed), CD Bonn (short dashed), IIB (short dot-dashed), and
W16 (long dot-dashed).  }
\label{uwr}
\end{figure} 

The normalization condition 
\begin{equation}
\int d^3r\,\Psi^{+\dagger}_{abm'}({\bf r})\,\Psi^+_{abm}({\bf
r})=\delta_{m'm}
\end{equation}
implies
\begin{equation}
1=\int_0^\infty dr\left[u^2(r) + w^2(r)\right]=\int_0^\infty p^2\,dp
\left[u^2(p) + w^2(p)\right] 
\end{equation}
The $D$-state probability, 
\begin{equation}
P_D=\int_0^\infty dr\, w^2(r)
\end{equation}
is an interesting measure of the strength of the tensor 
component of the $NN$ force, even though it is a model
dependent quantity with no unique measurable value \cite{Fr79}.

\begin{figure}[t]
\begin{center}
\mbox{
   \epsfxsize=6.4in
\epsffile{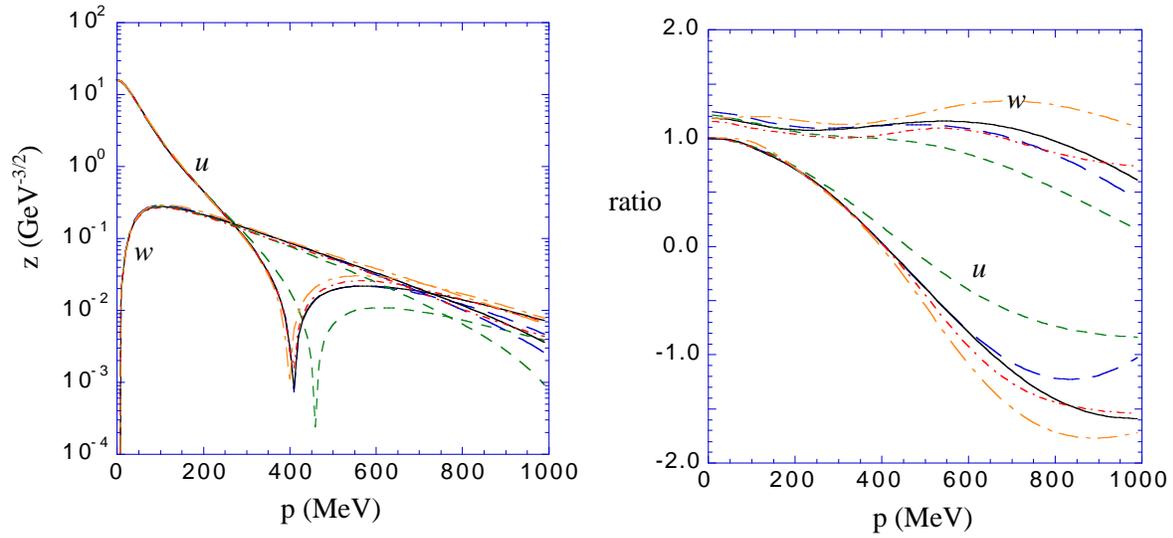}
}
\end{center}
\caption{Momentum space wave functions for five models 
discussed  in the text (see the caption to Fig.~1).  The wave
functions in the right panel have been divided by the scaling 
functions $u_s(p)$ and $w_s(p)$.}
\label{uwp}
\end{figure} 

The best nonrelativistic wave functions are calculated from the
Schr\"odinger equation using a potential adjusted to fit the 
$NN$ scattering data for lab energies from 0 to 350 MeV.  The
quality of realistic potentials have improved steadily, and now
the best potentials give fits to the $NN$ data with a
$\chi^2$/d.o.f $\simeq1$.  The Paris potential \cite{Paris} was
among the first potentials to be determined from such realistic
fits, and it has since been replaced by the Argonne V18
potential (denoted by AV18) \cite{AV18}, the Nijmegen potentials
\cite{Nijmegen}, and most recently by the CD Bonn potentials
\cite {CDBonn,CDBonn2000}.  The $S$ and $D$-state wave functions
determined from three of these models are shown in
Figs.~\ref{uwr} and
\ref{uwp}.  These figures also show $S$ and $D$ wave functions
from two relativistic models to be discussed shortly.  In the 
right panel of the second figure we plot the dimensionless
ratios $u(p)/u_s(p)$ and $w(p)/w_s(p)$, where the scaling
functions, in units of GeV$^{-3/2}$, are
\begin{eqnarray}
u_s(p)&=& \frac{16m\epsilon}{(m\epsilon + p^2) (1 +
p^2/p_0^2)}\nonumber\\ w_s(p)&=& \frac{16m\epsilon\;
p^2/p_1^2}{(m\epsilon + p^2)(1 + p^2/p_0^2)^2}\, .
\label{scale1}
\end{eqnarray}
Here $\epsilon$ is the deutreron binding energy and $m$ the 
nucleon mass (we  used $m\epsilon=940\times 2.224$ MeV$^2$), 
$p_0^2=0.15$ GeV$^2$, and $p_1^2=0.1$ GeV$^2$. We emphasize that
these scaling functions {\it have absolutely no theoretical
significance\/} and were introduced merely to remove the most
rapid momentum dependence so that the percentage difference
between models can be more easily read from the ratio graph. 
We conclude that the  five models shown are almost identical
(i.e.variations of less than 10\%) for momenta below about 400
MeV, and that they vary by less than a factor of 2
as the momenta reaches 1 GeV (except near the zeros).


\subsection{Relativistic wave functions}

The definition of the relativistic deuteron wave function 
depends in large part on the formalism used to treat
relativity.  In formalisms based on hamiltonian dynamics
(discussed in Sec.~\ref{Hexamples}) the wave function in the
deuteron rest frame can be taken to be identical to the
nonrelativistic wave function, and no further discussion is
necessary until the wave function in a moving frame is needed. 
In formalisms based on the  Bethe-Salpeter
equation~\cite{BS51}, the covariant spectator
equation~\cite{Grosseq},  or on  some other quasi-potential
equation~\cite{BSLT,PW}, the wave functions usually have
additional components which do not vanish in the rest frame. 

In the relativistic spectator formalism~\cite{Grosseq}, where
one of the two bound nucleons is off-mass shell, the wave
function is a sum of a positive energy component  and a
negative energy component~\cite{BG79}
\begin{equation}
\Psi_{a \alpha m}({\bf p})= \sum_b\left\{\Psi^+_{abm}({\bf p})\,
u_\alpha(-{\bf p},b) + \Psi^-_{abm}({\bf p})\,v_\alpha({\bf p},
b)\right\}\,  ,
\label{relwf}
\end{equation}
where $u_\alpha(-{\bf p},b)$ and $v_\alpha({\bf p}, b)$ are 
nucleon spinors  for the off-shell particle (particle 2 in this
case) with Dirac index
$\alpha$.  The positive energy part has the same structure as
the nonrelativistic wave function, with an $S$ and $D$-state
component.  The new negative energy wave function has the form
\begin{eqnarray}
\fl\Psi^-_{abm}({\bf p})&= v_t(p)\sum_{m_s}
Y_{1\,m-m_s}(\hat{\bf p}) 
\;\chi^{1m_s}_{ab} \left<1\,1\,m-m_s\, m_s|1\,m\right> -
v_s(p) \, Y_{1\,m}(\hat{\bf p}) \;\chi^{0}_{ab}
 \label{pspacePwf} \nonumber\\
\fl&=\sqrt{\frac{3}{4\pi}}\frac{1}{2}\left[\frac{v_t(p)}{\sqrt{2}}\,(\sigma_1
+
\sigma_2)\cdot \hat{\bf p} + v_s(p)\,(\sigma_1 -
\sigma_2)\cdot \hat{\bf p}  \right]\,\chi^{1m}_{ab}\, , \label{Pwf}
\end{eqnarray}
where $v_t(p)$ and $v_s(p)$ are two additional $P$-state
components of the wave function, and 
\begin{equation}
\chi^{0}_{ab} = \frac{1}{\sqrt{2}}
\Biggl\{\left|+\right>_a\left|-\right>_b -
\left|-\right>_a\left|+\right>_b\Biggr\}
\end{equation}
is the nuclear spin 0 wave function.  The names of the 
$P$-state wave functions follow from the fact that $v_t$ couples
to the spin triplet ($S=1$)  and $v_s$ to the spin singlet
($S=0$) wave function.  The equivalence of the two forms given in
Eq.~(\ref{Pwf}) follows from identities like those given in
Eq.~(\ref{identity}).  

Two model relativistic $S$ and $D$-state wave functions 
were shown in Figs.~\ref{uwr} and \ref{uwp}.  Both models
are based on relativistic one boson exchange model
developed in Ref.~\cite{gvoh92}.  Model IIB is a revised
version of the model of the same name originally
described in Ref.~\cite{gvoh92} and  Model W16 is one of
a family of models with varying amounts of off-shell
sigma coupling that were introduced in connection the
relativistic calculations of the triton binding energy
described in Ref.~\cite{sg97}.  These models are
described further in a number of conference talks
\cite{Pstate}.  The relativistic $P$-state components are small,
but can make important contributions to the deuteron magnetic
form factor.  As Figs.~\ref{uwr} and
\ref{uwp} show, the large $S$ and
$D$-state components of these relativistic wave functions are
very close to their nonrelativistic counterparts.


\section{Elastic Electron Deuteron Scattering}
\label{sec:elasticscat}

\subsection{Deuteron Form Factors and Structure Functions}
\label{dffdef} 

Because of the very small value of the electromagnetic fine 
structure constant ($\alpha=e^2/4\pi\hbar c\simeq 1/137$),
elastic electron--deuteron scattering is described to high
precision by assuming that the electron exchanges a single
virtual photon when scattering from the deuteron.  In 
this one-photon exchange approximation \cite{twogamma} 
elastic scattering is fully described by three deuteron
form factors \cite{G65,acg80,donn86}.   In its most
general form, the relativistic deuteron current can be
written \cite{HJ62,G65}
\begin{eqnarray}
\left<d'|J^\mu|d\right>=-\Biggl(&&\left\{G_1(Q^2)\;
[\xi'^*\cdot\xi]-G_3(Q^2)\frac{(\xi'^*\cdot q)(\xi\cdot q)}{2m_d^2}\right\}
\,(d^\mu+d'^\mu)  \nonumber\\
 &&+ G_M(Q^2)\;[\xi^\mu(\xi'^*\cdot q)
-\xi'^{*\mu}(\xi\cdot q)]\Biggr)\, ,\label{deutcurrent}
\end{eqnarray} 
where the form factors $G_i(Q^2)$, $i=1-3$, are all functions 
of $Q^2$, the square of the four-momentum transferred by the
electron, with $q=d'-d$ and $Q^2=-q^2$.  [In most of the
following discussion we will suppress the explicit
$Q^2$ dependence of the form factors.]  In practice,
$G_1$ and $G_3$ are replaced by a more physical choice of form 
factors
\begin{eqnarray}
G_C&&=G_1 +\frac{2}{3}\,\eta\,G_Q\nonumber\\ 
G_Q&&=G_1-G_M+(1+\eta)G_3\, ,
\end{eqnarray}
with $\eta=Q^2/4m_d^2$.  At $Q^2=0$, the form factors $G_C$, $G_M$, and $G_Q$
give the charge, magnetic and quadrupole moments of the deuteron
\begin{eqnarray}
G_C(0) &= 1 &\qquad ({\rm in\ units\ of}\, e )\nonumber\\ 
G_Q(0) &= Q_d &\qquad ({\rm in\ units\ of}\, e/m_d^2)\nonumber\\
G_M(0) &=\mu_d &\qquad ({\rm in\ units\ of}\, e/2m_d) \, .     
\end{eqnarray}

The form factors can also be related to the helicity amplitudes 
of the deuteron current (where helicity is the projection of
the spin in the direction of the particle three-momentum).  In
the Breit frame (where the energy  transfer
$\nu$ is zero) the polarizations of the incoming ($\xi$) and
outgoing ($\xi'$) deuteron are 
\begin{eqnarray}
\xi_\lambda=\cases{(0,\pm1,- i,0)/\sqrt{2} & $\lambda=\pm$\cr
 (-Q/2,0,0,D_0)/m_d & $\lambda=0$}\nonumber\\ 
\xi'_{\lambda'}=\cases{(0,\mp1,- i,0)/\sqrt{2} & $\lambda'=\pm$\cr
\;\,(Q/2,0,0,D_0)/m_d & $\lambda'=0$\; ,}     
\end{eqnarray}
(where the phases for the incoming deuteron follow the 
conventions of Jacob and Wick
\cite{JW59} for particle 2) and the virtual photon
polarization is
\begin{eqnarray}
\epsilon_{\lambda_\gamma}=\cases{(0,\mp1,- i,0)/\sqrt{2} &
$\lambda_\gamma=\pm$ \cr\;\; (1,0,0,0) & $\lambda_\gamma=0$\; .}     
\end{eqnarray}
Hence, denoting the helicity amplitudes by
$G^{\lambda_\gamma}_{\lambda'\lambda}$, the three independent amplitudes
are 
\begin{eqnarray}
G^0_{00}(Q^2)=2D_0\left(G_C +\frac{4}{3}\eta G_Q\right) \nonumber\\ 
G^0_{+-}(Q^2)= G^0_{-+}(Q^2) = 2D_0\left(G_C -\frac{2}{3}\eta G_Q\right) 
\nonumber\\ 
G^+_{+0}(Q^2)=-G^+_{0-}(Q^2)
=G^-_{-0}(Q^2)=-G^-_{0+}(Q^2) =2D_0\,\sqrt{\eta}\,G_M \, ,
\label{dhelicity}    
\end{eqnarray}
where $2D_0=\sqrt{4m_d^2 + Q^2}$.

The scattering amplitude in the one-photon approximation is
\begin{eqnarray}
{\cal M}=-\frac{e^2}{Q^2}\,\Bigl[\bar u(k',\lambda')\,\gamma^\mu\,u(k,\lambda)
\Bigr]\,\left<d'|J_\mu|d\right>  \, , \label{matrix}
\end{eqnarray}
where $u$ and $\bar u$ are electron spinors with $k,\lambda$ ($k',\lambda'$)
the momentum and helicity of the incoming (outgoing) electron,
respectively.  Squaring (\ref{matrix}), summing over the final spins, and
averaging over initial spins give the following result for the unpolarized
differential cross section   
\begin{equation}
{ {d\sigma}\over{d\Omega} } = {{d\sigma}\over{d\Omega}} \Bigg|_{NS} 
\Bigl[ A(Q^2) + B(Q^2) \tan^2(\theta/2) 
\Bigr] = {{d\sigma}\over{d\Omega}} \Bigg|_{NS}\,S(Q^2,\theta)
\label{edxsect}
\end{equation}
where $S(Q^2,\theta)$ is defined by this relation, and
\begin{equation}
{{d\sigma}\over{d\Omega}} \Bigg|_{NS} = { {\alpha^2 E^{\prime}
\cos^2(\theta/2)} \over
 {4 E^3 \sin^4(\theta/2)} }=\sigma_M {E'\over E}=\sigma_M \left(
1+{2E\over m_d}\sin^2{\textstyle{1\over2}}\theta\right)^{-1}\, ,
\label{mott}
\end{equation}
is the  cross section for scattering from a particle without internal
structure ($\sigma_M$ is the Mott cross section), and $\theta$, $E, E'$, 
and $d\Omega$ are the electron scattering angle, the  incident and final
electron  energies, and the solid angle of the scattered electron, all in
the lab system.
The structure functions $A$ and
$B$ depend on the three electromagnetic form factors 
\begin{eqnarray}
A(Q^2) &&= G_C^2(Q^2) + {{8}\over{9}} \eta^2 G_Q^2(Q^2) + 
 {{2}\over{3}} \eta G_M^2(Q^2)\nonumber\\
&&\equiv A_C(Q^2) + A_Q(Q^2)+A_M(Q^2) \nonumber\\
B(Q^2) &&= {{4}\over{3}} \eta(1+\eta) G_M^2(Q^2)\, ,
\label{AandB}
\end{eqnarray}
where the definitions of $A_C$, $A_Q$, and $A_M$ should be 
clear from the context.

While cross section measurements can determine $A$, $B$, and 
$G_M$, separating the charge $G_C$ and quadrupole $G_Q$ form
factors requires polarization measurements.  The polarization 
of the outgoing deuteron  can be measured in a second,
analyzing scattering.  The cross section for the double
scattering process can be written \cite{acg80}
\begin{eqnarray}
\fl{ {d\sigma}\over{d\Omega d\Omega_2} } =  { {d\sigma}\over{d\Omega
d\Omega_2} }\Bigg|_0 \Bigl[1 &&+ {\textstyle{3\over2}}h\,p_xA_y\sin \phi_2
\nonumber\\ &&+ {\textstyle{1\over\sqrt{2}}} t_{20}A_{zz} -
{\textstyle{2\over\sqrt{3}}} t_{21}A_{xz}\cos\phi_2 +
{\textstyle{1\over\sqrt{3}}}t_{22}(A_{xx}-A_{yy})\cos2\phi_2\Bigr]\, ,
\label{eq:ds}
\end{eqnarray}
where $h=\pm1/2$ is the polarization of the incoming electron 
beam, 
$\phi_2$ the angle between the two scattering planes (defined in the same
way as the $\phi$ shown in Fig.~\ref{fig_cross}), and
$A_y$ and the $A_{ij}$ are the vector and tensor analyzing powers of the
second scattering.
Although there is a $p_z$ component to the vector polarization,
the term is omitted from Eq.~(\ref{eq:ds}) as there is no longitudinal vector analyzing power;
without spin precession, this term can not be determined.
The polarization quantities $p_i$ and $t_{2m_\ell}$
(sometimes denoted $T_{2m_\ell}$, but we will reserve capital letters for
target asymmetries) are functions of the form factors and the
electron scattering angle
\begin{eqnarray}
S\,p_x=-{\textstyle{4\over3}}\,\Bigl[\eta(1+\eta)\Bigr]^{1/2}G_M(G_C+
{\textstyle{1\over3}}\eta G_Q)\,\tan{\textstyle{1\over2}}\theta\nonumber\\
S\,p_z=\;\;\,{\textstyle{2\over3}}\eta\Bigl[(1+\eta)(1+\eta\sin^2 
{\textstyle{1\over2}}\theta)\Bigr]^{1/2}G^2_M  \tan
{\textstyle{1\over2}}\theta\,\sec{\textstyle{1\over2}}\theta\nonumber\\
- \sqrt{2} S\, t_{20} ={\textstyle{8\over3}} \,\eta\, G_C G_Q +
{\textstyle{8\over9}}\,\eta^2 G_Q^2 +
{\textstyle{1\over3}} \eta \Bigl[1 + 2(1+\eta)\tan^2
{\textstyle{1\over2}}\theta\Bigr] G_M^2 \nonumber\\
\sqrt{3}\, S \,t_{21} = 2 \,\eta \Bigl[\eta + \eta^2\sin^2
 {\textstyle{1\over2}}\theta\Bigr]^{1/2} G_M G_Q\, \sec
{\textstyle{1\over2}}\theta \nonumber\\
-\sqrt{3} S \,t_{22} = {\textstyle{1\over2}} \,\eta \,G_M^2
\end{eqnarray} 
The same combinations of form factors occur in the tensor polarized target 
asymmetry as in the recoil deuteron tensor polarization.

Of these quantities, $t_{20}=T_{20}$ has been most extensively measured; it
does not require a polarized beam or a measurement of the out of plane
angle
$\phi_2$.  For measurements of $A$ and $T_{20}$ at forward electron
scattering angles, the $G_M$ terms are very small, and one may approximate
$A$ and $T_{20}$ by
\begin{eqnarray}
S \rightarrow  \tilde{A}\equiv &&  G_C^2 + {{8}\over{9}} \,\eta^2\,
G_Q^2\nonumber\\
- \sqrt{2} S \,\tilde{T}_{20} = && { {8}\over{3} } \,\eta\, G_C G_Q + 
{{8}\over{9} } \,\eta^2\, G_Q^2 \label{tildeAapprox}
\end{eqnarray}
Introducing $y = 2 \eta G_Q / 3 G_C$ gives
\begin{equation}
\tilde{T}_{20} = - \sqrt{2}\; { {y(2+y)}\over{1 + 2 y^2} }
\end{equation}
The minimum of $T_{20} \approx \tilde{T}_{20} = -\sqrt{2}$ 
is reached for $y=1$.  The node in the charge form factor, $G_C =
0$, occurs when $S \rightarrow {{8}\over{9}} \eta^2 G_Q^2$, and
$- \sqrt{2}\, S \,\tilde{T}_{20} \rightarrow { {8}\over{9} } \eta^2 G_Q^2$,
giving $\tilde{T}_{20} = - 1/\sqrt{2}$.

This approximation also makes it clear that $\tilde{T}_{20}$ largely
depends on the deuteron structure, rather than the nucleon
electromagnetic form factors.
In the nonrelativistic limit (to be discussed shortly), both $G_C$ and $G_Q$
are a product of the nucleon isoscalar electric form factor multiplied by
the {\it body form factor\/}, which is an integral over products of the
deuteron wave functions weighted by spherical Bessel functions.  Hence, in
this approximation, the nucleon electric form factor cancels in the ratio
$y\propto G_Q/G_C$, and $\tilde{T}_{20}$ depends only on the
deuteron wave function.

We note that the relations above between the form factors
and the observables are model independent, so it is possible
to extract form factors from the data and compare directly to 
theoretical calculations.
The most complete form factor determination
appeared recently in Ref.~\cite{abbottphen00}
(see also the analysis in Ref.~\cite{Si01}).  We will discuss
the data below in Sec.~\ref{thyandexp}, after we have reviewed
the experiments and the theory. 


\subsection{Experimental overview}

The initial measurements of elastic $ed$ scattering
were by McIntyre and Hofstadter in the mid 
1950s~\cite{mcintyre55}. Since then many experiments have run at
several laboratories; the fits of Ref.~\cite{abbottphen00}
include 269 cross sections from 19 references, dating from 1960
to the present\footnote
{An
important feature of the recent fits of the world data is that
the measured cross sections were refit rather than using
extracted structure functions or form factors.  This is
necessary since most extractions of $A$ ($B$) used corrections
for contributions of $B$ ($A$) to their cross sections from
earlier data. In some cases alternative definitions (or
incorrect formulas) have been given. A minor point is the
definition of $\sigma_{M}$; in some cases  the recoil factor
$E^{\prime}/E$ is included, while in our  definition
Eq.~(\ref{mott}) it is not.  The magnetic form factor
$G_M$ can be in units of $e/2m_d$ (our convention), 
$e/2m_p$, or  dimensionless, with magnitude of 1.714,
0.857, or 1.0,  respectively,  at $Q^2$ = 0, and leading
to modified coefficients in Eq.~(\ref{AandB}). Buchanan
and Yearian~\cite{buchanan65} have an alternate
definition of $G_C$, and $G_Q$, with $A \propto
(1-\eta)^2(G^2_C + G^2_Q)$.  Benaksas
\etal~\cite{benaksas66} and Galster \etal~\cite{galster71} 
include an extra factor $1+\eta$ in the magnetic terms in 
$A$ and $B$,  which changes the $Q^2$ dependence of the
magnetic form factor,  though not its value at $Q^2$ $=$
0.   Ganichot
\etal~\cite{ganichot72} and Grosset\^ete
\etal~\cite{grossetete66}  both use a factor of $e^2$,
rather than $\alpha^2$, in their  definitions of the Mott
cross section.  Cramer \etal~\cite{cramer85} give a
dimensionally incorrect formula for their $\sigma_0$ ($=$
$d\sigma / d\Omega|_{NS}$), with explicitly stated energy
factors of 
$E^{\prime} / ( E E^{\prime}) = 1 / E$, 
as opposed to our factor of $E^{\prime} / E^3$.
}.  
Polarization experiments are much more difficult.  The first
results were published in 1984 and there are now only
20 published $t_{20}$ data points, and 19 points for other
polarization observables.
The fits of Ref.~\cite{Si01} include a slightly larger data base,
with 340 points for momentum transfers up to $Q$ of about 1.6 GeV;
this misses only a handful of the largest momentum transfer
SLAC and JLab data points. 
 
Forward-angle cross section measurements suffice to determine
$A$, both because $B$ is small and because of the $\tan^2
\theta/2$ dependence.
The magnetic form factor $G_M$ is determined from large angle 
measurements of $B$, since the $A$ contribution vanishes
as $\theta$ $\rightarrow$ 180$^\circ$.
With $Q^2 \simeq 4 E E^{\prime} \sin^2(\theta/2)$ and
$E^{\prime} = E ( 1 + (2E/m_d) \sin^2(\theta/2))^{-1}$,
we obtain the following relations at $\theta$ $=$  180$^{\circ}$:
\begin{eqnarray}
Q^2 &=& 4 E^2 \biggl( 1 + 2E/m_d \biggr)^{-1}\, \\
E   &=& \biggl( Q^2 + \sqrt{Q^2(Q^2+4m^2_d)} \  \biggr) / 4m_d  .
\end{eqnarray}
One can see that the beam energies needed for high $Q^2$ 
measurements of $B$ are quite low, with
$E$ $=$ 0.65, 1.02, 1.35, 1.67, and 1.97 GeV corresponding to
$Q^2$ $=$ 1, 2, 3, 4, and 5 GeV$^2$, respectively.   
Note that throughout this review we use $Q$ $=$ $\sqrt{Q^2}$ to avoid
confusion with the magnitude of the three momentum transfer 
${\bf q}$, and we use units of GeV and GeV$^2$, not fm$^{-1}$ or
fm$^{-2}$.

Accurate measurements require that $Q^2$ be known accurately 
since $A$ and $B$ vary rapidly with $Q^2$.
Energy or angle offsets of a few times 10$^{-3}$
could lead to $Q^2$ being off by up to 0.5\%.
For both $A$ and $B$, this leads to offsets that increase with $Q^2$, 
reaching about 2\% at $Q^2$ $=$ 1 GeV$^2$ and
4\% at $Q^2$ $=$ 6 GeV$^2$.

While cross section measurements can determine $A$, $B$, and $G_M$,
separating the charge $G_C$ and quadrupole $G_Q$ form
factors requires polarization measurements,
most often $t_{20}$.
Coincidence detection of the scattered electron and deuteron,
which suppresses the background and allows experiments to be
performed with moderate resolution, is a common technique.

\subsubsection{ Experimental status of $A$}   

\begin{figure}
\begin{center}
\mbox{
   \epsfxsize=6.0in
\epsffile{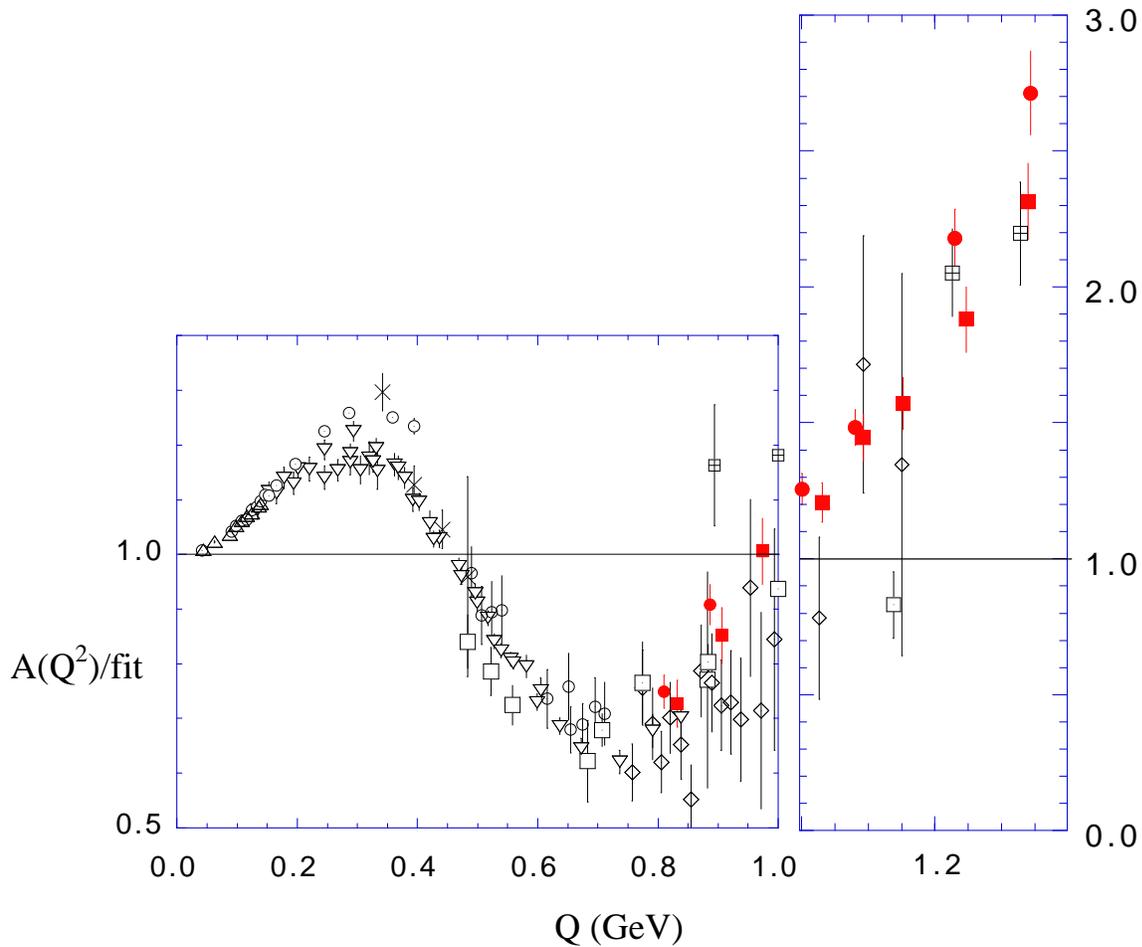} 
}
\end{center}
\vspace*{-0.2in}
\caption{The data for $A$ at low and moderate $Q$,
normalized as explained in Section~\ref{compareNR}. 
The data sets are
described in Table \ref{Adat}.  Note that the right and left
panels have different vertical scales.  All data
referred to in Table \ref{Adat} are shown except the
highest $Q$ points from Refs.~\cite{alexa99}
and\cite{arnold75}.}
\label{Adata}
\end{figure}

Several experiments have measured the structure function
$A$ at small $Q$.  Of particular note are the high
precision, 1 - 2\%  measurements from Monterey 
\cite{berard73}, Mainz \cite{simon81}, and
Saclay ALS \cite{plat90}. The only measurements at moderately
large $Q$ are from SLAC E101 \cite{arnold75}, Bonn
\cite{cramer85} and CEA \cite{elias69}, plus the two recent
JLab experiments in Halls  A \cite{alexa99} and C
\cite{abbott99}. Data for several experiments are shown in
Fig.~\ref{Adata} and summarized in Table~\ref{tab:edelast};
see Refs.\ \cite{Si01}, \cite{abbottphen00} or \cite{sick96} for 
more extensive listings of data.

 
\begin{table}
\caption{\label{tab:edelast} Some measurements of $A$.   
Symbols are given for data shown in the figures.}  
\begin{indented}
\item[]\begin{tabular}{@{}llccl} 
\br
Experiment & $Q$ (GeV) & symbol & \# of  &
Year and\\
&&& points & Reference\\
\mr
Stanford Mark III & 0.48 - 0.88$\quad$ & $\opensquare$  &
5 &  1965 \cite{buchanan65} \\ 
Orsay             & 0.34 - 0.48 & $\times$ & 4 &
1966 \cite{benaksas66} \\ 
CEA               & 0.76 - 1.15 & $\opendiamond$   &  18 & 
1969 \cite{elias69} \\
DESY              & 0.49 - 0.71 & $\;\opencircle$  & 10 & 
1971 \cite{galster71} \\
Monterey          & 0.04 - 0.14 & $\opentriangle$  & 9 &
1973 \cite{berard73} \\
SLAC E101         & 0.89 - 2.00 &
$\opensquare\!\!\!\!\!\!\:\!\:+$  & 8 &
1975 \cite{arnold75} \\
Yerevan           & 0.12 - 0.19  & {\it not shown$^*$\/} & 25 
& 1979 \cite{akimov79}\\ 
Mainz             & 0.04 - 0.39 & $\odot$   & 16 &
1981 \cite{simon81} \\
Bonn              & 0.71 - 1.14 & 
$\!\!\opensquare\!\!\!\cdot$ & 5 &
1985 \cite{cramer85} \\
Saclay ALS        & 0.13 - 0.84 & $\opentriangledown$  & 43 &
1990 \cite{plat90} \\
JLab Hall A      & 0.83 - 2.44 & $\!\fullsquare$  & 16 &
1999  \cite{alexa99} \\
JLab Hall C      & 0.81 - 1.34 & $\;\fullcircle$  & 6 &
1999  \cite{abbott99} \\
\br
\label{Adat}
\end{tabular}
\vspace*{-0.15in} 
\item[] $^*$Have larger errors and are
consistent with the other data sets.
\end{indented}
\end{table}

Fig.~\ref{Adata} reveals an unfortunate history of 
certain measurements not agreeing to within the stated uncertainties.
For example, at low $Q$ the Monterey and Mainz data overlap well, but
the overlap of Mainz and Saclay ALS data indicates problems.
The four largest $Q$ Mainz points used Rosenbluth separations, with 
$A$ largely determined from forward angles of 50$^{\circ}$, 
60$^{\circ}$, 80$^{\circ}$, and 90$^{\circ}$ at 298.9 MeV.
Saclay $A$ data were extracted from measured cross sections using 
previous $B$ data.
The closest corresponding Saclay points, for the same scattering
angles at a beam energy of 300 MeV, have cross section about 7\%
smaller; the difference is beyond the quoted experimental uncertainties.
Significant differences such as this are often obscured by semilog plots
or not plotting all data sets. 
The body of data, aside from the lowest $Q$ Orsay point, suggests the
correctness of the Saclay measurments.
Theoretical predictions span the range between the two data sets,
and do not help to determine which is correct.
Thus, a new high precision experiment in this $Q^2$ range appears desirable.

The agreement between data from CEA, SLAC E101, and Bonn
near 1 GeV was also unsatisfactory.
In discussing these measurements, we will compare to the trend of the
data as determined by the Saclay and JLab measurements.
The CEA data have large uncertainties, and are systematically
low by about 1$\sigma$.
This experiment measured scattered electrons in a shower 
counter and deuterons in a spectrometer that used a
quadrupole magnet with a stopper blocking out the central weak
field region.  In such a case it is difficult to determine
the solid angle precisely, and this uncertaintly might
introduce systematic errors into this data.
Alternatively, since the spectra
were not significantly wider than the elastic peak, it has
been suggested that over-subtraction of background was a 
problem. However, the background rates were determined to be
consistent  with expected rates from random coincidences and
target cell walls.  Bonn measured coincidence cross
sections at large electron scattering angles, 
$\theta_e$ $\approx$  80$^\circ$ - 140$^\circ$.
Using forward angle data from SLAC E101, CEA, and Orsay,
Bonn determined $A$ and $B$. 
Slightly inconsistent results from the other experiments
led to a small uncertainty on the Bonn determination of $A$.
Thus, it is only the largest $Q^2$ point, for which there was only the
large angle Bonn data, that has very significant disagreement
with other determinations of $A$. 
Finally, the lowest $Q^2$ SLAC point is high. 

The disagreements between the CEA, SLAC E101, and Bonn data were part
of the motivation for two JLab experiments that determined A.
Hall A experiment E91-026 \cite{alexa99} measured $A$ for $Q^2$ 
from 0.7 to 6.0 GeV$^2$.
Hall C experiment E94-018 \cite{abbott99} measured $A$ 
in the same kinematics as its $t_{20}$ points, from
0.7 to 1.8 GeV$^2$.
The main advantages of these experiments over previous work 
include the continuous beam, large luminosities, and modern
spectrometers. The Hall A measurements \cite{alexa99} used
$>$ 100 $\mu$A beams on a 15 cm cryogenic
LD$_2$ target, to achieve a luminosity of 
approximately 5 $\times$ 10$^{38}$/cm$^2$/s,
and two approximately 6 msr spectrometers.
The Hall C measurements used the HMS spectrometer along with 
the deuteron channel built to measure $t_{20}$ with the recoil
polarimeter POLDER. A feature of this system is that the solid
angles of the two spectrometers were well matched, to within a
few percent. 

In the overlap region, the two JLab experiments show better
precision than the earlier data and generally good agreement;
comparisons of theory to data should focus on these results, rather than
the older data.
However, these measurements also show a significant
disagreement with each other.
Uncertainties in each experiment are dominated by systematics of 
approximately 5 - 6\%,
with statistical precisions near 1\%.
The Hall C data are systematically larger than the Hall A
data by just over 2$\sigma$, slightly over 10\%,
and there appears to be a tendency of the data sets to diverge with
increasing momentum transfer.
This discrepancy will be decreased by a few percent, but not eliminated, 
by a correction \cite{GVO01} to a lower, more accurate,
beam energy in Hall C during the experiment.
It is unclear if the discrepancy can be further resolved.

An important experimental point is the use in these experiments,
and in many earlier ones, of $ep$ elastic scattering
to calibrate the solid angle acceptance; a 
fit to the world $ep$ cross section data is often used \cite{bosted95}.
However, recent high precision polarimetry results
\cite{jones99,gayou01,gayou02}
imply that $G_E^p/G_M^p$ is significantly smaller
than previously believed, with $G_E^p/G_M^p$ dropping
nearly linearly for $Q^2$ from about 0.5 to 5.6 GeV$^2$.
Refitting the world cross section data, with the JLab
data for the form factor ratio, decreases $G_E^p$ but
enhances $G_M^p$ by about 2\% \cite{jonespc}.
The new fits imply that the $ep$ cross section is generally a few percent
larger than would have been calculated previously, 
less than the systematic uncertainties
of most experiments, and too small to affect comparisons
of measurements of the $ed$ cross sections and $A$.
The effects on the theoretical deuteron form factor predictions
will be addressed below.

In summary, the structure function $A$ is reasonably well 
determined up to $Q^2$ $=$ 6 GeV$^2$, if one neglects several
poorer data points.
There remain regions in which there are up to about 10\%
systematic discrepancies between data of different experiments; 
the resolution of these 
problems is at present unclear. 

\subsubsection{ Experimental status of $B$ }   

\begin{table}
\caption{
\label{tab:edelastb} 
Some measurements of $B$. 
Symbols are given for data shown in the figures.}  
\begin{indented}
\item[]\begin{tabular}{@{}llccl} 
\br
Experiment & $Q$ (GeV) & symbol & \# of  & Year and\\ &&& points & 
Reference\\  \mr
Stanford Mark II  & 0.10 - 0.13$\quad$ & $^*$  & 2 &  
1964   \cite{goldemberg64} \\  
Stanford Mark III & 0.48 - 0.68 & $\opensquare$ & 4 & 1965
  \cite{buchanan65} \\ 
Orsay             & 0.20 - 0.28 & $^*$  &  3 & 
  1966 \cite{grossetete66} \\ 
Orsay             & 0.34 - 0.44 & $^*$  &
  3 &  1966 \cite{benaksas66} \\ 
Stanford Mark III & 0.44 - 0.63 &
  $^*$   & 5 & 1967 \cite{rand67} \\ 
Orsay             & 0.14 - 0.48 & $^*$
    & 4 & 1972 \cite{ganichot72} \\ 
Naval Research Lab & 0.11    &  $^*$ & 1 & 1980
  \cite{jones80}\\  
Mainz             & 0.25 - 0.39 & $\;\opencircle$  & 4 & 1981
  \cite{simon81} \\ 
Bonn              & 0.71 - 1.14 & $\opentriangledown$
  & 5 & 1985 \cite{cramer85} \\ 
Saclay ALS        & 0.51 - 1.04 & $\opendiamond$  & 13 &
  1985 \cite{auffret85} \\ 
SLAC NPAS NE4     & 1.10 - 1.66
& $\opensquare\!\!\!\!\!\!\:\!\:+$  &  9
  & 1987  \cite{arnold87,bosted90} \\ 
JLab Hall A      & 0.7 - 1.4 &
  $\!\fullsquare$  & 6 & unpublished  \\
\br
\end{tabular}
\item[] $^*$These data sets are not shown ($B$ must
be inferred from the publication).
\end{indented}
\end{table}

The highest $Q^2$ measurements of the $B$ structure 
function come from SLAC NPAS experiment NE4
\cite{arnold87,bosted90},  which covered the $Q^2$ range
of 1.20 to 2.77 GeV$^2$. These measurements extended the
range of previous data from Saclay \cite{auffret85} (which went
to 1.1 GeV$^2$),  and from Bonn \cite{cramer85} (which went to
1.3 GeV$^2$, and gave the results for $A$ discussed
above).
There is good overlap in all but a few of the earliest 
$B$ measurements. Measurements of $B$ were taken as part
of E91-026 for $Q^2$ $=$ 0.7 to 1.4 GeV$^2$, but analysis
is not yet final.

\subsubsection{ Experimental status of polarization measurements }

A summary of the world data is shown in Table~\ref{t20tab}.
The first polarization measurements were from 
an Argonne/Bates recoil polarimeter experiment
\cite{schulze84} and  a Novosibirsk VEPP-2 experiment
\cite{dimitriev85,wojtse86} using a polarized gas jet target.
In the gas jet experiment, a polarimeter measured the 
gas polarization after it passed through the interaction region.

\fulltable{\label{t20tab}
World data for tensor polarization observables.}
{\small
\begin{tabular}{@{}lllllll}
\br
Experiment  & Type & $Q$ (GeV) & Observables & Symbol & \# of  &
Year and  \\
            &      & &       & &  points   &  Reference  \\
\mr
Bates              & polarimeter  & 0.34, 0.40    & $t_{20}$
   & $\quad\times$ & 2 & 1984 \cite{schulze84} \\
Novosibirsk VEPP-2 & atomic beam  & 0.17, 0.23   & $T_{20}$
   & $\quad\opensquare\!\!\!\!\!\!\:\!\:+$  & 2 & 1985
\cite{dimitriev85,wojtse86} \\ Novosibirsk VEPP-3 & storage cell
& 0.49, 0.58 & $T_{20}$ 
   & $\quad\triangle$ & 2 & 1990 \cite{gilman90} \\ 
Bonn               & polarized target  &  0.71        & $T_{20}$
   & $\quad\opendiamond$  & 1 & 1991
\cite{boden91} \\ Bates              & polarimeter  & 0.75 -
0.91 & $t_{20}$,$t_{21}$,$t_{22}$ 
   & $\quad\fullcircle$ & 3 & 1991 \cite{the91,garcon94} \\
Novosibirsk VEPP-3 & storage cell & 0.71        & $T_{20}$ 
   & $\quad\,^*$ & 1 & 1994
\cite{popov94}
\\  NIKHEF             & storage cell & 0.31    &
   $T_{20}$,$T_{22}$
   & $\quad\opencircle$ & 1 & 1996 \cite{ferro96} \\
NIKHEF             & storage cell & 0.40 - 0.55 & $T_{20}$ 
   & $\quad\opensquare$ & 3 & 1999 \cite{bouwhuis99} \\
JLab Hall C 94-018 & polarimeter & 0.81 - 1.31 & $t_{20}$,$t_{21}$,$t_{22}$
   & $\quad\fullsquare$ & 6 & 2000 \cite{abbott00} \\
Novosibirsk VEPP-3 & storage cell & 0.63 - 0.77   & $T_{20}$
   & $\quad\odot^{**}$ & 5 & 2001 \cite{nikolenko01} \\ 
\br
$^*$ Not shown in the figures.\\
$^{**}$ Preliminary data.
\end{tabular}}
\endfulltable 
 
There were three second generation experiments.
An Argonne/Novosibirsk VEPP-3 measurement \cite{gilman90} 
pioneered the use of storage cells, increased the internal
target density about a factor of 15 over the gas jet alone, and
pushed out to 0.58 GeV, near the minimum in
$t_{20}$. Because the polarization of the gas varies in
the cell, due to  wall and beam interactions, it was
decided to normalize the gas polarization by setting the
lowest $Q$ datum, at 0.39 GeV, to theory where
the uncertainties are small. Such internal targets in
storage rings are now common. A Bonn polarized target
experiment \cite{boden91} had large  uncertainties. At
Bates, the AHEAD deuteron polarimeter was used
\cite{the91,garcon94} to determine $t_{20}$ in the range
just past the minimum of $t_{20}$ to just past the node in
$G_C$. A continuation of the Novosibirsk experiment had large
uncertainties \cite{popov94}, and was never published.
Note that, to facilitate comparison between different
experiments,  the data are often ``corrected'' to an electron 
scattering angle of 70$^{\circ}$, but this adjustment 
and the uncertainty it introduces are small. 
 
Over the past several years, internal target experiments 
at NIKHEF \cite{ferro96,bouwhuis99} and 
Novosibirsk \cite{nikolenko01}
have improved the precision of the lower ${Q^2}$ data, 
over a range of $Q\approx$ 0.3 - 0.8 GeV.
The improvements in Novosibirsk include higher luminosity
resulting from an improved atomic beam source
and a modified beam tune that allows use of a higher impedance 
storage cell.
JLab Hall C E94-018 \cite{abbott00} used the recoil polarimeter POLDER to 
measure to the highest $Q^2$, 1.72 GeV$^2$.

The overlap of the data is good, but apparent systematic shifts can be
seen, as the NIKHEF and Bates measurements are more negative 
than the JLab and Novosibirsk measurements;
note that this is not a difference between polarized targets 
and recoil polarimeters.
The issue of determing at what $Q^2$ $G_C$ $=$ 0 is affected by this difference.
The Bates data \cite{the91,garcon94} suggest
a larger $Q^2$ than do the Novosibirsk \cite{nikolenko01} and
JLab \cite{abbott00} data.
The fits of Ref.~\cite{abbottphen00} do not include 
the unpublished Novosibirsk data \cite{popov94,nikolenko01},
and average between the Bates  and JLab points.

We do not discuss the data for $t_{21}$ and $t_{22}$.
Because of their dependence on $G_M$, they have not been
as useful in providing new information as has $t_{20}$.
To test time reversal invariance, one measurement of the {\em
induced} vector polarization was made \cite{prepost68}.
The observed result was consistent with zero.


\subsection{Nonrelativistic calculations without interaction 
currents}

 
\subsubsection{Theory}

It is instructive to see how the deuteron form factors
are related to the free nucleon form factors and the
deuteron wave function in the nonrelativistic limit.
Because the deuteron is an isoscalar target, only the isoscalar nucleon form
factors 
\begin{eqnarray}
G_{E}^s &=& G_{E}^p + G_{E}^n \nonumber\\
G_{M}^s &=& G_{M}^p + G_{M}^n \label{nnff}
\end{eqnarray}
will contribute to the form factors.  In the nonrelativistic theory, {\it
without exchange currents or $(v/c)^2$ corrections\/}, the deuteron form
factors are 
\begin{eqnarray}
G_C  &=& G_{E}^sD_C \nonumber\\
G_Q  &=& G_{E}^sD_Q \nonumber\\
G_M  &=& \frac{m_d}{2m_p}\Bigl[G_{M}^s D_M + G_{E}^s D_E\Bigr]\, ,
\label{body}
\end{eqnarray}
where the {\it body form factors\/} $D_C$, $D_Q$, $D_M$, and $D_E$ are all
functions of $Q^2$.  If we choose to evaluate these in the Breit frame,
defined by 
\begin{eqnarray}
\fl q_0=\nu=0 \, ,\quad
D_0=\sqrt{m_d^2+{\textstyle{1\over 4}}{\bf q}^2}\, ,\quad
d^\mu=\left\{D_0, -{\textstyle{1\over 2}}{\bf q}\right\}\, ,
\quad d'^\mu =\left\{D_0,\;\; {\textstyle{1\over 2}}{\bf
q}\right\} \, ,
\end{eqnarray}
then the relativistic and nonrelativistic momentum transfers 
are identical, $Q^2={\bf q}^2$, and the relativistic nucleon
form factors can be used without corrections.  Note that, in
this nonrelativistic limit, only the nucleon electric form
factors contribute to the deuteron charge and quadrupole
structure, while both nucleon form factors contribute to the
deuteron magnetic structure.

The nonrelativistic formulae for the body form factors $D$ 
involve overlaps  of the wave functions, weighted by spherical
Bessel functions
\begin{eqnarray}
D_C(Q^2) &=& \int_0^{\infty} dr \Bigl[u^2(r) + w^2(r)\Bigr]
j_0\left(\tau \right)\nonumber\\ 
D_Q(Q^2) &=& { {3}\over{\sqrt{2}\eta} } 
\int_0^{\infty} dr\; w(r) \left[u(r) - { {w(r)}\over{\sqrt{8}} } \right] 
j_2\left(\tau \right)\nonumber\\
D_M(Q^2) &=&  \int_0^{\infty} dr \,\Bigl[ 2\,u^2(r) - w^2(r) \Bigr] j_0(\tau)
+ \Bigl[\sqrt{2}\,u(r)w(r) + w^2(r)\Bigr] j_2(\tau)\nonumber\\
D_E(Q^2) &=& { {3}\over{2} }  \int_0^{\infty} dr\, w^2(r)\,
\Bigl[ j_0(\tau) + j_2(\tau) \Bigr] \label{bodyff}
\end{eqnarray}
where $\tau=qr/2=Qr/2$.  At $Q^2=0$, the body form factors 
become
\begin{eqnarray}
D_C(0) &=& \int_0^{\infty} dr \Bigl[u^2(r) + w^2(r)\Bigr]=1\nonumber\\ 
D_Q(0) &=& { {m_d^2}\over{\sqrt{50}} } 
\int_0^{\infty} r^2\,dr\; w(r) \left[u(r) - { {w(r)}\over{\sqrt{8}} } \right] 
\nonumber\\
D_M(0) &=&  \int_0^{\infty} dr \,\Bigl[ 2\,u^2(r) - w^2(r) \Bigr]=2-3P_D
\nonumber\\ 
D_E(0) &=& { {3}\over{2} }  \int_0^{\infty} dr\, w^2(r)= { {3}\over{2} }
P_D
\end{eqnarray}
giving the nonrelativistic predictions
\begin{eqnarray}
Q_d &&=D_Q(0)\nonumber\\
\mu_d&&=\mu_s\,(2-3\,P_D)+1.5\,P_D= 1.7596-1.1394\,P_D\, ,
\end{eqnarray}
with $\mu_s$ = 0.8798 the isoscalar nucleon magnetic moment.
The experimental value of the deuteron magnetic moment (in these 
units) is 1.7139, leading to a predicted $D$-state probability of
$P_D=4.0\%$.  However this estimate cannot be taken too seriously
because the magnetic moment is {\it very sensitive to relativistic
corrections and interaction currents\/} which can easily alter this
result significantly.  These contributions will be reviewed
qualitatively later in this review.

\begin{figure}[t]
\begin{center}
\mbox{
   \epsfxsize=4.0in
\epsffile{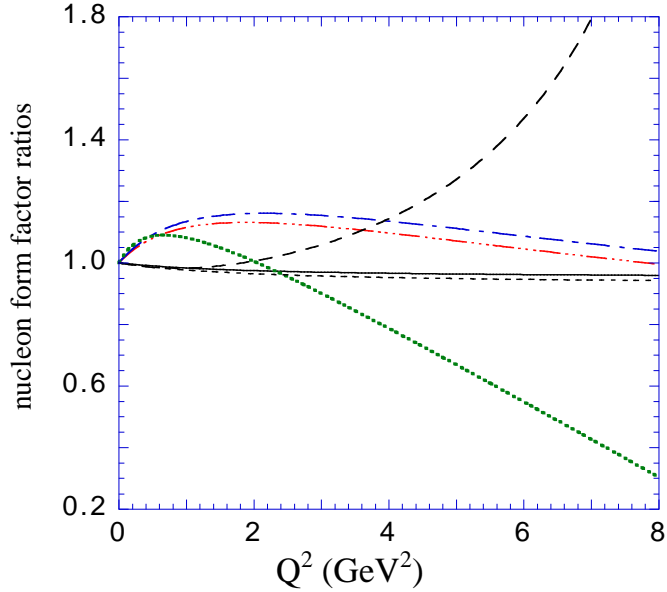} 
} 
\end{center}
\caption{Ratios of the nucleon isoscalar form factors: 
$G^s_{E}$(MMD)$/F_D$ (triple-dot-dashed line), $G^s_{M}$(MMD)/
$(\mu_sF_D)$ (dot-dashed line), $\mu_s G^s_{E}$(MMD)/$G^s_{M}$(MMD)
(Solid line),
$G^{s\;\rm JLab}_{E}/F_D$ (dotted line), $F_1^s/G^s_E$ (MMD) (short
dashed line), and $F_1^s/G^s_E$ (JLab) (long dashed line). } 
\label{Ges-study}
\end{figure}

The study of deuteron form factors is complicated by the fact 
that they are a {\it product\/} of the {\it nucleon
isoscalar\/} form factors, $G^s$, and the {\it body\/} form
factors, $D$.  The dependence of the deuteron form  factors on
older models of the nucleon form factors is well discussed in
Ref.~\cite{GVO01}. A year ago the  model of Mergell, Meissner
and Drechsel \cite{MMD} (referred to as MMD) gave a good fit,
and could have been adopted as a standard.  Figure
\ref{Ges-study} shows the MMD isoscalar electric and magnetic form factors
divided by the familiar dipole form factor   
\begin{equation}
F_D(Q^2)= {\displaystyle\left(1+{Q^2\over0.71}\right)^{-2}}
\label{dipole}
\end{equation} 
(with $Q^2$ in GeV$^2$).  Note that the MMD model does not 
differ by more than 20\% from the dipole over the entire $Q^2$
range, suggesting that the dipole approximation works very well
(on the scale of the experimental errors -- see below). 
However, recent measurements of the proton charge form factor
are producing a surprising result, and at the time this review
was being completed the picture was begining to change.

The recent JLab measurements of both the neutron and proton 
charge form factors now suggest that the isoscalar charge form
factor may be well approximated by   
\begin{equation}
G^{s\,\rm JLab}_{E}(Q^2)=\left\{{1.91\tau\over(1+5.6\tau)}
+(1.0-0.1262\,Q^2) \right\}F_D(Q^2)\, . \label{JLabges} 
\end{equation} 
where $F_D$ is the dipole form factor and $\tau = Q^2 / 4 m_p^2$.
This JLab model is a 
sum of the old Galster \cite{galster71} fit for the nucleon
charge form factor (supported by the recent
measurements \cite{Genmeas}) and a linear approximation to the
new JLab $G^p_E/G^p_M$ data \cite{jones99} (from which the
charge form factor is obtained by assuming that
$G^p_M=\mu_pF_d$).   Figure
\ref{Ges-study} shows that this form factor differs significantly from
the dipole (and also the previously favored MMD model), and may have a
significant affect on the theoretical interpretation of the data.  This
will be discussed in Sec.~\ref{thyandexp} below.  [The $F_1/G_E$ ratios
shown in the figure will be discussed in Sec.~\ref{Hexamples}
below.]         
 
%

\begin{figure}[t]
\begin{center}
\mbox{
   \epsfxsize=6.0in
\epsffile{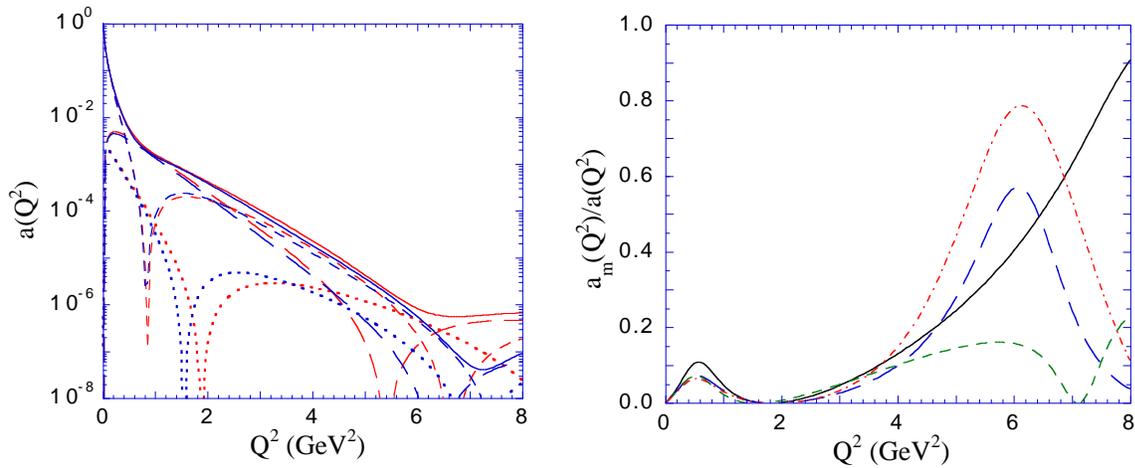}  
}
\end{center}
\caption{Plots of $a(Q^2)$ and the ratio $a_m(Q^2)/a(Q^2)$ discussed in the
text.  Left panel shows contributions for models AV18 and
IIB-nonrelativistic: $a$ (solid lines), $a_q$ 
(long dashed lines), $a_c$ (short dashed lines), $a_m$ (dotted
lines).  In these plots the ratio $G^s_M/G^s_E$ is fixed at
$\mu_s$.  In each case IIB decreases more rapidly than AV18
at low $Q^2$.   Right panel: IIB-RIA (solid line), AV18 (long
dashed line),  IIB-nonrelativistic (short dashed line), all with
$G^s_M/G^s_E=\mu_s$, and AV18 with the ratio of $G^s_M/G^s_E$ given by
Eq.~(\ref{JLabges}) (dot-dashed line).}
\label{a-study}
\end{figure}

Dividing the individual factors
$A_C$, $A_Q$, and $A_M$ [introduced in Eq.~(\ref{AandB})] by $(G^s_E)^2$
gives reduced quantities that are (except for the weak dependence on the
ratio of $G_{Ms}/\mu_sG_{Es}$) independent of the choice of nucleon
form factor.  The contribution of these reduced quantities, which
we denote by $a_C$, $a_Q$, and $a_M$, to the total
$a=A/(G^s_E)^2$ is shown in  Fig.~\ref{a-study}.  The figure shows
that the contribution of the magnetic term, $a_M$, to the total $a$
is small for $Q^2<
4$ GeV$^2$ (for most of the $Q^2$ range it is less than a few
percent, reaching 10\% at $Q^2\simeq0.5$ and also near 4 GeV$^2$). 
Above 
$Q^2$ of 4 GeV$^2$ it is larger, and very model dependent.  This
justifies the observation that the $A$ structure function can be well
approximated by $\tilde A$, as stated eariler in
Eq.~(\ref{tildeAapprox}).  [Note that the new JLab data for $G^p_E$,
discussed briefly above, may enhance the magnetic contributions to $A$
above 4 GeV$^2$, but will not change these conclusions qualitatively].

\begin{figure}
\begin{center}
\mbox{
   \epsfxsize=6.3in
\epsffile{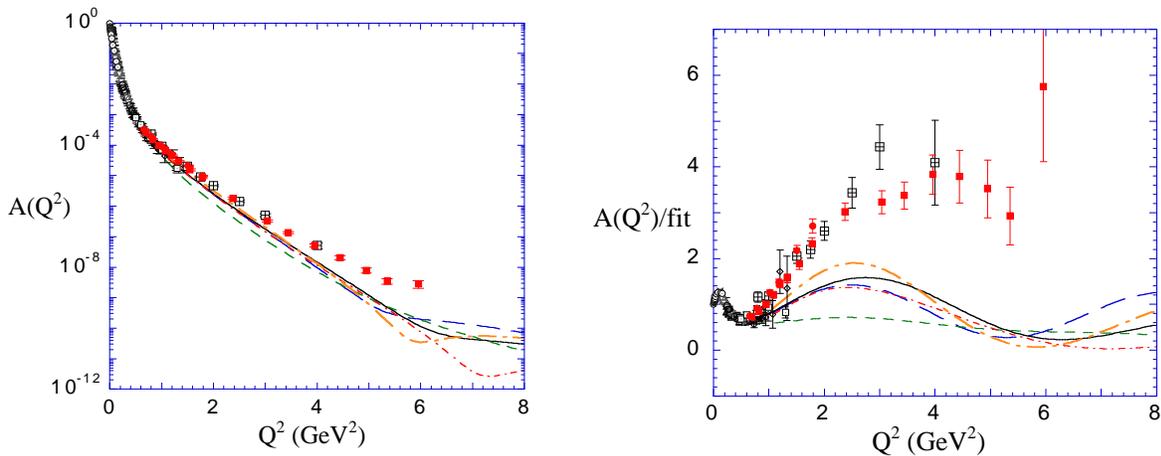} 
} 
\end{center}
\caption{The structure function $A$ for the five nonrelativistic
models discussed in the text, calculated using the MMD
nucleon form factors.  The models are labeled as in Figs.~[1]
and [2].  The left panel shows data and models divided by a
``fit'' described in the text. 
See Table~\ref{Adat} for references to the data.
}
\label{AhighQ2}
\end{figure}

\begin{figure}[t]
\begin{center} 
\mbox{
   \epsfxsize=5.5in
\epsffile{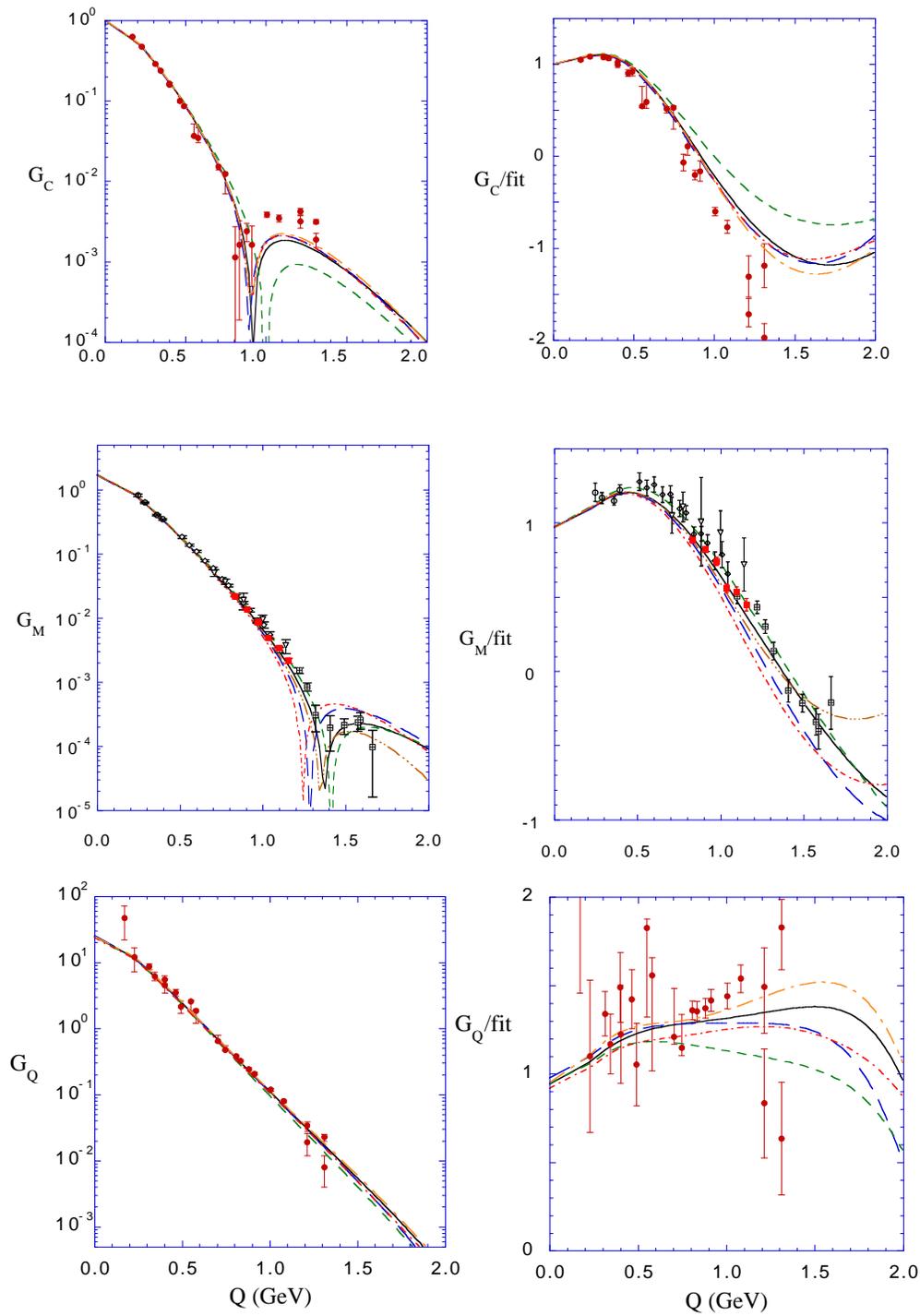} 
}
\end{center}
\caption{Deuteron form factors for the five nonrelativistic 
models  compared to data.  The data for $G_C$ and $G_Q$ are
from analysis of the complete $A$ and
$t_{20}$ data sets \cite{abbottphen00}.  The data for $G_M$
were extracted from the experimental measurements of $B$,
referenced in Table \ref{tab:edelastb}.  MMD nucleon form
factors have been used with the nonrelativistic models. }
\label{Gfactors}
\end{figure}

\begin{figure} 
\begin{center}
\vspace*{-0.1in}
\mbox{
   \epsfxsize=3.5in
\epsffile{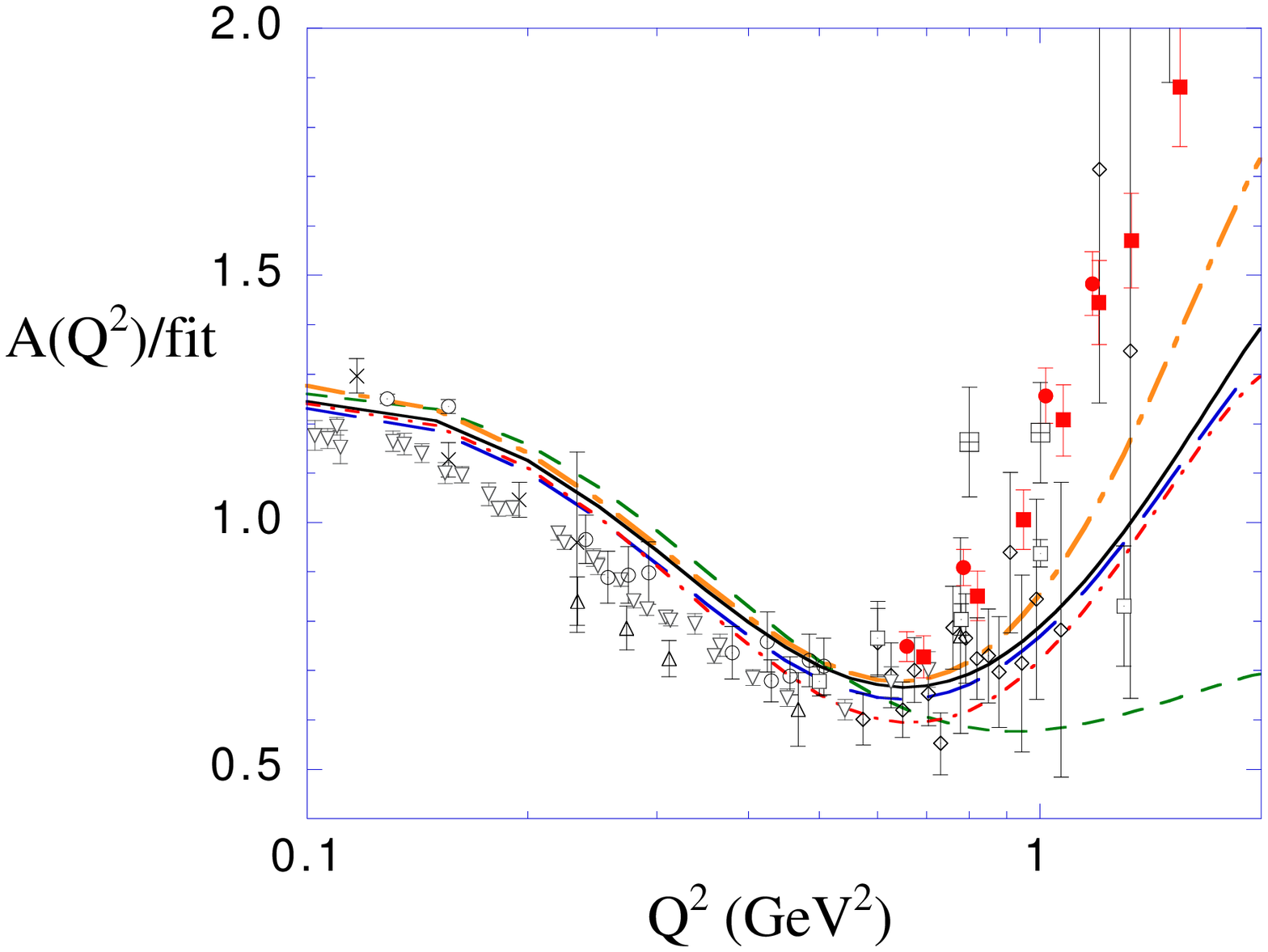} 
} 
\end{center}
\caption{The structure function $A$ for the five nonrelativistic
models discussed in the text.  The models are labeled as in
Figs.~[1] and [2].  
See Table~\ref{Adat} for references to the data.
}
\label{AlowQ2}
\end{figure}

\subsubsection{Comparison to data}
\label{compareNR}
 
How well does this simple nonrelativistic theory explain the 
data? The high $Q^2$ data for $A$ provide the most stringent
test.   In Fig.~\ref{AhighQ2} we compare the data for
$A$ with calculations using the five nonrelativistic wave 
functions shown in Figs.~\ref{uwr} and \ref{uwp}.  The
calculations use Eq.~(\ref{body}) with MMD isoscalar nucleon
form factors and nonrelativistic body form factors given in
Eq.~(\ref{bodyff}).  In the right panel the data and models have
been divided by the ``fit'' described in Eq.~(\ref{scale})
below.    

It is easy to see that the nonrelativistic models {\it are a 
factor 4 to 8 smaller than the data for $Q^2>
2$ GeV$^2$\/}.  Furthermore,
since the  difference between different deuteron models is
substantially smaller than this discrepancy, it is unlikely 
that any {\it realistic\/} nonrelativistic model can be found
that will agree with the data.  If the nucleon isoscalar charge
form factor were larger than the MMD model by a factor of 2 to
3 it might explain the data, but this is also unlikely since
the variation between nucleon form factor models is
substantially smaller than this.  [If we use the fit
Eq.~(\ref{JLabges}) to the JLab $G_{Ep}$ measurements the
discrepancy will be even larger.]  We are forced to conclude
that these high $Q^2$ measurements {\it cannot be explained by
nonrelativistic physics and present very strong evidence for
the presence of interaction currents, relativistic effects, or
possibly new physics.\/}

A detailed comparison of the nonrelativistic models with the 
three deuteron form factors, $G_C$, $G_M$, and $G_Q$ is given in
Fig.~\ref{Gfactors}.  The functions used to scale the data and theory in the
right-hand panels of the figure are
\begin{eqnarray}
G_C&&=e^{-Q^2/3.5}\;\left(1+{Q^2\over
m\epsilon}\right)^{-1}\left(1+{Q^2\over0.71}\right)^{-2}\nonumber\\
G_M&&=1.7487\,e^{-Q^2/2.5}\;\left(1+{Q^2\over
m\epsilon}\right)^{-1}\left(1+{Q^2\over0.71}\right)^{-2}\nonumber\\
G_Q&&=\frac{25.8298}{1.01}\left(e^{-Q^2}+0.01\,e^{-Q^2/100}\right)\left(1+{Q^2\over
m\epsilon}\right)^{-1}\left(1+{Q^2\over0.71}\right)^{-2}\, .
\label{scale}
\end{eqnarray}
where $Q^2$ is in GeV$^2$ and $m\epsilon=0.936\times0.0022246/0.197^2$. 
While some of the factors in these expressions are 
theoretically motivated (note the presence of the dipole form
of the nucleon form factor) we do  not attach {\it any\/}
theoretical significance to these functional forms; they merely
provide a reasonably simple way to scale out the rapid
exponential decreases from the form factors.  Figure
\ref{Gfactors} shows that the nonrelativistic models do a good
job of predicting the form factors to a momentum transfer
$Q\simeq0.5$ GeV, beyond which departures  from the data and
variations of the models make the agreement increasingly
unsatisfactory.\footnote
{
We are not inclined to take the
discrepancy between theory and the first $G_Q$ point
seriously; kinematic factors make it difficult to extract
this point accurately and it is only one standard
deviation from the theory. 
The large $G_Q$ and small $G_C$ values for the  points at
0.55 GeV and 0.58 GeV result from the
$t_{20}$  data points, from \cite{bouwhuis99} and
\cite{gilman90} respectively, being about 1 standard
deviation more negative than calculations and overlapping
the negative limit for $t_{20}$ of $-\sqrt{2}$.  Note that
the tabulated uncertainty of $G_Q$ at $Q=$ 0.55 GeV in
Ref.~\cite{abbott00} should be asymmetric, +0.075/$-$0.713,
(as shown in the Fig.\ \ref{Gfactors}).} 
 
However, careful comparison reveals that there are still
(small) discrepancies between the data and the nonrelativistic
theory,  even at low $Q^2$.  The data and curves from the lower
panel of Fig.~\ref{AhighQ2} are shown on an expanded
logarithmic scale  in Fig.~\ref{AlowQ2}.  In the lowest $Q^2$
range from about 0.15 to 0.4 GeV$^2$ the data lie {\it
below\/} the nonrelativistic theory, and are larger than the
nonrelativistic theory only for $Q^2$ above 1 GeV$^2$.  The
very low $Q^2$ discrepancy seems to be due in part (but not
entirely) to the Columb distortion corrections that have been
used recently to explain the deuteron radius \cite{sick96}.  We
will discuss these corrections in Sec.~\ref{nucsec} below.

Before we turn to a detailed discussion of the
possible explanations for the failure of nonrelativistic
models to explain the form factors at high $Q^2$, we discuss the
low momentum transfer results from the perspective of effective
theories.


\subsection{Effective field theory}
\label{sec:EFT}

The recent development of effective field theory
provides a powerful method for theoretical study of low $Q^2$
physics. We will briefly review these results here, and return
to the discussion of the high
$Q^2$ results in the next section.   

Effective field theory techniques exploit the fact that the
physics at low energies $E<\!\!<M_0$ (or large distances
$\lambda>\!\!>\lambda_0=1/M_0$) cannot be sensitive to the 
{\it details\/} of the interactions at very high energies 
$E>\!\!>M_0$ (or short distances).  For example, a low energy
long wavelength probe may detect the presence of a
small scattering center, but cannot resolve its
structure (much as the far-field of a collection of electric
charges depends on only one parameter, the total charge).  The
parameters that depend on the short-range physics may be
very important, but they cannot be calculated and must be
determined by a fit to the data.  
 
Effective field theory works best if the distance scales of the
(unknown) short-range physics and the (known) long-range physics
are clearly separated.  Then for energies well below the scale
of the short-range physics (which we take to be $M_0$), the
short-range physics is treated systematically by  expanding in
powers of $E/M_0$.  In applications to the $NN$ system, two
scales have been discussed.  The so-called ``pionless
theory'' chooses $M_0\sim m_\pi$, and therefore requires no
theory of the $\pi N$ interaction.  This approach can work only
at {\it very\/} low energies.  The chiral theory chooses
$M_0\sim m_\rho$ and attempts to describe $NN$ scattering up to
the $\rho$ mass scale using the known pion-nucleon interaction
as given by chiral symmetry.  (More precisely, if the magnitude
of the center of mass relative momentum $|{\bf p}|<M_0/2$, the
nucleon lab kinetic energy will $E_{\rm lab}<M_0^2/2m$, which
is $E_{\rm lab}<10$ MeV for the pionless theory, and $E_{\rm
lab}<320$ MeV for chiral perturbation theory.)  

The effective range theory introduced by Bethe
\cite{Bethe49} is an early version of what we now call
the pionless effective theory.  Weinberg \cite{weinberg90} first
applied modern chiral perturbation theory to $NN$ scattering.  He
proposed making a chiral expansion of the $NN$ potential,
and then inserting this potential into the Schr\"odinger
equation.  Later Kaplan, Savage, and Wise (KSW) \cite{KSW98}
criticized the consistency of this approach, and introduced an
alternative organizational scheme, sometimes referred to as $Q$
counting, in which the pion interaction is to be included as a
perturbative correction (as opposed to including it as part of
the potential, and counting it to all orders, as proposed by
Weinberg).  KSW applied this method to calculation of the
deuteron form factors \cite{KSW99}. It is now known that
the tensor part of the one pion exchange interaction is too
strong to be treated perturbatively, and recent work has focused
on how to include the singular parts of one pion exchange in the
most effective manner \cite{PRS00,phillips00}.  In the
following discussion we review the recent results from
Phillips, Rupak, and Savage (PRS), who give a nice account of
the calculations of the deuteron form factors in a pionless
theory \cite{PRS00}.

The effective Lagrangian density for a pionless effective theory
of the $NN$ interaction in any channel (the coupled $^3S_1-^3D_1$
for example) is
\begin{eqnarray}
{\cal L}=&-C_0 \left(\psi^{\rm T}  \,\psi\right)^\dagger
\left(\psi^{\rm T} \,\psi\right)\nonumber\\
& -{\textstyle{1\over2}}\,C_2
\left[\left(\psi^{\rm T}  \,\psi\right)^\dagger
\left(\nabla^2\psi^{\rm T}  \,\psi + \psi^{\rm T} 
\,\nabla^2\psi -2\vec{\nabla} \psi^{\rm T} \cdot
\vec{\nabla} \psi \right) + {\rm h.c.} \;
\right] \nonumber\\
&+ \cdots \, , \label{EFT1}
\end{eqnarray}
where $\psi$ is a (nonrelativistic) nucleon field operator and
$C_0$, $C_2$, and the general coefficient $C_{2n}$ (which fixes
the strength of the terms with $2n$ derivatives) are determined
from data.  The coefficients $C_{2n}$ parameterize the
strength and shape of the short range interaction.  The
scattering amplitude predicted by (\ref{EFT1}) is a sum of
bubble diagrams which can be regularized using the KSW
dimensional regularization scheme with power law divergence
subtraction \cite{KSW98}.  In lowest order (LO) this bubble sum
is
\begin{eqnarray}
{\cal M}&= {-C_0(\mu)\over
1+{\displaystyle {m\over4\pi}}C_0(\mu)(\mu + ip)} =
{4\pi\over m}\left({1\over p\cot\delta_0-ip}\right)\nonumber\\
&= -{4\pi\over m}\left({Z_d\over \gamma+ip}\right) +R(p) 
\end{eqnarray}
where $p$ is the magnitude of the nucleon three-momentum in the
c.m.\ system,  $R(p)$ is regular at the pole $p=i\gamma$,
$\gamma=\sqrt{m\epsilon}$ with $\epsilon$ the deuteron binding
energy, $Z_d$ is related to the asympotic normalization of the
deuteron wave function, and the dependence of $C_0$ on the
(arbitrary) renormalization point $\mu$ is dictated by the
requirement that the overall result be {\it independent\/} of
$\mu$.  The LO result is 
\begin{equation}
C_0(\mu)={4\pi\over m}\left({1\over \gamma-\mu}\right)\, ,\qquad
Z_d=1\, ,\qquad R(p)=0\, . \label{zeq}
\end{equation}  
In terms of the effective range expansion 
\begin{equation}
p\cot\delta_0 = -{1\over a} + {\textstyle{1\over2}}\rho_d
(p^2+\gamma^2) + w_2 (p^2+\gamma^2)^2 + \cdots
\end{equation}  
with $a$ the scattering length and $\rho_d$ the effective range,
the LO calculation gives
\begin{equation}
a=1/\gamma\, , \qquad\rho_d = 0\,. \label{req}
\end{equation}

Contributions from the next to leading order (NLO) term
$C_2$ changes the relations in (\ref{zeq}) and (\ref{req}); in
particular, the wave function renormalization constant
$Z_d$ begins to differ from unity and the effective range
$\rho_d$ to differ from zero.  PRS point out that the most
stable results are obtained by constraining $C_0$ and $C_2$ to
give the experimental values of the deuteron parameters
$\gamma^{-1}=4.319$ fm and $Z_d-1=0.690$ instead of $\gamma$
and $\rho_d=1.765$ fm.  This is because the asympotic deuteron
wave function is fixed by $\gamma$ and $Z_d$ 
\begin{equation}
\psi_{\rm tail}(r) = \sqrt{{\gamma Z_d}\over{2\pi}}
\,{e^{-\gamma r}\over{r}} \label{asywave}
\end{equation}
and {\it it is the wave function and not the scattering\/} that
largely determines the deuteron form factors and other
electromagnetic observables. 

\begin{figure} 
\mbox{ \epsfxsize=2.5in
\epsffile{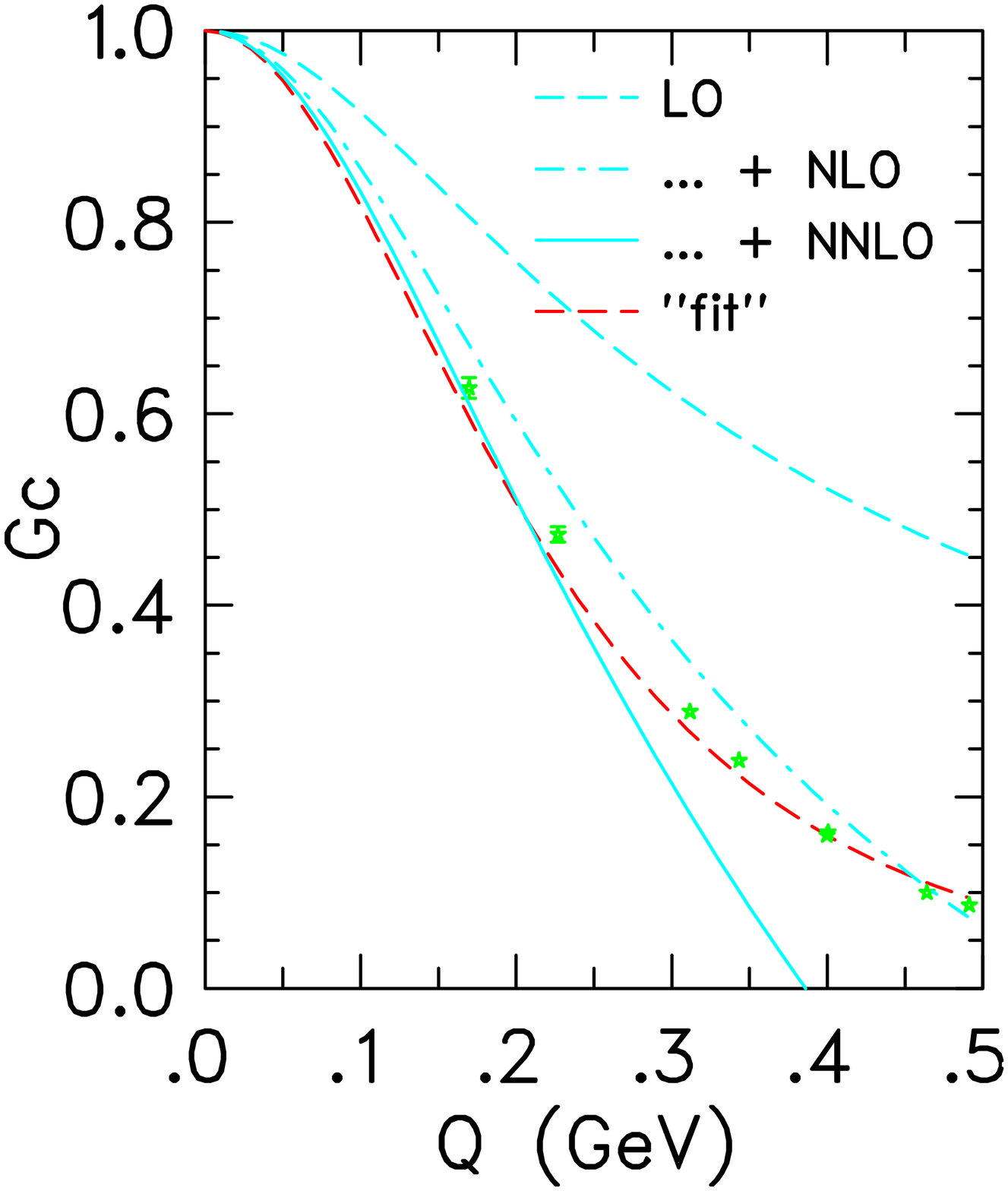}}$\qquad\qquad$
\mbox{ \epsfxsize=2.5in
\epsffile{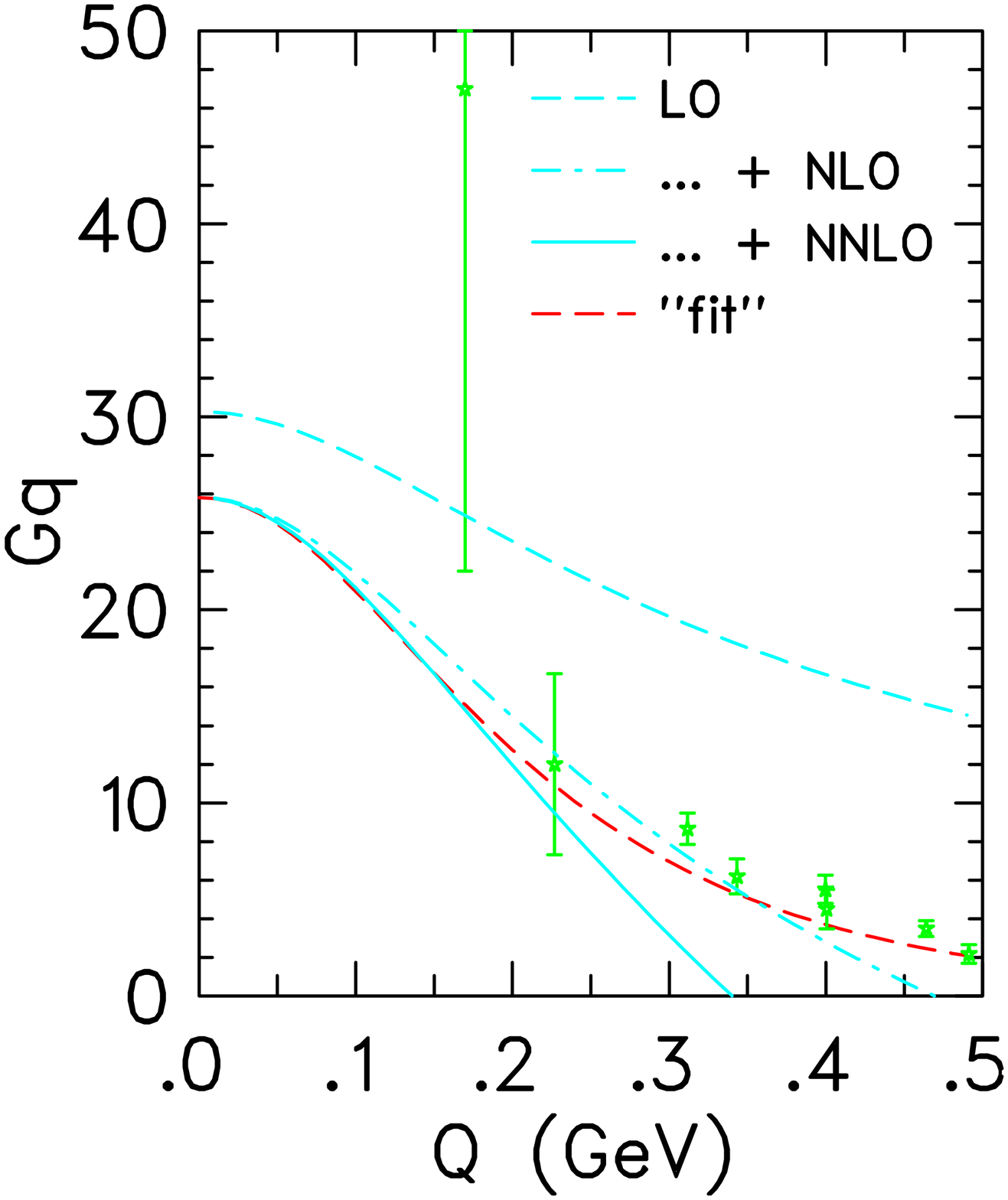}}
\caption{\label{fig:EFT} Predictions of the charge and quadruple
deuteron form factors from the pionless effective theory, as
developed by PRS \cite{PRS00}. The data were extracted in 
Ref.~\cite{abbottphen00} and previously shown in
Fig.~\ref{Gfactors}. The ``fits'' are from  Eq.~(\ref{scale}).}  
\end{figure} 

Using this approach, the LO charge form factor is given
entirely by the asymptotic wave function (\ref{asywave})
\begin{equation}
G_C^{(0)}(Q^2) = {{4\gamma}\over{q}}
\tan^{-1}{{q}\over{4\gamma}}\, ,
\end{equation}
where $q=|{\bf q}|$ is the magnitude of the three momentum
transferred by the electron, and working in the Breit frame
(where the differences between relativistic and nonrelativistic
theory is a minimum) is also equal to $\sqrt{Q^2}$.
PRS show that expansion of the charge form factor up to NNLO
terms is 
\begin{eqnarray}
G_C(Q^2) =& G_C^{(0)}(Q^2) - (Z_d-1)\left[1-G_C^{(0)}(Q^2)
\right] \nonumber\\
&- {{1}\over{6}}\,r_N^2\,Q^2\,G_C^{(0)}(Q^2) + \cdots
\end{eqnarray}
and because the wave function is correctly normalized, there are
no wave function effects beyond NLO (the second term).
At NNLO effects from the finite nucleon size, $r_N$, appear.
Similarly, the LO quadrupole form factor obtained by PRS is  
\begin{equation}
G_Q^{(0)}(Q^2) =  {{3m_d^2\,\eta_{sd}}\over{2\sqrt{2}\gamma q^3}}
\left[ 4 q\gamma - (3q^2+16\gamma^2)\tan^{-1}{{
q}\over{4\gamma}}\right] 
\end{equation}
with $\eta_{sd}$ the asymptotic $S/D$-state ratio,
and the LO quadrupole moment, $Q_d^{LO}$, equal to 
$\eta_{sd}/\sqrt{2}\gamma^2$
$=$  0.335 fm$^2$.  The expansion of $G_Q$ to NNLO is then
\begin{eqnarray}
G_Q(Q^2)  =& G_Q^{(0)}(Q^2) 
+ m_d^2 \,\Delta Q_d + (Z_d-1)\left[ G_Q^{(0)}(Q^2) - m_d^2
Q_d^{LO} \right] \nonumber\\
&- {{1}\over{6}} \,r_N^2\, Q^2  G_Q^{(0)}(Q^2) 
+ \cdots
\end{eqnarray}
Note that the quadrupole moment at NLO includes a contribution
$\Delta Q_d$ from a four-nucleon-one-photon contact term, not
determined by  $NN$ scattering, and is used to fit the
experimental value of $Q_d$. PRS suggest that the absence of
this piece of short distance physics in conventional
calculations may explain their underprediction  of the
quadrupole moment. The finite size of the nucleon again comes in
at NNLO. 

With parameters largely set by other data, the deuteron
charge, quadrupole, and magnetic form factors are well predicted
up to about $Q=$ 0.2 GeV, as shown in Fig.~\ref{fig:EFT}.  The
approach seems to converge well, but beyond NNLO more parameters
enter, and there is less predictive power.  The great strength
of the pionless effective theory is that strips away complexity,
revealing the essential physics required to understand the low
$Q$ results, and showing (for example) the central
importance of the asympotic S-state normalization $Z_d$. 
However, as expected, it clearly does not work for $Q$
much beyond 0.4 to 0.5 GeV.  The theory with pions
(sometimes referred to as a ``pionful'' theory) will work
to higher $Q^2$
\cite{phillips00}.  Removal of divergences from these theories
is under active study.  

We now return to discussion of the reasons for the failure
of nonrelativistic theory at high $Q^2$.


\subsection{Alternative explanations for the failure of
nonrelativistic models}

In Sec.~\ref{compareNR} we showed that the naive
nonrelativistic theory cannot explain the deuteron form factor 
data for $Q\ge 0.5$ GeV.  In this section we classify the
possible explanations for this failure, preparing the way for
detailed discussions to follow in Secs.\ref{nucsec} and
\ref{quarksec}.

The differences between the data and the nonrelativistic theory can
only be explained by a combination of the following effects

\begin{itemize}
\item interaction (or meson exchange) currents;
\item relativistic effects; or
\item new (quark) physics.
\end{itemize}
The only possibilities excluded from this list are
variations in models of the nucleon form factors, or model dependence
of the deuteron wave functions.  In the previous section we argued
that {\it neither\/} the current uncertainty in our knowledge of the
nucleon form factors, {\it nor\/} the model dependence of the
nonrelativistic deuteron wave functions are sufficient to provide
an explanation for the discrepencies.

\begin{figure} 
\begin{center}
\vspace*{-0.4in}
\mbox{
   \epsfxsize=3.5in
\epsffile{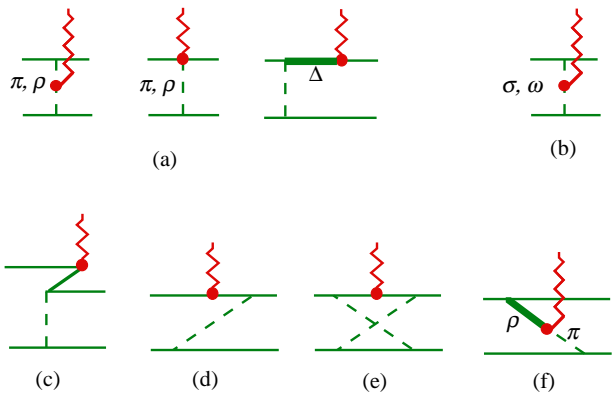} 
} 
\end{center}
\caption{Exchange currents that might play a role in meson theories. 
(a) Large $I=1$ $\pi, \rho,$ and $\Delta$ currents that do not
contribute to the deuteron form factors, and (b) possible $I=0$
currents that are identically zero.  The currents that do
contribute to the deuteron form factors are shown in the second
row: (c) ``pair'' currents from nucleon
$Z$-graphs; (d) ``recoil'' corrections; (e) two pion exchange
(TPE) currents; and (f) the famous $\rho\pi\gamma$ exchange
current.}
\label{MEC}
\end{figure}

Possible interaction currents that might account for the 
discrepency are shown in Fig.~\ref{MEC}.  Because the deuteron
is an isoscalar system, the familiar large $I=1$ exchange
currents are ``filtered'' out and only $I=0$ exchange currents
can contribute to the form factors. The $I=0$ currents tend to
be smaller and of a more subtle origin.  The nucleon
$Z$-graphs, Fig.~\ref{MEC}c, and the recoil corrections,
Fig.~\ref{MEC}d, are both of relativistic origin.  (The recoil 
graphs will give a large, incorrect answer unless they are
renormalized \cite{Gr91,JL75,TH73}.)  The two-meson exchange
currents should be omitted unless the force also contains these
forces.  The famous
$\rho\pi\gamma$ exchange current is very sensitive to the
choice of $\rho\pi\gamma$ form factor, which is hard to
estimate and could easily be a placeholder for new physics
arising from quark degrees of freedom.  
 
In most calculations based on meson theory, the two pion 
exchange (TPE) forces and currents arising from crossed boxes
are excluded, and, except for the $\rho\pi\gamma$ current (which
we will regard as new physics), the exchange currents are of
relativistic origin.  Additional relativistic effects arise from
boosts of the wave functions, the currents, and the potentials,
which can be calculated in closed form or expanded in powers of
$(v/c)^2$, depended on the method used.  At low $Q^2$
calculations may be done using effective field theories
(discussed in Sec.~\ref{sec:EFT}) in which a small parameter is
identified, and the most general (i.e. exact) theory is expanded
in a power series in this small parameter.  In these
calculations relativistic effects are automatically included (at
least in principle) through the power series in
$(v/c)^2$.   {\it Hence any improvement on
nonrelativistic theory using nucleon degrees of freedom leads us
to relativistic theory.\/}
 
Alternatively, one may seek to explain the discrepancy using
quark degrees of freedom (new physics).  When two nucleons 
overlap, their quarks can intermingle, leading to the creation
of new $NN$ channels with different quantum numbers (states
with nucleon isobars, or even, perhaps, so-called ``hidden
color'' states).  These models require that assumptions be made
about the behavior of QCD in the nonperturbative domain, and
are difficult to construct, motivate, and constrain.  At very
high momentum transfers it may be possible to estimate the
interactions using perturbative QCD (pQCD).  Very little has
been done using other approaches firmly  based in QCD, such as
lattice gauge theory or Skrymions (but see
Ref.~\cite{Nyman:1986nm}).

We are thus led to two different alternatives for 
explanation of the failure of nonrelativistic models.  In one 
approach the nucleon (hadron) degrees of freedom are retained,
and relativistic methods are developed that treat boost and
interaction current corrections consistently.  In another
approach, quark degrees of freedom are used to describe the
short range physics, and techniques for handling a multiquark
system in a nonperturbative (or perturbative) limit are
developed.  These two approaches will be reviewed in the next
two sections.  While the discussion appears to be focused on 
the deuteron form factors, it is actually more general, and
will be applied later to the treatment of deuteron
photodisintegration.

Are these two approaches really different?  Superficially, of
course, the answer must be: Yes!  However, QCD tells us
that all physical states must be color singlets, and a basis of states
that describes any color singlet state can be constucted from  {\it
either\/} quarks (and gluons) {\it or\/} hadrons (this would not be
true if colored states were physical).  So at a deeper level it
appears that either approach (hadrons or quarks) should work, and
the best choice is the system that can describe the relevant
physics more compactly.  Further discussion of this issue is clearly
beyond the scope of this review.


\subsection{Relativistic calculations using nucleon degrees of 
freedom}
\label{nucsec}  

This long section is divided into six parts as follows: (i)
Introduction, (ii) Overview of propagator dynamics, (iii)
Choice of propagator and kernel, (iv) Examples of propagator
dynamics, (v) Overview of hamiltonian dynamics, and (vi) Examples
of hamiltonian dynamics. 

 
\subsubsection{Introduction} 

\begin{figure} 
\begin{center}
\mbox{
   \epsfxsize=5.2in
\epsffile{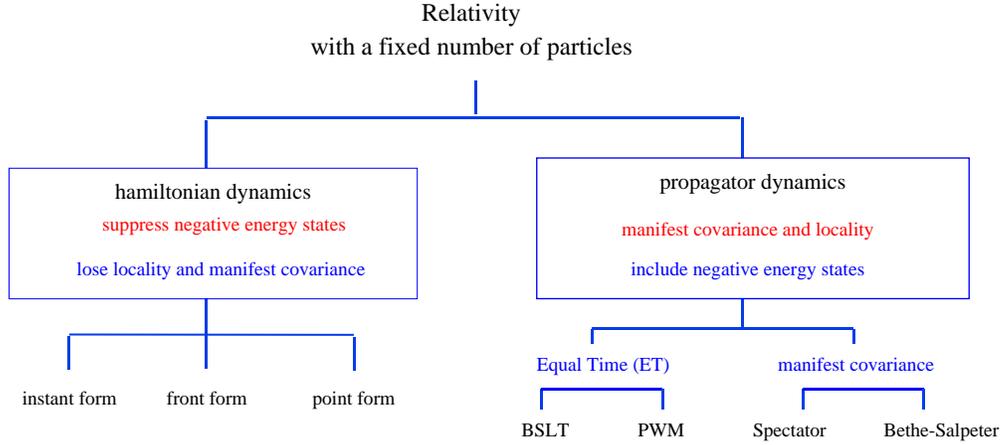} 
} 
\end{center}
\caption{The relativistic decision tree discussed in the text. }
\label{decision}
\end{figure}

The inhomogeneous Lorentz group, or the Poincar\'e group, is 
described by 10 generators: three pure rotations, three 
pure  boosts, and four pure translations.  If we require
the interactions to be local and manifestly covariant
under the Poincar\'e group, we are led to a local 
relativistic quantum field theory with particle
production and annihilation \cite{Weinbergbook}.  In this
case the Poincar\'e transformations of all matrix
elements can be shown to depend only on the kinematics
(i.e.\ they depend only on the masses and spins of the
external particles).  The disadvantage is that the number
of particles is not conserved. If perturbation theory can
be used, this approach is very successful, but in the
nonperturbative regime of strong coupling meson theory it
leads to an infinite set of coupled equations that cannot
be solved in closed form.  Numerical, nonperturbative
solutions of field theory can be obtained in Euclidean space for
a few special cases \cite{Cetin}.  Methods that limit the
intermediate states to a {\it fixed number of
particles\/} (two nucleons in this case) are more
tractable, and all modern calculations are based on the
choices depicted in the decision tree shown in
Fig.~\ref{decision}.    

In deciding which method to use, if is first necessary to decide
whether or not to allow {\it antiparticle, or negative
energy\/} nucleons to propagate as part of the virtual intermediate
state.  Since nucleons are heavy and composite, so that their
antiparticle states are very far from the region of interest, some
physicists believe that intermediate states should be built only
from positive energy nucleons, and that all negative energy effects
(if any) should be included in the interaction.  These methods are
referred to collectively as {\it hamiltonian dynamics\/} and are
represented by the left hand branch in the figure.  Unfortunately,
it turns out that this choice precludes the possibility of retaining
the properties of locality and manifest covariance enjoyed by field
theory.    Alternatively, in order to keep the locality and manifest
covariance of the original field theory, other physicists are
willing to allow negative energy states into the propagators. These
methods, represented by the right-hand branch of the figure, are
referred to collectively as {\it propagator dynamics\/}. 
Including negative energy states tends to make
calculations technically more difficult and harder to interpret
physically, and those who advocate the use of hamiltonian
dynamics do not believe the advantages of exact
covariance justify the work it requires.  

Unfortunately, these two methods are so fundamentally 
different that  many physicists do not realize that the
limitations of one may not apply to the other.  For
example, for some choices of propagator dynamics all 10
of the generators of the Poincar\'e group will depend
only on the kinematics, and the Poincar\'e
transformations of {\it all amplitudes can be done
exactly\/}.  With hamiltonian dynamics this is not the
case; some of the 10 generators must  depend on the
interaction, and transformation of matrix elements under
these ``dynamical'' transformations must be calculated. 
Comparison of the two methods is therefore very
difficult; the language and issues of each are very
different and one can be easily misled by the different
appearance of the results.  We cannot discuss these
issues in detail in this review, and refer the reader to
two recent references that survey the subject
\cite{Gr91,Keisterpoly}.  Here we will give a short
review of some recent calculations, and explain these
differences as we go along. 

\subsubsection{Overview of propagator dynamics}

Propagator calculations all start from the field theory 
description of two (in this case) interacting particles. 
While some may prefer to express the field theory as a
path integral, it is also possible to adopt a more
intuitive approach and imagine expanding the path
intergral as a sum of Feynman diagrams (ignoring issues
of convergence for the moment).  In order to generate the
deuteron bound state, which produces as a pole in the
scattering matrix, it is necessary to sum an infinite
class of diagrams, written as
\begin{eqnarray}
\fl {\cal M}(p,p';P)=&{\cal V}(p,p';P)
+\int\frac{d^nk}{(2\pi)^n}\,{\cal V}(p,k;P)G(k,P){\cal V}(k,p';P)
\nonumber\\
&+\int\frac{d^nk}{(2\pi)^n}\int\frac{d^nk'}{(2\pi)^n}\,{\cal
V}(p,k;P)G(k,P){\cal V}(k,k';P)G(k',P){\cal
V}(k',p';P)\nonumber\\
&+\;\cdots
\label{sumdiag} 
\end{eqnarray}  
where ${\cal V}(p,p';P)$ is the {\it kernel\/} being iterated, 
$G(k,P)$ the two body propagator, ${\cal M}(p,p';P)$ the
scattering amplitude, and the other quantities are defined below. 
This sum is
obtained in closed form by solving the integral equation  
\begin{equation}
{\cal M}(p,p';P)={\cal V}(p,p';P)
+\int\frac{d^nk}{(2\pi)^n}\,{\cal V}(p,k;P)G(k,P){\cal M}(k,p';P)\,
. \label{rel1} 
\end{equation}  
If the series (\ref{sumdiag}) is compared to a geometric 
series $1 + z+z^2 +\cdots$, then the solution to the
integral equation (\ref{rel1}) can be compared to the sum
of the geometric series 
$1/(1-z)$.  The geometric series converges only when $|z|<1$,
but its unique  analytic continuation, $1/(1-z)$, is  valid for
all $z$.  Similiarly, it is assumed that the solution to
(\ref{rel1}) is valid even when the series (\ref{sumdiag})
diverges.  And just as the geometric series has a pole at
$z=1$, the solution to (\ref{rel1}) will have a pole at
$P^2=m_d^2$, the square of the deuteron mass.


\begin{figure} 
\begin{center}
\vspace*{.5in}
\mbox{
   \epsfxsize=4.0in
\epsffile{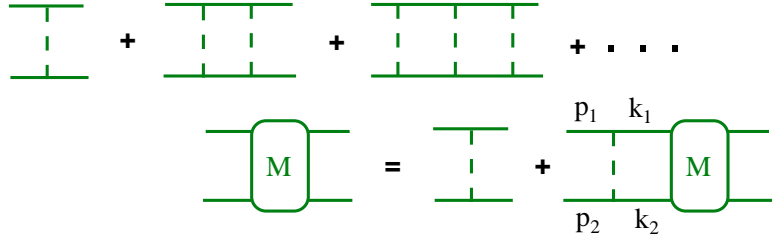} 
} 
\end{center}
\caption{The summation of ladder diagrams leads to the covariant
scattering equation. }
\label{feynmanladder}
\end{figure}

The amplitudes ${\cal V}$, $G$, and ${\cal M}$ are all matrices in
the $NN$ spin-isospin space, and are functions of the four-momenta 
$P=p_1+p_2$ and $p=(p_1-p_2)/2$, with $p_1$ and $p_2$ the momenta
of the two particles (labeled in Fig.~\ref{feynmanladder}).  The
dimension of the volume integration is
$n$, normally either 3+1=4 (3 space + one time dimensions) for the
Bethe-Salpeter method, or 3+0=3 for the quasipotential methods described
below.  

\begin{figure} 
\begin{center}
\mbox{
   \epsfxsize=3.5in
\epsffile{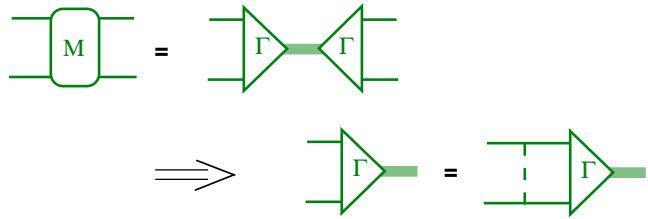} 
}  
\end{center}
\caption{The bound state equation holds near the pole in the 
scattering amplitude.}
\label{feynmanbound}  
\end{figure}

If Eq.~(\ref{rel1}) has a homogenous solution at some external four
momentum
$P_0^2=m_d^2$, the scattering matrix will have an $s$ channel pole
(represented in Fig.~\ref{feynmanbound}), signifying the
existence of a deuteron bound state.  The {\it vertex\/}
function for the  deuteron bound state satisfies the
equation
\begin{equation}
\Gamma(p;P_0)=\int\frac{d^nk}{(2\pi)^n}\,{\cal V}(p,k;P)
G(k,P_0)\Gamma(k;P_0)
\label{bound1} 
\end{equation} 
with covariant normalization condition
\begin{eqnarray}
\fl1=&-\int\frac{d^nk}{(2\pi)^n}\,\Gamma(k;P_0)\,\frac{\partial
G(k,P_0)} {\partial P_{0}^2}\,\Gamma(k;P_0)\nonumber\\
\fl&-\int\frac{d^nk}{(2\pi)^n}\int\frac{d^nk'}{(2\pi)^n}
\,\Gamma(k';P_0)G(k',P_0)\,\frac{\partial V(k',k;P_0)}  
{\partial P_{0}^2}\,G(k,P_0)\Gamma(k;P_0)\, .
\label{bound2} 
\end{eqnarray} 
The covariant bound state wave function is defined by
\begin{equation}
\Psi(p;P_0)=G(p;P_0)\,\Gamma(p;P_0)\, .
\label{wfdef} 
\end{equation} 
%

\begin{figure} 
\begin{center}
\mbox{
   \epsfxsize=3.5in
\epsffile{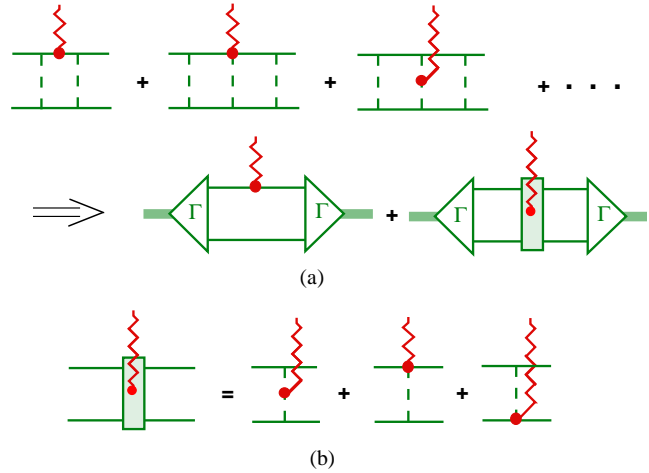} 
}  
\end{center}
\caption{The current operator follows from the
electromagnetic interaction of all the consituents in the ladder sum.  For
bound states, all of these interactions can be collected 
into the two covariant diagrams shown in panel (a), with
the interaction current, shown in panel (b), constructed from
$\gamma \phi NN$ and $\gamma \phi \phi$ couplings (where the
exchanged meson is denoted by $\phi$).  }
\label{feynmancurrent}  
\end{figure}

One of the advantages of the propagator approach is that 
the construction of the current operator is comparatively
straightforward.  It follows (at least in principle) from 
summing all electromagnetic interactions with all the
consituents everywhere in the ladder sum.  For bound 
states described by the Bethe-Salpeter or Spectator
formalisms (see the discussion below) there are two
diagrams, illustrated in Fig.~\ref{feynmancurrent}, that
can be written  
\begin{eqnarray}
\fl{\cal J}^\mu(P_0',P_0)=e\int\frac{d^nk_2}{(2\pi)^n}
\,\Psi^\dagger\left({\textstyle{1\over2}}P'_0-k_2;P'_0\right)
\,J_N^\mu(k'_1,k_2;k_1,k_2)\,
\Psi\left({\textstyle{1\over2}}P_0-k_2;P_0\right)   \nonumber\\
+\,e\int\frac{d^nk}{(2\pi)^n}\int\frac{d^nk'}{(2\pi)^n}
\,\Psi^\dagger(k';P'_0)\;I^\mu(k'_1,k'_2;k_1,k_2)\;\Psi(k;P_0)\, .
\label{current1} 
\end{eqnarray} 
In the first term, $J_N^\mu$ is the sum of the neutron and 
proton currents [recall Eq.~(\ref{nnff})] and we have
chosen particle 1 to interact with the photon (always
possible because of the antisymmetry of the wave
function).  The interaction current is $I^\mu$, and
assumes a comparatively simple form if the kernel is a
sum of single particle exchanges.  This case is
illustrated in Fig.~\ref{feynmancurrent}.  

Current conservation, 
\begin{eqnarray}
q_\mu{\cal J}^\mu(P_0',P_0)=0\, , 
\label{current2} 
\end{eqnarray} 
follows automatically \cite{GR87} from the bound state 
equation (\ref{bound1}) if the nucleon and interaction
currents satisfy the following two-body Ward-Takahashi
(WT) identities
\begin{eqnarray}
q_\mu \;J_N^\mu(k'_1,k'_2;k_1,k_2)
&&={\textstyle{1\over2}}[1+\tau_3]\,\Bigl\{G^{-1}(k;P_0)-G^{-1}(k';P'_0)\Bigr\}
\nonumber\\ q_\mu \;I^\mu(k'_1,k'_2;k_1,k_2)&&= 
{\cal V}(k',k;P'_0)\,{\textstyle{1\over2}}[1+\tau_3] -
{\textstyle{1\over2}}[1+\tau_3]\,{\cal V}(k',k;P_0)\, .
\label{current3} 
\end{eqnarray} 
Note the appearance of
${\textstyle{1\over2}}[1+\tau_3]$, the isoscalar charge 
operator in isospin space. The $J_N$ identity is the
two-body version of the familiar one-body WT identity
\begin{eqnarray}
q_\mu j_N^\mu(k'_1,k_1)=e_N\left\{S^{-1}(k_1)-S^{-1}(k'_1)\right\}\, ,
\label{current4} 
\end{eqnarray} 
with $e_N=e$ (0) for the proton (neutron) and the 
undressed  nucleon propagator normalized to
$S_0^{-1}(k_1) = m -\not\!\!k_1$.  Note that the
constraint on the interaction current is {\it not\/} zero
(and hence the interaction current is {\it not\/} zero)
if the kernel depends on the isospin or the total
four-momentum $P_0$.


\subsubsection{Choice of propagator and kernel}   

To fully specify a propagator dynamics, one must choose a 
propagator,
$G$, a kernel, ${\cal V}$, and current operators $J_N^\mu$ and 
$J_I^\mu$.   

Four different progagators have been used in
the study of the deuteron form factors.   The Bethe-Salpeter 
(BS) equation \cite{BS51} uses a fully off shell propagator for
two nucleons
\begin{eqnarray}
 G_{BS}(p,P)={\displaystyle S_1(p_1)S_2(p_2)}   
\to {\displaystyle\frac{\Lambda_1^+\left({W\over2}+p\right)\,
\Lambda_2^+\left({W\over2}-p\right)}{\left[E_p^2
-\left({W\over2}+p_0\right)^2\right]  
\left[ E_p^2-\left({W\over2}-p_0\right)^2\right]}}\, ,
\end{eqnarray}
where $\Lambda_i(p)/2m=(m_i+\not\!p)/2m$ is the 
(off-shell) positive energy projection operator and the
right hand expression is the propagator  for identical
particles in the rest frame (with $P=\{W,{\bf 0}\}$, and
$W$  used as a shorthand for the four-vector $P$ in the rest
frame).  This choice of relativistic equation was the first to be
introduced and is perhaps the best known.  It retains the
full integration over all components of the relative
four-momentum
$p$, and all of the off-shell degrees of freedom (2 for 
spin $\times$ 2 for ``$\rho$-spin'', where
$\rho=+$ are positive energy $u$ spinor states and
$\rho=-$ are negative energy $v$ spinor states) of both 
of the  propagating nucleons, for a total of
$4\times4=16$ spin degrees of freedom.  The equation has
inelastic cuts arising from the production of the
exchanged mesons (when energetically possible) and
additional singularities when the nucleons are
off-shell.  These can be removed by transforming the
equation to Euclidean space. However, the BS equation,
when used in {\it ladder\/} approximation, does not have
the correct one-body limit.  Numerical comparisons of 
solutions obtained from the sum of {\it all\/} ladder and
crossed ladder exchanges with ladder solutions of the BS
equation, carried out for scalar theories, have shown
that the ladder sum is inaccurate and that the one-body
limit requires inclusion of crossed exchanges
\cite{TjonFS}.   The BS equation has been solved in
ladder approximation by Tjon and his collaborators
\cite{TjonBS}, and used to calculate the deuteron form 
factors \cite{Tjonff}.  The fits to the $NN$ phase shifts
originally obtained from these works are unsatisfactory
by today's standards.

The Spectator (or Gross) equation (denoted by $S$) 
\cite{Grosseq} restricts one of the two nucleons to its
positive energy  mass-shell.  If particle one is
on-shell, the spectator propagator is
\begin{eqnarray}
G_S(p,P)&={\displaystyle
{2\pi\,
\delta_+(m^2-p_1^2)\,\Lambda_1(p_1)\,S_2(p_2)}}\nonumber\\
&\to
{\displaystyle\frac{2\pi \,\delta\left(E_p-{W\over2}-p_0\right) 
\Lambda_1^+(\hat {p}_1)\,
\Lambda_2^+(W-\hat {p}_1)}{2E_pW (2E_p-W)}}
\end{eqnarray}
where $\hat p_1=\left\{E_p,\,{\bf p}\right\}$.  This has 
the effect of fixing the relative energy in terms of the
relative three-momentum so as to maintain covariance and
reduce the four dimensional integration to three
dimensions [$n=3$ in Eq.~(\ref{rel1})].  Identical
particles are treated by properly (anti)symmetrizing the
kernel.  The restriction of one of the particles to its
positive energy mass-shell also removes the
$\rho=-$ states of one of the nucleons, reducing the 
number of spin degrees of freedom to 2$\times$4=8.  A
primary motivation and justification for this approach is
that it has the correct one-body limit, and the three
body generalization satisfies the cluster property
\cite{sg97}.  The equation also has a nice nonrelativistic
limit that can be easily interpreted.  Numerical studies 
of scalar field theories 
\cite{TjonFS} show that the  exact ladder and crossed 
ladder sum is better approximated by the ladder
approximation to this equation than it is by the ladder
approximation to the BS equation.   The method can be
extended to include gauge invariant electromagnetic
interactions \cite{GR87}.  Its principle drawback is that
the kernel has unphysical singularities which can only be
removed by an {\it ad-hoc\/} prescription.  Results from
this method will be reviewed in the next section. 

The internal momentum integration can also be restricted 
to three dimensions in such a way that, for equal mass
particles in the rest frame, the relative energy is zero
and the particles are equally off shell. The
Blankenbecler-Sugar-Logunov-Tavkhelidze (BSLT) equation
\cite{BSLT} can be defined so that the two propagating
particles are on their positive energy mass-shell,
reducing the number of spin-$\rho$-spin variables to
2$\times$2=4. However, this equation does not satisfy the
cluster property.  The approach of Phillips, Wallace, and
Mandelsweig \cite{PW,MW}, which we denote by PWM, also
puts the particles equally off-shell, but includes all
negative energy contributions.  Setting $n=3$ in
Eq.~(\ref{rel1}), the PWM propagator for equal mass
particles in the c.m.~system is
\begin{eqnarray}  
\fl G_{\rm PWM}(p,P)=\int dp_0\,{\cal G}_{\rm
PWM}(p,P){\displaystyle =\int
dp_0\left\{\,S_1(p_1)S_2(p_2) +
 G_C(p,P)\right\}}\nonumber\\
\fl\qquad\to
{\displaystyle\Biggl\{\frac{ \Lambda_1^+(\hat {p}_+)\,
\Lambda_2^+(\hat {p}_-)}{(2E_p-W)} +\frac{ \Lambda_1^-(\hat {p}_+)\,
\Lambda_2^-(\hat {p}_-)}{(2E_p+W)}}+\frac{ \Lambda_1^+(\hat
{p}_+)\,
\Lambda_2^-(\hat {p}_-)}{2E_p}+\frac{ \Lambda_1^-(\hat {p}_+)\,
\Lambda_2^+(\hat {p}_-)}{2E_p}\Biggr\} \label{PWeq}
\end{eqnarray}
with $4\times4=16$ spin degrees of freedom.  This 
propagator differs from BSLT primarily by the presence of
the additional $G_C$ term [which contributes the last
two terms in the curly brackets involving the ``mixed''
$\Lambda^+\Lambda^-$ projection operators] that includes
contributions from crossed graphs approximately, and
correctly builds in the one body limit.   The retarded 
kernel to be used with this propagator, in ladder
approximation, is
\begin{eqnarray}  
\fl {\cal V}_{\rm PWM}(p,P)=G_{\rm PWM}^{-1}(p,P)
\left\{\int dp_0 {\cal G}_{\rm PWM}(p,P) {\cal V}_{BS}(p,P)  {\cal
G}_{\rm PWM}(p,P) \right\} G_{\rm PWM}^{-1}(p,P)
\label{PWkernel}
\end{eqnarray}

Perhaps the principle obstacle to implementing this 
method is that construction of current operators is
problematic, and manifest Poincar\'e invariance is lost
(but Wallace \cite{W01} has recently shown how to compute
boosts for  scalar particles exactly). Calculations using
this method will be described in next section.


\subsubsection{Examples of propagator dynamics}
\label{pexamples}

We now turn to a description of two examples of propagator 
dynamics.

\begin{figure} 
\begin{center}
\mbox{
   \epsfxsize=4.5in
\epsffile{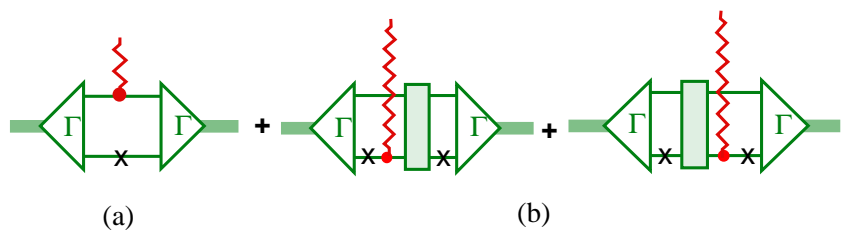} 
}  
\end{center}
\caption{The complete impulse approxiation (CIA) to a ladder
Spectator theory.  The $\times$ denotes the on-shell particle. 
In diagrams (b) the kernel connects the region where one of the two
particles is on-shell to the region were {\it both\/} particles are
off-shell.  For identical particles, it can be shown that (b)
$\simeq$ (a), so that the sum of these three diagrams is
approximately equal to 2$\times$(a), which is referred to as the
relativistic impulse approximation (RIA). }
\label{Scurrent}  
\end{figure}

{\it Van Orden, Devine, and Gross\/} [VOG].  The
Spectator equation has been used to successfully describe
$NN$ scattering and the deuteron bound state
\cite{gvoh92}, and this work uses these results to
describe the deuteron form factors \cite{vdg95}.  The
relativistic kernel used to describe the $NN$ system
consists of the exchange of 6 mesons [$\pi$, $\eta$,
$\sigma$, $\delta$ (or $a_0$), $\rho$ and $\omega$]. 
The model includes a form factor for the off-shell nucleon
\cite{gvoh92}, giving a ``dressed'' single nucleon 
propagator of the form
\begin{eqnarray}
S_d(p)={h^2(p)\over m-\not\!p}\qquad
h(p)={(\Lambda_N^2-p^2)^2
\over(\Lambda_N^2-p^2)^2+(m^2-p^2)^2}\, ,
\end{eqnarray}
where $\Lambda_N$ is one of the parameters of the model.
Coupling constants and form factor masses (13 parameters
in all) are determined by a fit to the data and the 
deuteron wave functions are extracted \cite{gvoh92}.  
To insure current conservation, the one-nucleon 
current must satisfy the Ward-Takahashi identity
\begin{eqnarray}
(p'-p)^\mu j_\mu(p',p) =h(p')\left\{S_d^{-1}(p)-
S_d^{-1}(p')\right\}h(p)\, , \label{WT}
\end{eqnarray}
and this requires an off-shell modification of the
single nucleon current.  The solution used by VOG is  
\begin{eqnarray}
\fl j_\mu(p',p) =
f_0(p',p)\left(F_1(Q^2)\,\gamma^\mu+{F_2(Q^2)\over2m}
i\sigma^{\mu\nu}q_\nu\right) +
g_0(p',p)\,F_3(Q^2){m-\not\!p'\over2m}\gamma^\mu
{m-\not\!p\over2m}  \label{onej}
\end{eqnarray}
where $F_3(0)=1$ but is otherwise undefined [in the applications
described below, $F_3=F_D$ where $F_D$ is the dipole form factor of
Eq.~(\ref{dipole})], and
\begin{eqnarray}
f_0(p',p)={h(p^2)\over h(p'^2)}\left[ {m^2-p'^2\over p^2-p'^2}
\right] +{h(p'^2)\over h(p^2)} 
\left[{m^2-p^2\over p'^2-p^2} \right]\nonumber\\
g_0(p',p)=\left({h(p^2)\over h(p'^2)}-{h(p'^2)\over h(p^2)}\right)
{4m^2\over p'^2-p^2}\, . 
\end{eqnarray}
While the on-shell form of the current (\ref{onej}) is 
fixed by the nucleon form factors, and the functions
$f_0$ and $g_0$ are fixed by the WT identity (\ref{WT}),
other aspects of the off-shell extrapolation of the
current (\ref{onej}) are {\it not unique\/}.  Using this
one-nucleon current, and recalling that there are no
currents of the type shown in Figs.~\ref{MEC}(a) or (b)
[we postpone discussion of the $\rho\pi\gamma$ current],
it was shown~\cite{GR87} that the  full two body current
to use with the Spectator equation is given by the
diagrams shown in Fig.~\ref{Scurrent}.  These diagrams
are manifestly covariant, and {\it automatically include
effects from Z-graphs or retardation\/} illustrated
in Figs.~\ref{MEC}(c) or (d).  They are referred to as
the complete impulse approximation (CIA) to distinguish
them from the relativistic impulse approximation (RIA),
an approximate current used in earlier calculations
\cite{acg80}.  The RIA is obtained by multiplying diagram
\ref{Scurrent}(a) by two, and is very close to the CIA.

{\it Phillips, Wallace, Divine, and Mandelsweig\/} [PWM].  This work
is based on the Mandelsweig and Wallace equation \cite{MW},
supplemented by contributions from the crossed graphs \cite{PW},
as described above.  It is sometimes referred to as the
equal-time approach.  A feature of  this equation is that it
includes the full strength of the $Z$-graphs; the
$\Lambda^+\Lambda^-$ contributions shown in Eq.~(\ref{PWeq}) are
roughly twice as strong as the $Z$-graph contributions included
in the Spectator equation.  The PWM propagator is also explicitly
symmetric, a convenience when applied to identical
particles.  In the published work reviewed here
\cite{PW}, the deuteron is described by a one boson
exchange force using the parameters of the Bonn-B
potential with the exception of the $\sigma$ meson
coupling, which is adjusted to give the correct deuteron
binding energy.  Lorentz invariance is broken by the
approximation; the boosts of the deuteron wave functions
from their rest frames are treated approximately.  The
PWM current is a modification of Eq.~(\ref{current1}). 
In the present work retardation effects [like those
illustrated in Fig.~\ref{MEC}(d)] are omitted from the
current operator; only one body terms and $Z$-graph
contributions are included. 
  
We now turn to a discussion of the other major approach
detailed in Fig.~\ref{decision}: hamiltonian dynamics.        
 

\subsubsection{Overview of hamiltonian dynamics}   

Approaches based on hamiltonian dynamics start from a 
very different  point than propagator dynamics, and this
is one reason it is difficult to compare the
two.  While propagator dynamics starts from field theory
(which can be described as a quantum mechanics with an
arbitrary number of particles), hamiltonian dynamics
starts from quantum mechanics with a
fixed number of particles.  For a detailed review, see
Ref.~\cite{Keisterpoly}. 

\begin{figure} 
\begin{center}
\mbox{
   \epsfxsize=3.0in
\epsffile{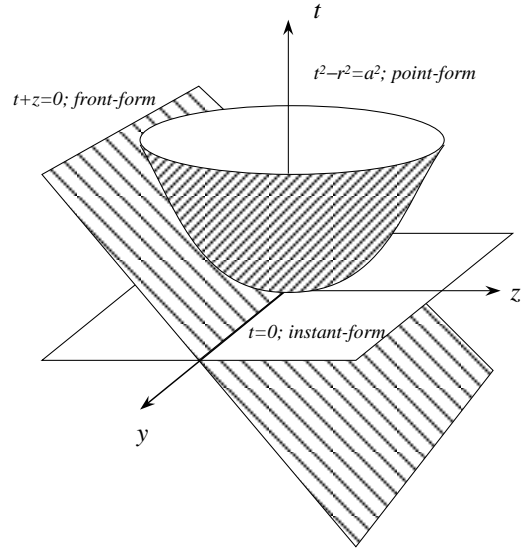} 
}  
\end{center}
\caption{Drawing of the three surfaces on which states can be
defined in quantum mechanics.  All points on the forward hyperbolid
are separated by a space-like interval.  }
\label{qmsurfaces}  
\end{figure}

Quantum mechanics begins with a Hilbert space of
states defined on a fixed space-like surface in 
four-dimensional space-time.  The various options for
choosing this space-like surface were classified by Dirac
in 1949 \cite{Dirac}.  The {\it instant-form\/}
corresponds to choosing to construct states at a
fixed time $t_0=0$, and is the choice usually made in
elementary  treatments.  Alternatively, {\it
front-form\/} quantum mechanics constructs states on a
fixed-light front, customarily defined to be
$t^+=t+z=0$.  (We use units in which the speed of light, $c$, is
unity.)  More generally, the light-front may be chosen in any
direction defined by $x^\mu n_\mu=0$, with
$n_\mu=\{1,{\bf n}\}$ and ${\bf n}^2=1$.   Finally, {\it
point-form\/} quantum mechanics constructs states on a 
forward hyperboloid, with $t^2-{\bf r}^2=a^2$; $t>0$ [the
limiting cases of $a=\infty$ gives the instant-form (with
$t_0=a$), and $a=0$ the front-form].  These three
surfaces are shown pictorially in Fig.~\ref{qmsurfaces}. 
[While the point $t=0$, ${\bf r}=0$ is not on the
hyperbolid, all distances between points on the
hyperbolid are space-like.] 

The Poincar\'e transformations are symmetries that leave 
all probabilities unchanged; they must be unitary
transformations  (with hermitian generators) on the space
of quantum states.   The 10 generators of the full
Poincar\'e group are the hamiltonian,
$H$, generator of time translations, three-momenta, 
$P^i$, generators of spatial translations, angular
momenta, $J^i$, generators of rotations, and $K^i$,
generators of boosts.  They satisfy the following
commutation relations:
\begin{eqnarray}
\fl&&[H,P^i]=[H,J^i]=[P^i,P^j]=0\, , \qquad[J^i,X^j]=i\epsilon_{ijk}
X^k\, , {\rm for}\; X^i=J^i, P^i, K^i\nonumber\\ 
\fl&&[K^i,K^j]=-i\epsilon_{ijk}J^k\, ,
\qquad[K^i,P^j]=-i\delta_{ij}H\, ,
\qquad[K^i,H]=-iP^i \label{ccr1}
\end{eqnarray}

For each of the forms of quantum mechanics there is a subgroup of
the Poincar\'e transformations that leave the states invariant
on the fixed surface associated with that form.  This is the
kinematic subgroup, and the transformations in this
subgroup will not depend on the dynamics (since the
dynamics describe how the states change away for the
fixed surface). 
In the {\it instant-form\/}, space translations and 
rotations clearly leave the surface $t=0$ invariant. 
Generators of these transformations form a subgroup of
the Poincar\'e group, with commutation relations
\begin{eqnarray}
&&[P^i,P^j]=0\, , \quad[J^i,J^j]=i\epsilon_{ijk}
J^k\, , \quad[J^i,P^j]=i\epsilon_{ijk} P^k \, .\label{ccr2}
\end{eqnarray}
Transformations of states under these transformations 
will not depend on the dynamics.  The hamiltonian $H$
carries the states away from the initial fixed $t=0$
surface, and contains the dynamics.  The other three
generators (the boosts) will also, in general, depend on
the dynamics because their commutators involve
$H$.

The {\it front-form\/} surface $t^+=0$ is left invariant 
by translations in the $x$, $y$, and $t^-$
directions [the generator of translations in the $t^-$ 
direction is $H^+$ because $Ht-P_zz=(H^+t^-+H^-t^+)/2$]. 
It is also left invariant by rotations and boosts in the
$z$ direction, and by the generalized boosts
$E^x=K^x+J^y$ and $E^y=K^y-J^x$.  These 7 generators form
a subgroup of the Poincar\'e group, with commutation
relations     
\begin{eqnarray}
\fl&&[P^i,P^j]=[P^i,H^+]=[J^z,H^+]=[E^i,E^j]=[E^i,H^+]=[J^z,K^z]=
[K^z,P^i]=0
 \nonumber\\
\fl&&[J^z,P^i]=i\epsilon_{ij}P^{j}\,
,\quad[J^z,E^i]=i\epsilon_{ij}E^{j}\, ,\quad 
[E^i,P^j]=-i\delta_{ij}H^+\nonumber\\ 
\fl&&[K^z,H^+]=-iH^+
\, ,\quad[K^z,E^i]=-iE^i\, ,\label{ccr3}
\end{eqnarray}
where $i$ and $j=\{x,y\}=\{1,2\}$, and  $\epsilon_{12}=1$,
$\epsilon_{21}=-1$, and
$\epsilon_{11}=\epsilon_{22}=0$.  The fact that the 
front-form kinematic subgroup includes {\it seven\/}
generators, including the boost $K^z$ and generalized
boosts $E^i$, makes the front-form popular.  But a
principle motivation for using the front-form  is that it
is a natural choice at very high momentum, where the
interactions single out a preferred direction (the beam
direction) and the dynamics evolves along the light-front
in that direction. The disadvantage is that the
generators that contain dynamical quantities are $H^-$ 
and $J^i$, and this means that angular momentum
conservation must  be treated as a dynamical constraint.

Finally, the {\it point-form\/} hyperbolid is left  
invariant by the homogeneous Lorentz group itself, with
commutation relations          
\begin{eqnarray}
[J^i,J^j]=i\epsilon_{ijk}
J^k\, ,\quad[J^i,K^j]=i\epsilon_{ijk}K^k\, ,\quad
[K^i,K^j]=-i\epsilon_{ijk}J^k\, . \label{ccr4}
\end{eqnarray}
The hamiltonian and the momentum operators $P^i$ all 
carry  the dynamical information. 

We see that each of the forms of quantum mechanics has a 
different set of kinematic generators, and in no 
case are they all kinematic.  Practitioners of
hamiltonian dynamics sometimes speak as if it were
impossible to treat the full Poincar\'e group
kinematically.  This is true only in the context of
hamiltonian dynamics; {\it all\/} of the generators are
kinematic in  the BS or Spectator forms of progagator
dynamics.    

Dynamics is introduced whenever the states are propagated 
away  from the surface on which they are initially
defined.  As in normal quantum mechanics, the deuteron
will be an eigenstate that propagates in ``time'' without
loss of probability; it will be an eigenstate of the
generalized hamiltonian.  In  the instant-form, the rest
state $\Psi_{\rm I}(0)$ is an eigenstate of the momentum
operators
\begin{eqnarray}
{\bf P}\Psi_{\rm I}(0)={\bf 0}\, , \label{instant2}
\end{eqnarray}
and the bound state equation in the rest frame is
\begin{eqnarray}
H\Psi_{\rm I}(0)=m_d\Psi_{\rm I}(0)\, . \label{instant}
\end{eqnarray}
In the front-form the rest state $\Psi_{\rm F}(0)$ is an 
eigenstate of the operators ${\bf P}_\perp=\{P^x,P^y\}$ and 
$H^+$  
\begin{eqnarray}
{\bf P}_\perp\Psi_{\rm F}(0)={\bf 0}\, ,\qquad H^+\Psi_{\rm
F}(0)=m_d\Psi_{\rm F}(0) \, ,
\label{front2}
\end{eqnarray}
and the dynamical bound state equation is
\begin{eqnarray}
H^-\Psi_{\rm F}(0)=m_d\Psi_{\rm F}(0)\, . \label{front}
\end{eqnarray}
Finally, in point-form the rest frame eigenfunction must 
satisfy the four dynamical equations
\begin{eqnarray}
P^\mu\Psi_{\rm P}(0)={P}_o^\mu\Psi_{\rm P}(0)\, , \label{point}
\end{eqnarray}
Where $P^\mu=\{H,P^i\}$ and $P_o^\mu=\{m_d,{\bf 0}\}$.
In applications, the dynamical equations
(\ref{instant}), (\ref{front}) and the $\mu=0$ component of
(\ref{point}) can all be taken to be the nonrelativistic
Schr\"odinger equation in the  rest frame, so the same
nonrelativistic phenomenology can be used for any of these
forms of mechanics. 

To complete the calculation of the deuteron form factors using
hamiltonian dynamics one must choose a current operator that
conserves current, and construct the proper matrix elements of this
operator between deuteron wave functions.  This will be discussed
next.


\subsubsection{Examples of hamiltonian dynamics}
\label{Hexamples}    

The steps taken to construct the current and calculate 
the form factors depend on the form of quantum mechanics
used, and the taste of the investigator involved.  Here
we briefly describe recent work by five groups.  

{\it Forest, Schiavilla, and Riska\/} [FSR]:  Based on 
the work of Schiavilla and Riska \cite{SR91}, Forest and
Schiavilla \cite{FS01} have done an instant-form
calculation of the deuteron form factors.  The original
work of Ref.~\cite{SR91} used on-shell matrix elements of
the one body charge and current operators
\begin{eqnarray}
\fl j^\mu_N(p',p)= \left({m^2\over E(p')E(p)}\right)^{1/2}\bar
u(p',s')\left[F_1(Q^2)\gamma^\mu
+{F_2(Q^2)\over2m}\,i\sigma^{\mu\nu}q_\nu\right]u(p,s)\label{1body1}
\end{eqnarray}
expanded in powers of $v/c$.  Here $F_1$ and $F_2$ are
the Dirac and Pauli form factors, usually replaced by the
familiar charge and magnetic form factors 
\begin{eqnarray}
G_E(Q^2)=F_1(Q^2)-{Q^2\over4m^2}F_2(Q^2)\nonumber\\
G_M(Q^2)=F_1(Q^2)+F_2(Q^2)\, .
\end{eqnarray}
In the recent unpublished work of Ref.~\cite{FS01} the 
calculations have been done in momentum space, where the
one body current operators have been evaluated without
making any $(v/c)$ expansions, the relativistic kinetic
energy $\sqrt{m^2+{\bf p}^2}$ has been used in place of
the usual nonrelativistic expansion $m+{\bf p}^2/(2m)$
[with the parameters of the AV18 potential refitted], and
the boost corrections to the deuteron wave functions have
been included.  This work also includes two-body charge
operators from  $\pi$ and $\rho$ exchange using methods
developed by Riska and collaborators \cite{SR90}.  

{\it Arenh\"ovel, Ritz, and Wilbois\/} [ARW]:  This
recent calculation \cite{ARW00} does a systematic $v/c$
expansion of relativistic effects that arise from the
one body current operator and from contributions from
meson exchanges.  The current operator (\ref{1body1}) is
approximated by  
\begin{eqnarray}
\fl j^{\mu}_N(p',p)\simeq\cases{ F_1\left(1-{{\bf
q}^2\over8m^2}\right)-F_2{{\bf
q}^2\over4m^2}
+\left(2G_M-F_1\right)\left[{i\sigma\cdot({\bf
q}\times{\bf p})\over4m^2}\right] & $\mu=0$\cr 
F_1\,{({\bf p}'+{\bf
p})^i\over2m} + G_M\, {i\,[\,\sigma\times {\bf q}\,]^i\over2m} 
+ {\cal O}\left[\left({v/c}\right)^3\right]   & $\mu=i$} 
\label{1body}
\end{eqnarray}
where $j^{\mu}_N=j^{\mu}_p$ or $j^{\mu}_n$ (with 
appropriate $F_1$ and $F_2$), $\sigma$ is the operator in
the nuclear spin space, and ${\bf q}={\bf p'}-{\bf p}$
is the {\it three\/}-momentum transferred by the
electron. This charge operator is correct to order
$(v/c)^2$, and the $(v/c)^3$ contribution to the current
operator is given in Ref.~\cite{RGWA97}.  The ${\bf
q}^2/8m^2$ correction term to the charge operator is
referred to as the Darwin-Foldy term.  The $\sigma\cdot
({\bf q}\times{\bf p})$ is the spin-orbit term. 

There are ambiguities in all calculations based on 
expansions in powers of $(v/c)^2$.  One ambiguity arises 
from the fact that the square of the three-momentum, ${\bf
q}^2$, depends on the frame in which it is evaluated.  
In the Breit frame, ${\bf q}^2=Q^2$, while in the
center of mass of the final deuteron (the frame preferred
by ARW), ${\bf q}^2=(1+\eta)Q^2$ (recall that $\eta$ was
defined in Sec.~\ref{dffdef}).  A second ambiguity
surrounds the choice of $G_E$ versus $F_1$.   Some
experts \cite{SR91} advocate using $G_E$ (because it is
the correct charge operator) in place of $F_1$.  The
difference between $F_1$ and $G_E$ is of higher order. 
These ambiguities introduce theoretical uncertainty into
any calculation.  The size of this uncertainty depends on
both the value of $Q^2$ and the choice of nucleon form
factors; for example the difference between using $F_1$ 
or $G_E$ can be inferred from the ratios shown in
Fig.~\ref{Ges-study} and is large for the recently
measured JLab form factors and small for the MMD form
factors.  Uncertainties of this kind do not arise if the
calculation is done covariantly, or to all orders in
$(v/c)^2$.    

ARW also include boost corrections originally derived to 
lowest order in $v/c$ by Krajcik and Foldy \cite{KF74}. 
The boost corrections can be written as an operation on
the wave function of the form
\begin{eqnarray}
B\Psi_0(r)=e^{-i\chi}\,\Psi_0(r)\simeq\left(1-i\chi\right)\,
\Psi_0(r)\, ,
\end{eqnarray}
where $\chi=\chi_0+\chi_V$ with   
\begin{eqnarray}
\chi_0= -\left({({\bf r}\cdot{\bf P}) ({\bf p}\cdot{\bf P})\over
16m^2}+{\rm h.c.}\right) + {[({\bf \sigma}_1-{\bf
\sigma}_2)\times{\bf p}]\cdot {\bf P}\over8m^2} \label{boost1} 
\end{eqnarray}
the boost associated with the kinetic energy and the spin 
and $\chi_V$ the boost associated with the potential.  In
(\ref{boost1}), ${\bf r}$ and ${\bf p}$ are the relative
coordinate and relative momentum of the nucleon pair, and
${\bf P}$ is the three-momentum of the moving deuteron.
ARW use the values of $\chi_V$ worked out by
Friar \cite{friar77}, and also include relativistic
effects from retardation, isobar currents, and meson
exchange.  To evaluate the latter a meson exchange model
is needed, and ARW use the interactions and parameters of
the  Bonn OBEPQ potentials [only results from the OBEPQ B
potential are presented in the next section, although
Ref.~\cite{ARW00} includes results from all three OBEPQ
potentials].  Friar has emphasized that relativistic
effects can be moved in and out of the wave functions and
currents by unitary
transformations \cite{friar77,friar80}, so that all of
these effects are ambiguous unless fully defined by the
theory.  Effects due to pair currents or recoil
corrections, shown in Fig.~\ref{MEC}(c) and (d), do not
appear to be included.    ARW state that their
calculations should be good only up to $Q^2\simeq$ 
1.2 GeV$^2$.
 
Corrections to the charge operator to order $(v/c)^2$ 
obtained  from instant-form dynamics and from the
Spectator form of propagator  dynamics have been
compared \cite{Gr91,friar77,friar80,gross78}.  In the 
cases studied, the same total result was obtained from
the sum of {\it all\/} of the corrections, but the
individual terms in the sum were found to have a very
different form even when they appeared to come from the
same physical effects.

{\it Carbonell and Karmanov\/} [CK]:  In this front-form
calculation \cite{CK99}  the direction of the light-front
[denoted by $\omega^\mu=\{1,{\bf n}\}$ where ${\bf
n}^2=1$]  is treated as an unphysical degree of freedom.
Wave functions and amplitudes may depend on $\omega$ but
only those components of scattering matrix elements
independent of $\omega$ will be physical.  It is argued
that this approach will give an explicitly covariant
front-form  mechanics \cite{KS92}.  When applied to the
deuteron form factors there are 11 spin invariants, three
that are physical and 8 that depend on $\omega$ and are
unphysical.  In an exact calculation the 8 unphysical
invariants would be zero, but in approximate
calculations, such as that carried out in
Ref.~\cite{CK99}, they will not be zero.  The deuteron
form factors can be extracted from the three physical
invariants by projecting them from the general result, as
derived in Ref.~\cite{KS92}.  For the choice
$\omega^\mu=\{1,0,0,-1\}$ (corresponding to choosing the 
front-form surface $t^+=0$) this method shows that the
charge and quadrupole form factors can be extracted from
the $J^+$ component of the current (in common with other
treatments), but also shows that the magnetic form factor
{\it cannot be obtained only from this component\/} and
requires a different projection (and includes
contributions from contact terms).  The rules for a
general graph technique for calculating amplitudes in
this formalism are given in Ref.~\cite{CDKM98}.   

Using this method the deuteron wave function will in 
general have 6 components, only three of which have been
found to be numerically large.  In addition to the
familiar $S$ and $D$-state components, the third large
component is proportional to a new scalar function $f_5$,
and adds the term
\begin{eqnarray}
\Psi^5_{abm}({\bf p})
=-\sqrt{3\over4\pi}{i\over2}f_5 \,(\sigma_1-\sigma_2)
\cdot(\hat{\bf p}\times{\bf n}) \;\chi_{_{12}}^{1m}
\end{eqnarray}
to the deuteron wave function displayed in 
Eq.~(\ref{ft1}) (to obtain this form we renormalized the
expression in Ref.~\cite{CK99} so that
$f_1=u$ and used the transformations in Ref.~\cite{BG79}).  In
Ref.~\cite{CK99} $f_5$ is calculated perturbatively using the Bonn
potential from Ref.~\cite{Bonn87} without change of 
parameters. They find that $f_5$ is the largest of the
three components for all momenta greater than 500 MeV,
and believe that the perturbative estimate  is accurate
to about 20\%.  The physical meaning of the $f_5$
contribution has been studied in threshold deuteron
electrodisintegration, where it contributes about 50\% of
important pair term contributions.

{\it Lev, Pace, and Salm\'e\/} [LPS]:  The LPS
\cite{LPS00} calculation is a recent version of a series
of light-front calculations that have assumed the
light-front is fixed (at $t^+=t+z=0$).  In the past,
calculations with fixed light fronts have run into a
problem with the loss of angular momentum conservation,
and before we review the LPS results we will discuss this
issue.

In calculating form factors with fixed light fronts it
has been conventional to choose a coordinate system where
$q^\pm=q^0\pm q^z=0$ and $q_\perp={\bf Q}$.  Current 
conservation is then satisfied if only one component of
the current ($J^+$) is non zero.  Consider the matrix
elements of the deuteron current,
$J^+_{\lambda'\lambda}$, where $\lambda'$ ($\lambda$) 
are the helicities of the outgoing (incoming)
deuterons.   One consequence of the loss of {\it
manifest\/} rotational invariance is that there are {\it
four\/} independent matrix elements of the
$J^+$ current related by the constraint
\begin{equation}
J^+_{00} + 2\sqrt{2\eta} J^+_{+0} - J^+_{+-} -
(1 + 2\eta) J^+_{++} = 0 \ .  \label{angular}
\end{equation}
This is a dynamical
constraint often referred to as the ``angular
condition'' \cite{GK84,chung88}.  The deuteron form 
factors can be extracted from {\it any\/} choice of three of the
matrix elements $J^+_{\lambda'\lambda}$, and if condition
(\ref{angular}) is not satisfied each choice will yield
different results.  The form factors will not be uniquely
determined unless the angular condition is satisfied.   

To avoid (or solve) this problem, LPS work in the Breit 
frame, where $q^\pm=\pm Q$ and $q_\perp=0$.  Current
conservation then requires that $J^+=J^-$.  A current
operator that satisfies these conditions was constructed
in Ref.~\cite{LPS98}.  For elastic scattering this
operator has the form 
\begin{equation} 
\fl J^\alpha={1\over2} \left\{J^{\alpha'}_{free}+ L^\mu_\nu\;
e^{i\pi S_x} \left(J^{\alpha'}_{free}\right)^* e^{-i\pi
S_x}\right\}\qquad
\cases{\alpha=\alpha' & for $\alpha=+, \perp$\cr
\alpha'=+ & for $\alpha=-$}
\end{equation}   
where $J_{free}$ is the free (one body) current operator, and
$L^\mu_\nu$ and $e^{i\pi S_x}$ are rotations by
$-\pi$ about the $x$ axis, $L^\mu_\nu$ in the vector space and
$e^{i\pi S_x}$ in the spinor space.  Note that the definition
insures that $J^+=J^-$ as required by current conservation.  
Using this current, LPS have calculated the deuteron quadrupole 
moment, $Q_d$, to 2\%  accuracy \cite{LPS99}.  The
calculation shown below in Sec.~\ref{thyandexp} uses MMD nucleon
form factors and the Nijmegen II deuteron wave functions. 

{\it Allen, Klink, and Polyzou\/} [AKP]:  The deuteron 
form factors have also been recently calculated using the
point-form of quantum mechanics \cite{AKP01}.  Here there
is no difficulity in writing down manifestly covariant
matrix elements, but there is some ambiguity in deciding
how to impose current conservation.  AKP work in the
Breit frame,  choose a one body impulse current to
describe the $\mu=0, 1$, and 2 components of the current,
and introduce a two body current
$J_2^z$ (which need not be calculated) to insure current
conservation.  New effects come from the way the wave functions
are constructed in point form (``velocity'' states are
constructed), and from the fact that momentum is now a 
dynamical generator, so that the momentum transferred to
the nucleon {\it inside\/} the deuteron is not equal to
the momentum transferred to the deuteron as a whole. 
They argue that the momentum transferred to each nucleon
inside the deuteron is
\begin{equation} 
Q^2_n=(p_1'-p_1)^2= Q^2{4(m^2+{\bf p}_\perp^2)\over m_d^2} \left(
1+\eta\right) \,>\, Q^2(1+\eta)\, .
\end{equation}   
At momentum transfers $Q^2\simeq4$ GeV$^2$ this is a 25\% increase,
and leads to a large suppression of the form factors.  This explains
part of the decrease in the size of the form factors predicted by
this model.  The results reported below use the MMD nucleon form
factors and the AV18 $NN$ potentials. 

We now turn to a brief review of methods using quark degrees 
of freedom.


\subsection{Calculations using quark degrees of freedom}
\label{quarksec}

Calculations based on quark degrees of freedom must 
confront  the fact that the deuteron is at least a six
quark system.  Since the six quarks are identical
(because of internal symmetries) the system must be
antisymmetrized, and it is not clear that the nucleon
should retain its identity when in the presence of
another nucleon.  How does the clustering of the six
quarks into the two three-quark nucleons appear at large
distance scales?  How do we treat the confining forces in
the presence of so many quarks?

The approach to these issues depends on whether of not
$Q^2$ is large enough to justify the use of perturbative 
QCD (pQCD). 

\subsubsection{Nonperturbative methods} 
\label{npQCD}

\begin{figure}
\begin{center}
\vspace*{-0.1in}
\mbox{
   \epsfxsize=4.0in
\epsffile{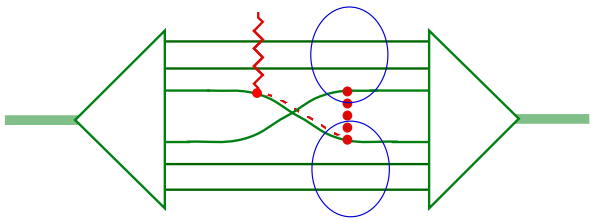} 
} 
\end{center}
\caption{Diagram showing a photon coupling to
a quark exchange diagram.  The exchanged gluon (or meson),
shown by the heavy dotted line, distributes the momentum
equally between the two three-quark clusters, suggesting
the result (\ref{exff}).}
\label{ffexchange}
\end{figure}

At modest $Q^2$ the momentum transferred by the gluons is
small and the QCD coupling is too large for perturbative
methods to be useful.  In the nonperturbative regime
calculations must be based on models.   Many papers
have been written addressing these issues, and a complete
review is beyond the scope of our discussion.  Here we mention
only two contributions that give the flavor of the discussion. 
Maltman and Isgur \cite{MI83,MI84} studied the ground
state of six quarks interacting through a
$qq$ potential previously used to explain the spectrum of
excited nucleons, and found that there was a natural
tendency for the quarks to cluster into two groups of
three (i.e. nucleons).  They obtained a reasonable
description of the deuteron, and confirmed that the short
range $NN$ repulsion could be largely understood in terms
of quark exchange. Later, de
Forest and Mulders \cite{FM87}, using a very simple model,
considered the effect of antisymmetrization on the structure
of the form factor.  Their calculations suggest that the zeros
seen in form factors could be a consequence of antisymmetration
alone. They also show that the factorized form of the impulse
approximation obtained from nonrelativistic (and some
relativistic) theories, which gives the deuteron form factors
as a product of a nucleon form factor and a nuclear (or body)
form factor
\begin{equation}
G_D(Q^2)=F_N(Q^2)\times D_1(Q^2)\, , \label{fac1}
\end{equation}
may not be a good description in the presence of
antisymmetration.  When quarks are exchanged between
nucleons it is no longer possible to separate the {\it
nucleon\/} structure from the {\it nuclear} structure. 
Consideration of the quark exchange diagram shown in
Fig.~\ref{ffexchange} suggests a factorization formula of the
form  
\begin{equation}
G_D(Q^2)=\left[F_N([Q/2]^2)\right]^2\times D_2(Q^2)\, .
\label{exff}
\end{equation}
Because nucleons are composite and identical, either of the
forms (\ref{fac1}) or (\ref{exff}) (or yet other relation) might
hold, and there is no clearly correct way to isolate the
structure of the nucleon from the structure of the bound state.

In model calculations these issues can be handled by
separating the problem into two regions: at large separations
($R>R_c$) it is assumed that the system separates into two
nucleons interacting through one pion exchange, and at small
distances ($R<R_c$) the system is assumed to coalesce into a
six-quark bag with all the quarks treated on an equal footing.
In this review we report the results of a calculation by Dijk
and Bakker \cite{DB88}, where references to other
calculations of this type can also be found (see also the work of
Buchmann, Yamauchi, and Faessler \cite{Bu89}).  

{\it Dijk and Bakker\/} [DB]: This calculation is based on the
quark compound bag model introduced by Simonov
\cite{S81}.  Here the six-quark wave function is assumed
to be the sum of a hadronic part and a quark part.   The
hadronic part is a fully antisymmetrized product of two
three-quark wave functions, each with the quantum numbers of a
nucleon, and a relative $NN$ wave function $\chi_{NN}({\bf
r})$
\begin{equation}
\left|\psi_h\right>={\cal A}\left[\psi_N(123)
\psi_N(456)\;\chi_{NN}({\bf r})\right]\, ,
\end{equation}
where ${\bf r}=({\bf r}_1+{\bf r}_2+{\bf r}_3-{\bf r}_4-{\bf
r}_5-{\bf r}_6)/3$ is the effective internucleon separation. 
The quark part is a sum of eigenstates $\left|\psi_\nu\right>$
of a confined 6-quark system
\begin{equation}
\left|\psi_q\right>=\sum_\nu a_\nu\left|\psi_\nu\right>\, ,
\label{cbstates}
\end{equation}
where the confined 6-quark states are zero outside of a
confining radius $r=b$, which is a parameter of the calculation. 
In the applications, only one term $\nu=1$ needs to be included
in the sum (\ref{cbstates}).  The dynamical quantities
determined by the calculation are the $NN$ wave function
$\chi_{NN}({\bf r})$ and the spectroscopic coefficient $a_1(E)$
which is a function of the energy $E$.  

The $NN$ scattering phase shifts and mixing parameters are
determined by replacing the spectroscopic coefficients by
boundary conditions on the surface $r=b$ and integrating the
Schr\"odinger equation for $r>b$.  The Paris
potential \cite{Paris} is used to describe the $NN$
interaction in the peripheral region and is set to zero
in the inner region.  Two models were developed; in this
review we report results from the fits to the Arndt
single energy 1986 \cite{Ar87} solutions, which DB denote
QBC86.  This fit finds $b=1$ fm.  Calculation of the form
factors requires an assumption about the form factor of
the internal compound bag part of the wave function. 
They use   
\begin{equation}
F_{c1}(Q^2)=\left(1+2Q^2/5\Lambda_1^2\right)^{-5} 
\end{equation}
with $\Lambda_1=1$ GeV obtained from a fit to the $A$ and
$B$ structure functions.  Results from this model are
reported in Sec.~\ref{thyandexp} below.

\subsubsection{Perturbative QCD}
\label{pertQCD}

If one believes the momentum transfer is high enough,
perturbative QCD (pQCD) may be used to study the deuteron form
factor and reactions.  Here it is assumed that the problem
naturally factors into a hard scattering process in which the
momentum transfer is distributed more or less equally to all of
the six quarks, {\it preceeded and followed\/} by soft,
nonperturbative scattering that sets the scale of the
interaction but does not strongly influence its
$Q^2$ behavior.  The $Q^2$ behavior is therefore determined by
the hard scattering, which can be calculated perturbatively. 
The formalism and method are reviewed in the seminal papers by
Brodsky and Farrar \cite{bf} and Lepage and Brodsky \cite{bl}.

\begin{figure}
\begin{center}
\vspace*{-0.1in}
\mbox{
   \epsfxsize=3.0in
\epsffile{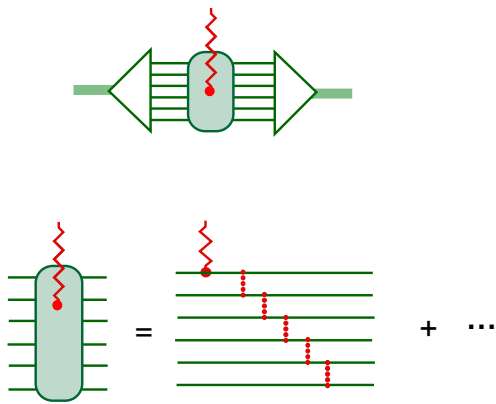} 
} 
\end{center}
\caption{Feynman diagrams that give the pQCD result.  Upper panel shows
the separation of the form factor into the hard scattering part and the
initial and final wave functions that contain the soft scattering.  Lower
panel shows one of the thousands (in this case) of hard scattering
diagrams that make up the hard scattering part.}
\label{pQCDdiagrams}
\end{figure}

These calculations of the form factor are all based on
the  diagrams shown in Fig.~\ref{pQCDdiagrams}.  In the
hard scattering, the momentum transfer $Q$ is distributed
to the six quarks through the five hard gluon exchanges,
the last of which carries a momentum of approximately
$Q/6$. If the spin factors are included with the quark 
propagators, the only large $Q^2$ dependence comes from
the $1/Q^2$ of each gluon propagator, giving the {\it
counting rule\/} for the hard scattering part
$F_h$  
\begin{eqnarray}
F_h(Q^2)\sim \left[Q^2\right]^{-(n_c-1)}\, ,
\label{counting}
\end{eqnarray}
where $n_c$ is the number of constituent quarks (6 for the 
deuteron) and $n_c-1$ the number of gluon propagators.  
This leads immediately to the prediction that the leading
contribution to the deuteron form factor should go like
$Q^{-10}$, or that $A\sim Q^{-20}$.  The argument also 
shows that the perturbative result cannot be expected to
set in until $Q/6>0.5$ to 1 GeV, somewhere in the region
of $Q^2$ from 9 to 36 GeV$^2$.  (All agree that pQCD must
give correct predictions at sufficiently high $Q^2$, but
how large this $Q^2$ must be is a topic of considerable
controversy \cite{ils,ar}.)  Note that this
simple argument does not set the scale of the form
factor; estimates can be obtained from detailed
evaluation of more than 300,000 diagrams that contribute
to the hard scattering \cite{Farrar}.   It turns out
that this leading twist pQCD estimate is 10$^3$ -- 10$^4$
times  smaller than the measured deuteron form factor,
implying large soft contributions to the form factor, in
agreement with \cite{ils,ar}, and suggesting that 
pQCD should not be used as an explanation  for the form
factor. The calculation is extremely complicated and a
confirmation, or refutation, is desirable.

Perturbative QCD also predicts the {\it spin dependence\/} of 
the hard scattering, and these predictions provide a more
stringent test of the onset of pQCD. These spin
dependent predictions have implications for the individual
deuteron form factors, and these were first presented in
Ref.~\cite{cg84}, and further developed in 
Refs.~\cite{brodskyhiller92,kobushkin94,caowu97}.  
Application of these rules to hadronic form
factors in general shows that
\begin{itemize} 
\item Hadrons with an {\it even\/} number of quarks will
be dominated by the {\it longitudinal\/} (charge)
currents, while these with an {\it odd\/} number of
constituents by {\it transverse\/} (magnet) currents. 
Hence, the dominant form factors at large $Q^2$ should be
the nucleon magnetic form factors and the deuteron charge
(or quadruple) form factors.

\item The dominant form factors at large $Q^2$ are those that 
conserve helicity. 

\end{itemize}
When applied to the deuteron, these rules lead to the conclusion that
the helicity amplitude $G^0_{00}$ [c.f. Eq.~(\ref{dhelicity})] dominates
at large $Q^2$; the others are smaller by at least a power of $Q$. 
In particular, this implies that $G^0_{+-}\sim G_C-2\eta
G_Q/3\to(\Lambda/Q)G^0_{00}$, where $\Lambda$ is the mass scale above which
the nonleading terms can be neglected.  Hence  
\begin{eqnarray}
\tilde t_{20}(Q^2)\to -\sqrt{2}\,\left\{1+{\cal
O}(\Lambda/Q)\right\}\, .
\label{t20prediction}
\end{eqnarray}
Unfortunately, this argument does not allow one to estimate the ratio
$B/A$, since $B$ is controlled by a different, independent helicity
amplitude. 

In Ref.~\cite{brodskyhiller92} an attempt was made to
improve on the constraint (\ref{t20prediction}).  These
authors used the front-form, and evaluated the current in
the light-front Breit frame where the plus component of
the momentum transfer $q^+ =0$.  In this frame  all three
deuteron form factors may be written in terms of matrix
elements of the $J^+=J^0+J^3$ component of the current 
\begin{eqnarray}
G_C= \kappa\left\{\left(1-{\textstyle{2\over3}}\eta\right)
G^+_{00} +{\textstyle{8\over3}}\sqrt{2\eta}  G^+_{+0}
+ {\textstyle{2\over3}}(2\eta-1) G^+_{+-}\right\}\nonumber\\
G_M= \kappa\left\{ 2 G^+_{00} + 2(2\eta-1){
G^+_{+0}\over\sqrt{2\eta}}-2 G^+_{+-}\right\}\nonumber\\ 
G_Q=\kappa\left\{ -G^+_{00} +2{
G^+_{+0}\over\sqrt{2\eta}}-{(\eta+1)\over\eta}G^+_{+-}\right\} \, ,
\label{frontff}
\end{eqnarray}
where $\kappa=[2p_+(2\eta+1)]^{-1}$.  Perturbative QCD predicts 
that $G^+_{00}$ will dominate at large $Q^2$, and if this happens at a
scale $\Lambda<<m_d$ it follows from (\ref{frontff}) that the form factors
go in the ratio of $G_C:G_M:G_Q=(1-2\eta/3):2:-1$.  This leads to a
prediction for $B/A$ and to a prediction for $\tilde t_{20}$ that
differs from (\ref{t20prediction}) at moderate $Q^2$.  However, rotational
invariance is not manifest in the light front, and there are {\it four\/}
nonzero components of the $J^+$ current corresponding to deuteron helicity
combinations of $00$, $++$, $+0$ and
$+-$ that are related by the angular condition discussed in
Sec.~\ref{Hexamples} above.  Carlson has recently shown
\cite{carlson01} that the angular condition places strong
constraints on the possible subleading behavior of the
helicity amplitudes.  Perturbative QCD predicts that the
subleading amplitudes will go like
\begin{equation}
\fl G^+_{+0} \to a (\Lambda/Q)G^+_{00}\ ,\qquad G^+_{+-} \to b
(\Lambda/Q)^2 G^+_{00}\ ,\qquad  G^+_{++} \to c (\Lambda/Q)^2G^+_{00}
\ , \label{BrodG}
\end{equation} 
where $a$, $b$, and $c$ are dimensionless constants of the order of
unity, and $\Lambda$ is the scale at which pQCD begins working for the
deuteron [the $G$s in Eq.~(\ref{BrodG}) are identical to the $J$s in
Eq.~(\ref{angular})].  Assuming that
$Q^2>>\Lambda^2$ (but making no assumption about the size of $\Lambda$)
the angular condition in leading order becomes:
\begin{equation}
\left\{1+\sqrt{2}a{\Lambda\over m_d}-{c\Lambda^2\over2m_d^2}
\right\}G^+_{00}=0\ .
\end{equation}
Solution of this equation therefore requires that $\Lambda\sim m_d$
and $Q^2>>m_d^2$.  Under these conditions the relations
(\ref{frontff}) again produce {\it only\/} the result 
(\ref{t20prediction}).


\subsection{Comparison of theory with experiment} 
\label{thyandexp}
  
In this section we compare theory with experiment and draw
conclusions from this comparison.  Our major conclusions will
be restated and summarized again in Sec.~\ref{conclusions}
below.

\begin{figure}[t]
\begin{center} 
\mbox{
   \epsfxsize=4.0in
\epsffile{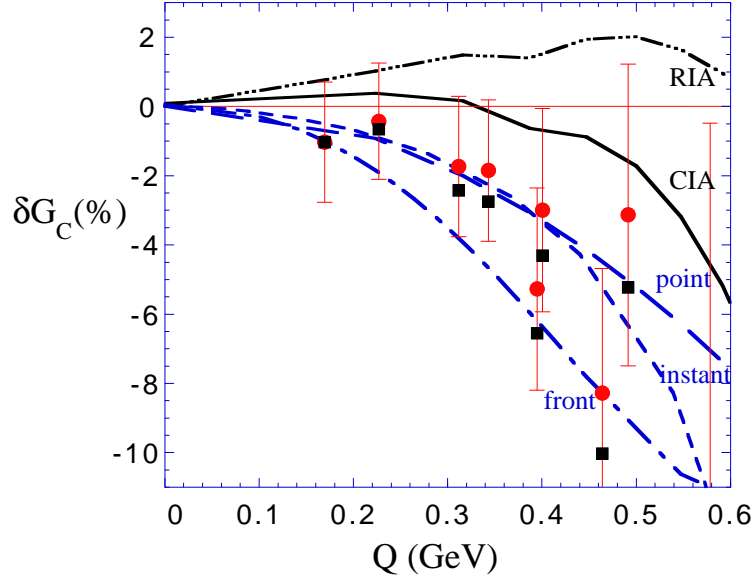} 
} 
\end{center}
\caption{Corrections to the charge form factor, $G_C$.  Each curve shows the
difference $G_C(full)-G_C(nr)$ expressed as a percentage of the
nonrelativistic result $G_C(nr)$.  The individual cases are discussed in the
text.  The squares without error bars are the extracted $G_C$ data from
Ref.~\cite{abbottphen00} and the circles with errors are the Coulomb corrected data as
discussed in the text.}
\label{GC-Relcorr}
\end{figure}

\subsubsection{The charge form factor at very small
$Q^2$}   
Figure \ref{GC-Relcorr} shows how coulomb distortion of the 
incoming and outgoing $ed$ plane waves effects the very low
$Q^2$ data  (extracted from Ref.~\cite{abbott00}) for the
charge form factor, $G_C$.  The figure also compares this data
with theory.

Corrections for Coulomb distortion change the deuteron radius
from an apparant 2.113 fm (as measured in $ed$ scattering) to
2.130 fm (after the correction) \cite{sick96}.  To
remove the distortions from the data of Ref.~\cite{abbott00},
we adjust $G_C$ by \cite{Sick01}
\begin{eqnarray}
\delta G_C\simeq -0.003+0.104\,Q^2
\, . \label{coulombdis}  
\end{eqnarray} 
Note that this decreases $G_C$ at very small $Q$ giving a
larger deuteron radius, but increases $G_C$ where the data
have been extracted.  The figure shows both the uncorrected
and the Coulomb  corrected data normalized to the
nonrelativistic AV18 calculation with the MMD  nucleon form
factor.  Note that the difference between the Coulomb
corrected and uncorrected data is about half of the
experimental error at
$Q\simeq$ 0.5 GeV.  

The figure also shows the size of relativistic and interaction
current corrections that arise from the instant-form
calculation ARW of Ref.~\cite{ARW00}, the front-form
calculation LPS of Ref.~\cite{LPS00}, the point form
calculation AKP of  Ref.~\cite{AKP01}, and the CIA and RIA
approximations from Ref.~\cite{vdg95}.  These calculations
were discussed in  Sec.~\ref{pexamples} and
\ref{Hexamples} above.  At the scale of the current 
experimental accuracy (a few percent), the 
relativistic treatments {\it differ noticeably\/}.  They
also differ from the EFT calculation (shown previously in
Fig.~\ref{fig:EFT}) which drops sharply below the data
for $Q>$ 0.2 GeV.  It is
important that these  calculations be systematically
compared and the different physical content of these
approaches be isolated and understood.  In particular, it
would be very interesting to know why the covariant CIA and
RIA have more positive corrections than those obtained from
the hamiltonian forms of dynamics. 

\begin{table}
\caption{\label{tab:theories}
Features of the eight theoretical models reviewed in
Secs.~\ref{pexamples}, \ref{Hexamples}, and \ref{npQCD}.}
\begin{indented}
\item[]\begin{tabular}{@{}llll}
\br model & dynamics & description & consistent current \\
\mr
VOG \cite{vdg95} & propagator & Spectator  & yes\\  
PWM \cite{PW} & propagator & modified
Mandelsweig-Wallace & no \\
FSR \cite{FS01} & hamiltonian & instant-form; no $v/c$
expansion & yes\\ 
ARW \cite{ARW00} & hamiltonian & instant-form; with $v/c$
expansion & yes \\
CK \cite{CK99} & hamiltonian & front-form; dynamical
light-front & no   \\ 
LPS \cite{LPS00} & hamiltonian & front-form; fixed light-front
& no \\
AKP \cite{AKP01} & hamiltonian & point-form & no\\ 
DB \cite{DB88} & nonrelativistic & quark-cluster  & 
yes \\
\br 
\end{tabular}
\end{indented}
\end{table}

\begin{figure}[t] 
\begin{center}
\mbox{
   \epsfxsize=6.5in 
\epsffile{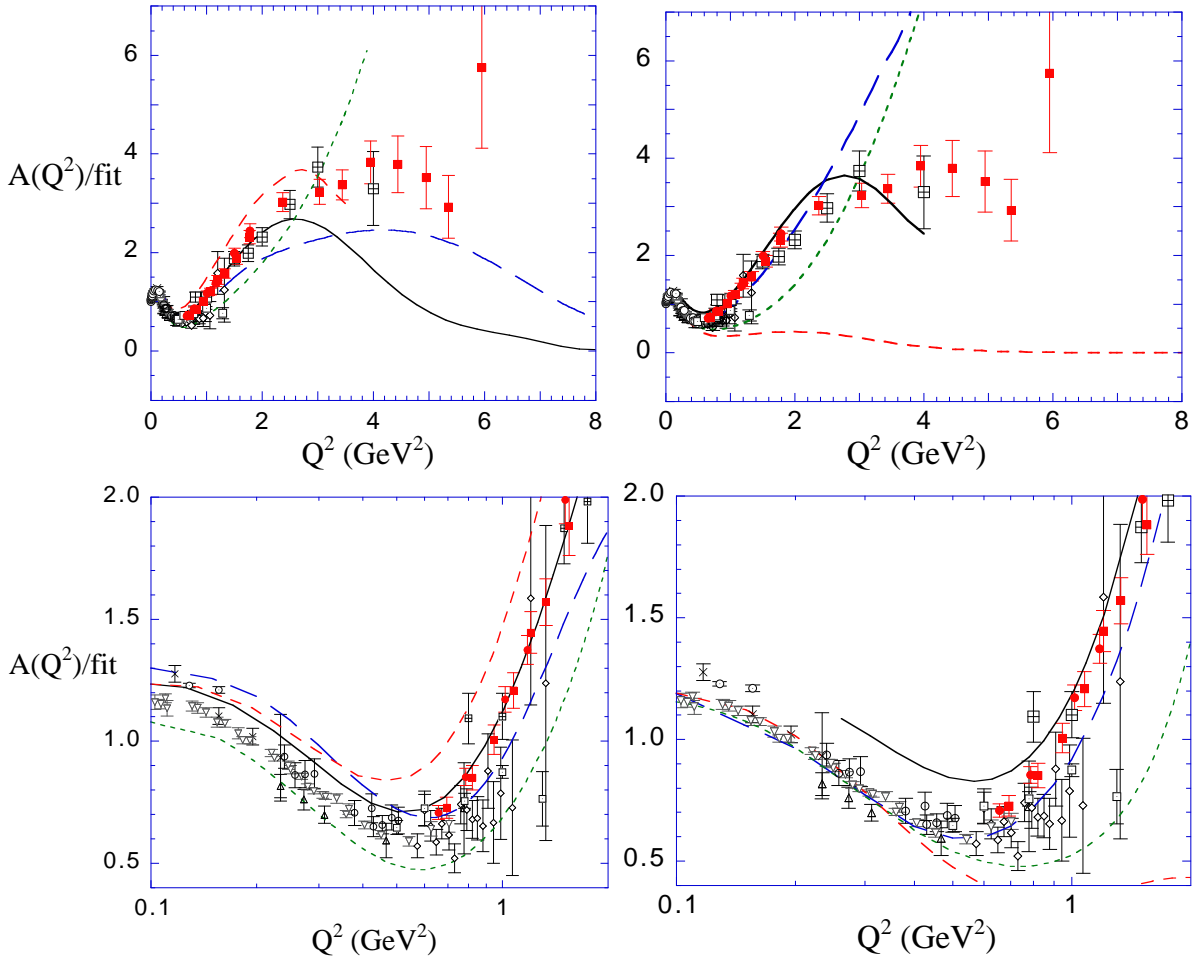}  
} 
\end{center}
\caption{The structure function $A$ for the eight
models discussed in the text. Left panels show the propagator
and instant-form results: FSR (solid line), VOG in RIA
approximation (long dashed line), ARW
(medium dashed line), and PWM (short dashed line).  Right panels
show the front-form CK (long dashed line) and LPS (short
dashed line), the point-form AKP (medium dashed line) and the
quark model calculation DB (solid line).  In every case the
calculations have been divided by $\tilde A$ calculated
from the fit (\ref{scale}).    See Table~\ref{Adat} for
references to the data.}
\label{AlowQ27}
\end{figure}

\subsubsection{Overview of the high $Q^2$ predictions}  
\begin{figure}[t]
\begin{center}
\mbox{
   \epsfysize=3.0in
\epsffile{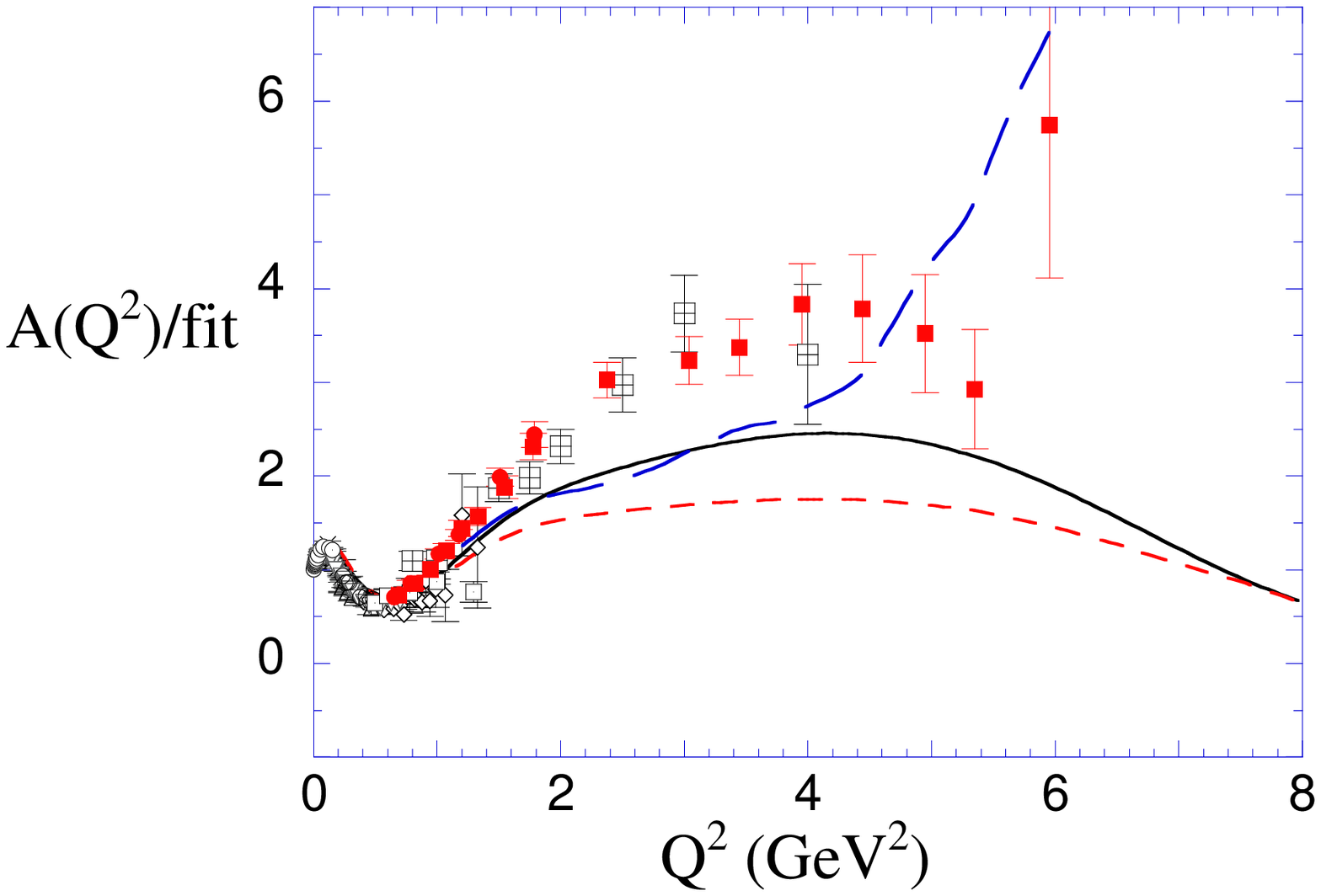} 
} 
\end{center}
\caption{Effects of the new JLab measurements and the
$\rho\pi\gamma$ exchange current on predictions for $A$.  The
solid line is the VOG model prediction (in RIA approximation)
with MMD nucleon form factors (identical to the long dashed
curves shown in Fig.~\ref{AlowQ27}).  The short dashed line is
the VOG prediction (in RIA approximation) with JLab model
nucleon form factors, Eq.~(\ref{JLabges}).  The long dashed
curve is the full VOG calculation (CIA, $\rho\pi\gamma$
exchange current with the form factor of Ref.~\cite{Rome2},
and MMD form factors).   }
\label{JLabA}
\end{figure}

\begin{figure}[t]
\begin{center}
\vspace*{-0.3in}
\mbox{
   \epsfysize=7.9in
\epsffile{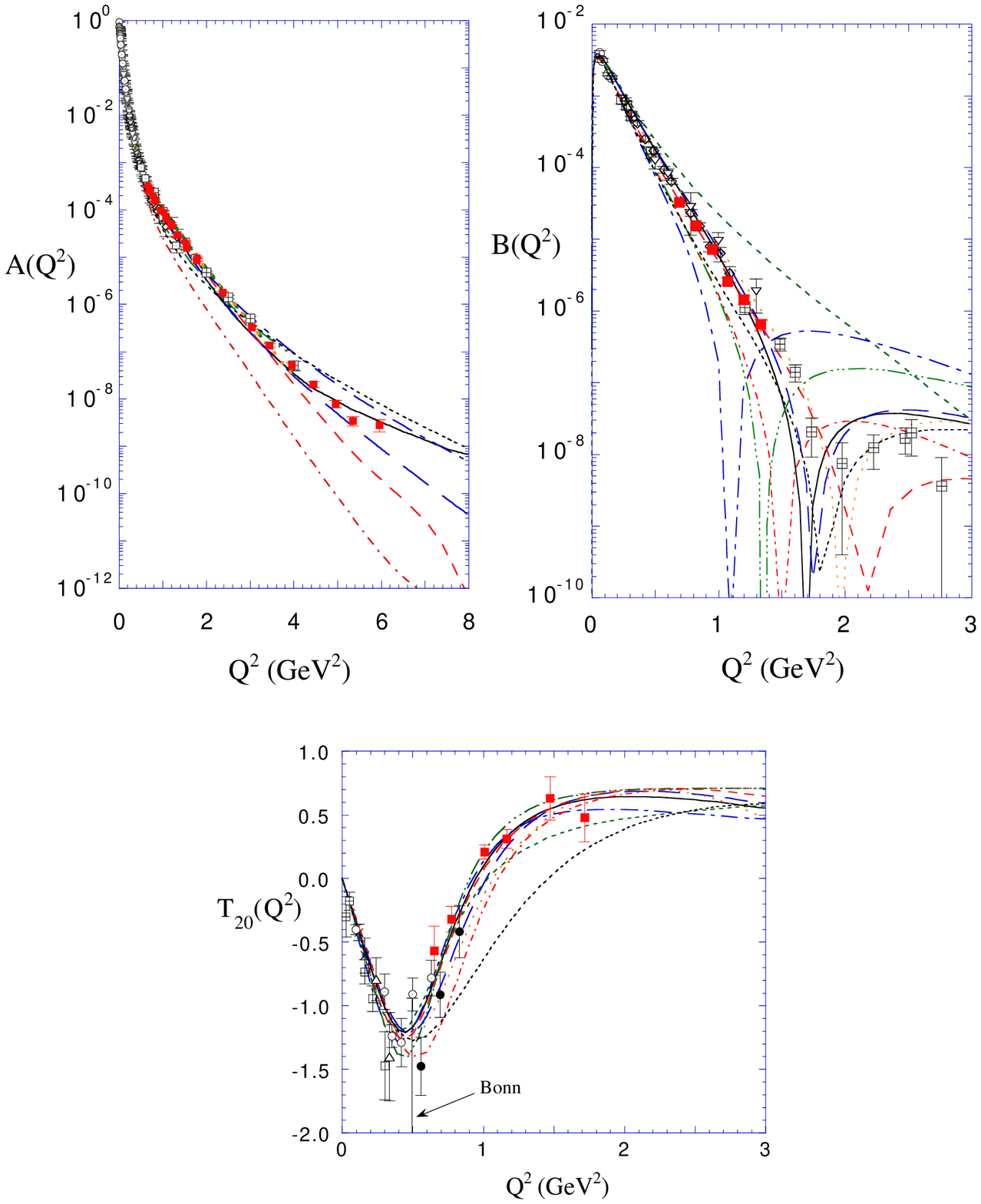} 
} 
\end{center}
\vspace*{-0.1in}
\caption{The structure functions $A$, $B$, and $T_{20}$.  
The models, in order of the $Q^2$ of their mimima in $B$, are:
CK (long dot-dashed line), PWM (dashed double-dotted line),
AKP (short dot-dashed line),
VOG full calculation (as shown in Fig.~\ref{JLabA} -- solid line),
VOG in RIA (long dashed line), LPS (dotted line),
DB (widely spaced dotted line), FSR (medium dashed line), and
ARW (short dashed line). See Tables~\ref{Adat}, \ref{tab:edelastb}, and
\ref{t20tab} for references to the data.  }
\label{RABT20}
\end{figure}

The high $Q^2$ predictions for the eight models reviewed
in Secs.~\ref{pexamples}, \ref{Hexamples}, and \ref{npQCD}
are shown in Figs.~\ref{AlowQ27}--\ref{RABT20}.  The models
are summarized in Table \ref{tab:theories}.

These calculations give very different results. 
Figure \ref{AlowQ27} shows the predictions for $A(Q^2)$,
with the model dependent $\rho\pi\gamma$ exchange current
intentionally omitted from all of the calculations.  All of
the models except the AKP point-form calculation give a
reasonable description of
$A$ out to $Q^2\sim 3$ GeV$^2$, beyond which they begin to
depart strongly from each other and the data.  Taking into
account that the $\rho\pi\gamma$ exchange current {\it could be
added to any of these models, and that this contribution
tends to increase A above $Q^2\sim 3$ GeV$^2$\/} 
(for the sign of the $\rho\pi\gamma$ coupling constant
used in
the discussion in the following paragraph), four models seem to
have the right general behavior: the VOG, FSR, ARW and the
quark model of DB (but there are no results for this model
beyond $Q^2=4$ GeV$^2$).  Ironically, none of the models favored by the
high $Q^2$ data does as well at low $Q^2$ as the three
``unfavored'' models shown in the right panels (unless
the Platchkov \cite{plat90} data are systematically too
low).  Another possibility suggested by effective field
theory \cite{PRS00} is that the asymptotic
normalization of the relativistic deuteron wave functions is
incorrect, and that a small adjustment in $NN$ parameters to
insure a good value for this constant would correct the 
problem. 
 
Figure \ref{JLabA} shows the effect of the new JLab
measurements of the nucleon form factors on predictions for
$A(Q^2)$.  These new form factors will {\it decrease\/}
predictions for $A$ for momentum transfers in a region around
$Q^2=4$ GeV$^2$ (by a factor of 2 for the model shown),
increasing the descrepancy between predictions and the data. 
[However, it may improve the prediction for those models
(PWM, CK, and LPS) that are currently too large in this
region.]  The figure also shows how the $\rho\pi\gamma$
exchange current could increase predictions at large $Q^2$. 
The difference between the solid and long dashed lines is
due largely to the effect of the $\rho\pi\gamma$
exchange current (but is also do in small part to the
fact that the CIA result is slightly smaller that the RIA). 
Unfortunately, the size of the $\rho\pi\gamma$ exchange current
is very sensitive to the $\rho\pi\gamma$ form factor, as
discussed in Ref.~\cite{Ito93}, and could even be too small to
see at these momentum transfers (if the current estimates of
the $\rho\pi\gamma$ form factor are too large).  This is our
reason for insisting that this contribution should be viewed
as new physics --- not readily predictable within a meson
model. 
 
Finally, Fig.~\ref{RABT20} shows the predictions for the
structure functions $A$, $B$, and $T_{20}$ for the eight
models discussed.  The LPS calculation shows a
large descrepency with the $T_{20}$ data, but the most
striking feature of these plots is the {\it large model
dependence\/} of the predictions for $B(Q^2)$.  The magnetic
structure function provides the most stringent test, and the
predictions are comparatively free of the
$\rho\pi\gamma$ exchange current (which gives only a small
contribution to $B$).  Examination of the figure shows
that the $B$ predictions of the PWM, ARW, AKP, CK models fare
the worst.  In all, taking the predictions for the three
structure functions together, the best results are obtained
with the FSR, VOG, and DB models.

\begin{figure}[t]
\begin{center}
\vspace*{-0.2in}
\mbox{
   \epsfysize=7.5in
\epsffile{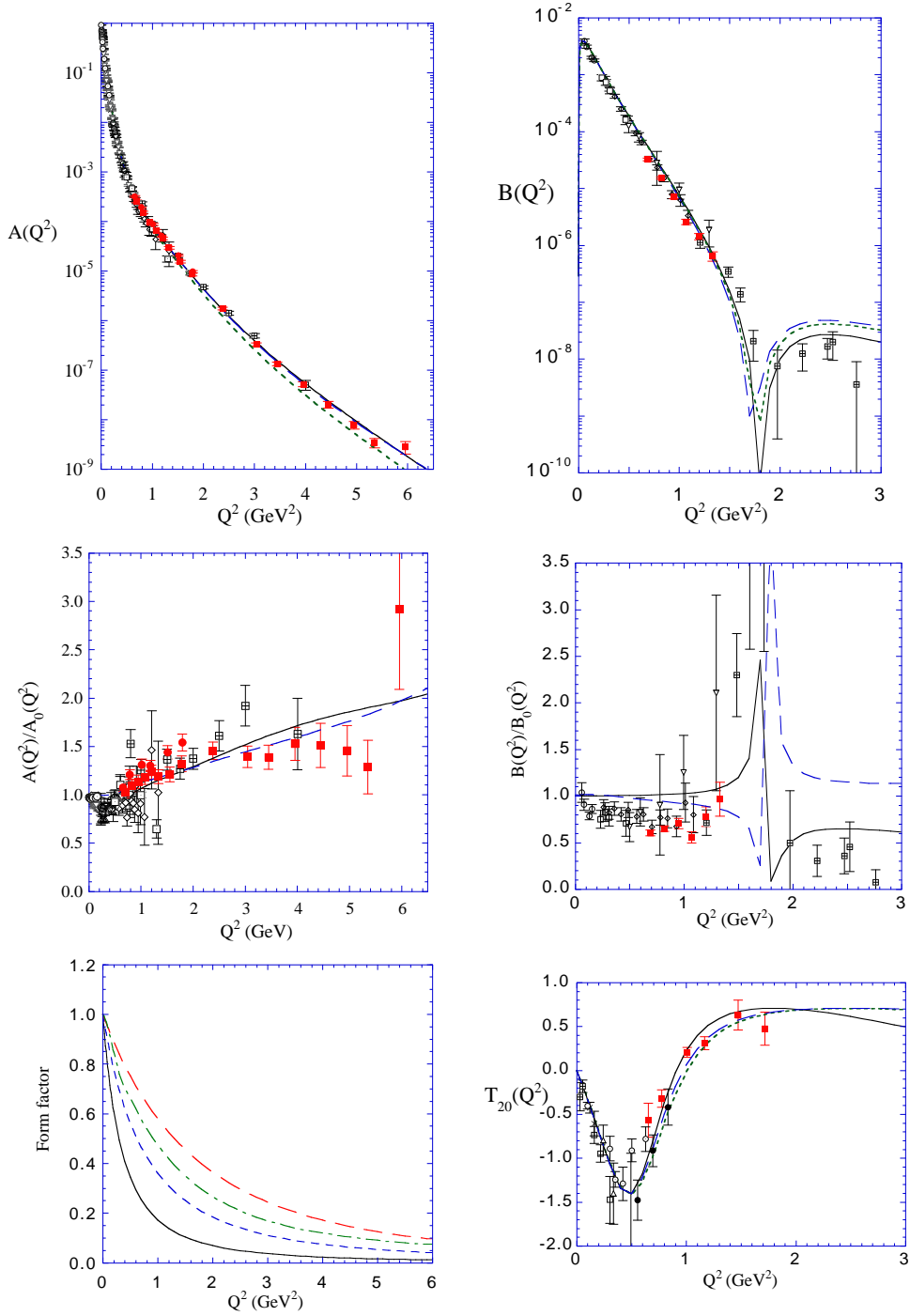} 
} 
\end{center}
\vspace*{-0.1in}
\caption{Upper panels ($A$ and $B$) and lower right panel
($\tilde T_{20}$) compare data to three theoretical
models based on VOG: (i) ``standard'' case referred to in
the text (dotted line), (ii) model with the tripole $F_3$
(solid line), and (iii) model with the dipole
$f_{\rho\pi\gamma}$ (dashed line).  The center two panels
show the data and models (ii) and (iii) divided by model
(i).  The lower left panel shows form factors: standard dipole
with $\Lambda^2=0.71$ (solid line), dipole with
$\Lambda^2=1.5$ (short dashed line), Rome
$f_{\rho\pi\gamma}$ (dot-dashed line) \cite{Rome2}, and the
tripole with $\Lambda^2=5$ (long dashed line).}
\label{fig:F3}
\end{figure} 
     
\subsubsection{Important lessons}

We conclude our comparison with theory by extending our
discussion somewhat beyond the limits of previously published
papers.  To make the point clearly, focus on the VOG
calculation using the Spectator equation, and recall that
current conservation {\it required\/} that the single nucleon
current in this approach, Eq.~(\ref{onej}), include a new
form factor, $F_3(Q^2)$.  This form factor must satisfy the
constraint $F_3(0)=1$, but is otherwise {\it completely
unspecified\/}.  In the published VOG calculation \cite{vdg95}
and in all of the plots shown so far, this form factor was
taken to be the standard dipole, $F_D$.    

Figure \ref{fig:F3} shows how agreement between theory and
experiment can be significantly improved by choosing a
different form for the unknown form factor $F_3$.  This
figure shows three theoretical predictions: (i) the
``standard'' VOG RIA prediction with $F_3=F_D$ and {\it no\/}
$\rho\pi\gamma$ exchange current, (ii) another model with 
{\it no\/} $\rho\pi\gamma$ exchange current, but with a
tripole $F_3$ of the form 
\begin{equation}
F_3(Q^2)={1\over (1+Q^2/5)^3} \, , \label{tripole}
\end{equation}   
and (iii) a model with $F_3=F_D$ and a dipole $\rho\pi\gamma$
form factor
\begin{equation}
f_{\rho\pi\gamma}(Q^2)={1\over (1+Q^2/1.5)^2} \, .
\label{rpgdipole}
\end{equation}   
In both form factors $Q^2$ is in GeV$^2$.  These form
factors are shown in the bottom left panel of the figure.

Figure \ref{fig:F3} shows that a good agreement between
theory and experiment can be obtained with the tripole
$F_3$ without any $\rho\pi\gamma$ exchange current, and that
to some extent the $\rho\pi\gamma$ exchange current can be
substituted for a hard $F_3$ (although $F_3$ is better than
$f_{\rho\pi\gamma}$ at improving all three observables
simultaneously).  We see that we have, in some sense, achived
the goals of a theory of elastic $ed$ scattering based on
nucleon degrees of freedom.  {\it With small adjustments of
unknown form factors associated with short range physics, the
NN theory can describe all three form factors quite well\/}.

It seems likely that any nucleon model with a {\it
consistent and complete\/} description of the current  (c.f.\
Table \ref{tab:theories}) can do as well.  The reasonable
results obtained from the FSR and DB models are probably due
to the fact that they have consistent, complete currents not
based on an expansion in powers of $(v/c)^2$ (which must fail
at high $Q^2$).

 
\subsection{Future prospects}

\subsubsection{ Future prospects for A }

From the discussions in Sec.~\ref{thyandexp} above, it is 
clearly of interest to extend measurements of $A$ to 
higher $Q^2$.  An $ed$ coincidence experiment is
straightforward, but prohibitive timewise with
present accelerators.  The proposed 12 GeV JLab
upgrade allows one to take advantage of the
approximate $E^2$ scaling of $\sigma_M$ at constant $Q^2$
and high energy \cite{ggppc00}. A large acceptance
spectrometer such as MAD would be very helpful. Depending
on the details of the upgrade, a one month experiment
could provide data to
$Q^2$ of 8 GeV$^2$.

It might also be desirable to do a new, high precision experiment
at {\em low} $Q^2$.  The goal of this experiment would be to
resolve the discrepancy between the data sets of
Refs.~\cite{simon81} and \cite{plat90}, and to check the low
$Q^2$ limit of the relativistic calculations.  These
measurements require little time, but do  require excellent
control of systematic uncertainties, at the level of 1 -- 2\%,
if they are to be meaningful.
 
\subsubsection{ Future prospects for B }

A hasty examination of Fig.~\ref{RABT20} might lead
one to believe that the problems with $B$ are mainly
theoretical, and that there is no need for new data.  We believe
this attitude would be inappropriate for two reasons.  First,
some of the calculations shown in the figures are in early
stages of development, and will improve before any  new
data are available.  The number of calculations is a
reflection of the challenge that the high $Q^2$ data
present, and a reflection of interest in, and knowledge
of relativistic methods that is emerging from the study
of these measurements.  Second, and most important,
measurements of
$B$ vary by 5 orders of magnitude.  The existence and position
of the first minimum has not yet been firmly established, and the
location and existence of a possible second minimum is
unknown. These minima in $B$ (if they exist) result from
cancellations of various physical effects and provide a
very precise test of any theory. 

As indicated above, measurements of $B$ do not require 
high energy, but do require large scattering angles, as
close as possible to 180$^\circ$.  At 180$^\circ$,
beam energies of 1 to 2 GeV cover a
$Q^2$ range from 1.4 GeV$^2$ to 5 GeV$^2$.
 
The SLAC NE4 experiment was a heroic effort, run with
$ed$ coincidences.
Energy resolution was limited by thick targets, 20 - 40 cm long.
There was a large but manageable
background of $\pi^-$ in the $e^-$ spectrometer, about 3
or 4 to 1, and an extremely large proton background from 
$\gamma d \rightarrow p n$ in the deuteron channel - up to two events 
per beam pulse were seen in the worst kinematics.
Since NE4 ran at beam currents of up to 50 mA instantaneous,
with a duty factor of about 0.3 $\times$ 10$^{-3}$,
corresponding to an average current of about 15 $\mu$A,
the JLab continuous beam structure essentially eliminates random coincidence
backgrounds such as these.
Background coincidence reactions included
$\gamma d \rightarrow \gamma d$ and
$\gamma d \rightarrow \pi^0 d$ at 180$^{\circ}$,
with the photon producing an electron detected in the electron channel.
The spectrum of these photoreactions ends near the elastic peak,
allowing the background contributions to be fit and determined. 
A JLab experiment would run with both larger spectrometer acceptance and
higher luminosity to increase rates, 
but with a shorter target to reduce these backgrounds.

The proposed configuration \cite{ggppc00} would use the Hall A
septum magnets to detect the forward-going deuterons at 
angles of 3 - 6$^{\circ}$,
along with special electron channels to detect scattered electrons
at about 160 - 170$^{\circ}$.
Based on the SLAC NE4 cross sections, a one month experiment
can map out $B$ to about 6 GeV$^2$.

\subsubsection{ Future prospects for $t_{20}$ }

Extending $t_{20}$ or other polarization observables
to higher $Q^2$ is quite difficult \cite{garcon??,gilman??,kox??}.
There are three obvious possibilites.

Recoil tensor polarimetry requires a well calibrated polarimeter
with a large figure of merit, but no such device exists.
For example, POLDER, used in the JLab Hall C $t_{20}$ experiment, 
relied on the $^1H(\vec{d},pp)n$ reaction,
for which the figure of merit decreases at larger energies,
leading to a practical upper limit in $Q^2$, that was reached in E94-018.
HYPOMME \cite{egle95,ladygin98,egle99} is promising, but not well enough
calibrated.

The combination of polarized electron beams with recoil vector 
polarimeters is an untested possibility \cite{acg81}.
With $A$, $B$, and $G_M$ known, $p_z$ is calculable and can calibrate the
polarimeter analyzing power, while $p_x$ determines the form factor
combination $G_C \, + \eta G_Q/3$.
(The ratio of the two polarization components depends on this combination
of form factors, times kinematic factors and divided by $G_M$.)
Since the JLab polarized source can provide 50 -- 100 $\mu$A
beams, there is no luminosity problem.
The difficulty with this measurement is that the polarization 
components are expected to be small in the $Q^2$ range of interest, 
$\approx$ 0.01, and $p_x/p_z$ $\sim$ 5 -- 10.
A one month measurement for one $Q^2$ of $\approx$ 2 - 2.5 GeV$^2$ 
can determine the {\em polarization components} well,
but if these are small as expected, 
the {\em form factors} can 
only be extracted with factor of two uncertainties.

An alternative is to use asymmetries from a polarized target.
However, the reduced currents that can be used with polarized targets
require a large acceptance detector such as CLAS, to make up for the 
lack of luminosity.
Also, although current polarized targets have moderately large
deuteron vector polarizations, tensor polarizations are small.
Furthermore, the asymmetry varies as 
$P_2(\hat{n}\cdot\hat{q})$, with $\hat{n}$ the polarization direction,
so it is desirable to have complete azimuthal coverage, with $\hat{n}$
in the direction of $\hat{q}$ at one azimuthal angle, 
rather than being purely transverse.
Extensive beam time would be needed, either 
as an external polarized target experiments at JLab, or 
as an internal polarized target experiment at HERMES.
We note that a series of moderate $Q^2$ measurements are
planned with the MIT Bates BLAST detector \cite{t20blast},
for $Q^2$ from 0.1 to 0.9 GeV$^2$. 

Proposed next generation colliders, such as EPIC/ERHIC,
are  promising due to large planned luminosities;
for this experiment the lower proposed
c.m.\ collision energies are desirable for ensuring exclusivity.
The spin direction of the polarized deuteron beam must be controllable.
In collider kinematics, the scattered electron and deuteron energies 
are close to their respective beam energies and are slow functions of $Q^2$,
while the scattering angles for fixed $Q^2$ vary slowly with the
beam energies.
Thus, if an experiment is possible, it would attempt large
azimuthal coverage of coincidence $e\vec{d}$ elastic scattering,
with the outgoing particles at angles from a few to 
about 20$^\circ$ from the beam line. 


\subsection{Conclusions to Section 3}
\label{conclusions} 

Comparison of theory and experiment leads to the
following conclusions:

\begin{itemize} 

\item Nonrelativistic quantum mechanics (without exchange 
currents  or relativistic effects) is ruled out by the $A(Q^2)$
data at high $Q^2$.  Reasonable variations in nucleon form
factors or uncertainties in the nonrelativistic wave functions
cannot  remove the discrepancies.
 
\item In some relativistic approaches using $NN$ degrees of
freedom only, short range physics not calculable within the
model ($F_3$ or $f_{\rho\pi\gamma}$, for example) can be
adjusted to give good agreement with all the data.  
 
\item Different ways of calculating relativistic effects
(or meson exchange currents) can give results that differ
substantially from each other.  Even at low $Q^2$, where all
calculations are constrained, these differences are larger
than errors in the data.  This is not understood, but may be
due to the failure of some models to use realistic currents.

\item The deuteron form factors provide
no evidence for the onset of perturbative QCD, but quark cluster
models could explain the data. 

\end{itemize}
Study of the experimental situation leads to the
following conclusions:

\begin{itemize}  

\item A good database of $A$, $B$, and $t_{20}$
measurements has been obtained; while discrepancies exist
they are generally not large enough to affect the
theoretical interpretation.

\item The minimum of $B$ is very sensitive to details of 
the models, and improved measurements of $B$
for $Q^2$ in the region 1.5 - 4 GeV$^2$ are particularly
compelling.  It is important to accurately map out the
zero in the $B$ structure function.

\item Detailed disagreements between theories and
different data sets suggests the need for precision
studies at low $Q^2$.
 
\end{itemize}



\section{ Deuteron Electrodisintegration } 
\subsection{Introduction}

As the deuteron has no excited bound states,
inelastic scattering experiments have largely
consisted either of (i) measurements in which the final
state mass, W, is very close to $2m$ (referred to as
{\it threshold electrodisintegration\/} even when $Q^2$ is
very large because the final state is close to the $NN$
scattering threshold), (ii) measurements near the quasifree
peak (defined by the condition that the ``spectator'' nucleon
remain at rest), or (iii) deep inelastic scattering in which
both $Q^2$ and $W$ become very large.  We will not discuss
deep inelastic scattering in this review.  For processes at
modest energies near the quasielastic peak, a rough estimate of
the cross section can be made using the {\em unrealistic}
plane wave impulse approximation (PWIA) in which all final
state interactions are ignored.  Denoting the momentum of the
outgoing struck nucleon by ${\bf p}_1$, the cross section in
PWIA is proportional to
\begin{equation}
d\sigma_{\rm PWIA} \simeq G^2_{N}(Q^2) \left<{\Psi}
({\bf p}_1-{\bf q})\right>^2 \label{PWIA}
\end{equation}   
where $G^2_{N}$ is some combination of the squares of the
electric and magnetic form factors of the nucleon, and
$\left<\Psi({\bf p})\right>^2$ is an average of the square of
the momentum space wave function of the deuteron with
internal relative momentum ${\bf p}$.  Were the PWIA
realistic, Eq.~(\ref{PWIA}) shows that inelastic scattering
in quasielastic kinematics would provide a direct measure of
the (square) of the deuteron wave function.  While the
PWIA is overly simplistic, it does illustrate (correctly) one
of the central justifications for quasielastic
measurements.   

Within the context of a more realistic dynamical theory, one
can use response function separations and polarization
observables to enhance the sensitivity to various
model dependent {\em nonobservables}, such as momentum
distributions, meson-exchange currents, and medium
modifications. One strong recent interest has been to choose
kinematics in which the unobserved nucleon has a
large momentum; the plane wave approximation shows that this
configuration enhances sensitivity to initial-state short
range correlations (i.e.\ the wave function) and possible
quark effects. A number of these experiments have been
carried out at various accelerators, but no experiments at
JLab have yet reported results. Thus, an experimental review
of this topic is  unwarrented at this time. However, because
photodisintegration, electrodisintegration,  and threshold
electrodisintegration are closely related theoretically, we
present the theoretical background for these processes below.

\subsection{Cross section and polarization observables}

In this subsection the cross section and polarization
observables for electrodisintegration and photodisintegration to
an $np$ final-state are reviewed briefly. (We do not discuss
pion and meson production.)  The electroproduction cross
sections will be obtained first, and photoproduction will then
be treated as a special case.

\begin{figure} 
\leftline{\epsfysize=3.0in \epsfbox{
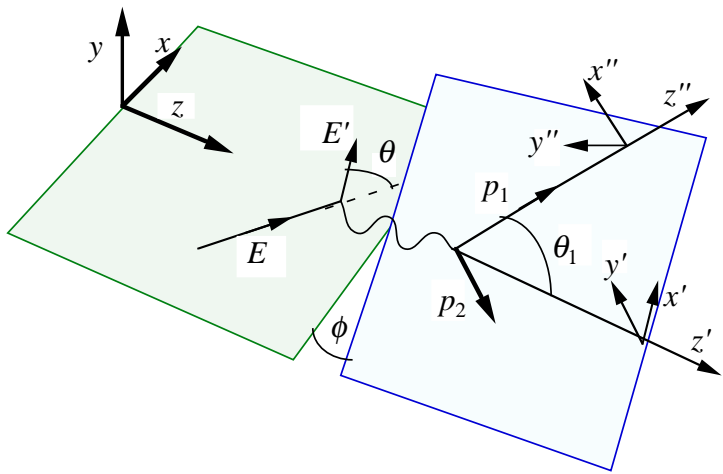}}
\caption{The kinematics of electron scattering when the initial 
hadronic state  is broken into two fragments with momenta $p_1$
and $p_2$.}
\label{fig_cross}
\end{figure}

\subsubsection{Electrodisintegration}
 
The most general decomposition of the $d(e,e'p)n$ coincidence cross section
was first discussed by Donnelly and Raskin \cite{raskin89}.  
Here we will follow the later work of Ref.~\cite{dg89}.  The
cross section can be shown to have the form (Eq.~(95) of
Ref.~\cite{dg89})
\begin{eqnarray}
{d^5\sigma\over d\Omega dE^{\prime}d\Sigma} &=& 
{\sigma_M\over 4\pi  m_d}\frac{Q^2}{{\bf q}_L^2}
\biggl\{\tilde{R}^{({\rm I})}_L +   s_T\tilde{R}^{({\rm I})}_T 
-\frac{1}{2}\left[\cos 2\phi \tilde{R}^{({\rm I})}_{TT}
+ \sin 2\phi 
\tilde{R}^{({\rm II})}_{TT}\right] \nonumber\\
&& + s_{LT}\left[\cos\phi \tilde{R}^{({\rm I})}_{LT} + 
\sin\phi\tilde{R}^{({\rm II})}_{LT}\right] + 2 h \,  
s_{T'}\tilde{R}^{({\rm II})}_{T^{\prime}} \nonumber\\
&& + 2hs_{LT'}\left[\sin\phi\tilde{R}^{({\rm I})}_{LT^{\prime}}
+\cos\phi\tilde{R}^{({\rm II})}_{LT^{\prime}}\right]\biggl\}\, ,
\label{crosssec}
\end{eqnarray}
where, in the lab frame, $q=\{\nu,{\bf q}_L\}$, and the nine
response functions are functions of
$Q^2=-q^2$, $\nu=Q^2/2mx$, and the
scattering angle
$\theta_1$ of the final-state proton (measured in a 
coincidence experiment and integrated over in an inclusive
measurement). The ejectile plane is  tilted at an angle
$\phi$ with respect to the electron scattering plane, as
illustrated in Fig.~\ref{fig_cross}.  The Mott cross section
and  other variables are as defined in Eq.~(\ref{mott}). The
electron kinematic factors are 
\begin{eqnarray}
s_T=\frac{1}{2}+\xi^2 \qquad\qquad s_{LT}=-\frac{1}{\sqrt{2}}\left(
1+\xi^2\right)^\frac{1}{2}\nonumber\\ 
s_{T'}=\xi\left( 1+\xi^2\right)^\frac{1}{2}\qquad 
s_{LT'}=-\frac{1}{\sqrt{2}}\xi \, ,
\end{eqnarray}
with $\xi=|{\bf q}_L|Q^{-1}\tan(\theta/2)$.  

One of the virtues of Eq.~(\ref{crosssec}) is that the 
response functions $\tilde{R}$ are {\it covariant\/}, and
hence (\ref{crosssec}) it  can be used to describe the cross
section in either the c.m.~of the outgoing $np$ pair or the
laboratory frame, provided we use the appropriate form of
$d\Sigma$:
\begin{eqnarray}
d\Sigma \vert_{c.m.}  = |{\bf p}_{1_{c.m.}}|
~d\Omega_{1_{c.m.}} \nonumber\\ 
d\Sigma \vert_{LAB}  = |{\bf p}_{1_L}|\,{\cal R}
~d\Omega_{1_L} \, , \label{sigma}
\end{eqnarray}
where ${\cal R}$ is the lab recoil factor
\begin{equation}
{\cal R} = {W\over  m_d}\left( 1 + 
\frac{\textstyle\nu p_1 - E_1 q \cos\theta_1}{\textstyle m_d p_1} 
\right)_L^{-1} 
\label{defrecoil}
\end{equation}
with $W$ the invariant mass of the outgoing pair. 

In any frame the nine response functions of
(\ref{crosssec}) are related to the components of
the deuteron response tensor
$R^{({\rm X})}_{\lambda_\gamma\lambda'_\gamma}$
\begin{equation}
\fl
R^{{\rm (X)}}_{\lambda_\gamma\lambda'_\gamma}=
\frac{m^2}{4\pi^2W}
\sum_{\lambda'_1\lambda_1}
\sum_{\lambda_2}\sum_{\lambda'_d\lambda_d} 
\rho_{\lambda_1\lambda'_1}^{N\,{\rm (X)}}
\left< \lambda'_1\lambda_2\left| J_{\lambda_\gamma}(q)
\right|\lambda_d\right>
\rho_{\lambda_d\lambda'_d}^d\left<\lambda'_d\left|
 J_{\lambda'_\gamma}^\dagger(q)\right|\lambda_1\lambda_2\right>
\label{resptens}
\end{equation}
with X = I or II, and
\begin{eqnarray}
\tilde{R}^{({\rm I})}_L=R_{00}\qquad\qquad& 
\tilde{R}^{({\rm I})}_T=R_{++}+R_{--} \nonumber\\
\tilde{R}^{({\rm I})}_{LT}=2{\rm Re}(R_{0+}-R_{0-}) \qquad&  
\tilde{R}^{({\rm I})}_{TT}=2{\rm Re}R_{+-} \nonumber\\
\tilde{R}^{({\rm I})}_{LT'}=2{\rm Im}(R_{0+}-R_{0-}) &
\tilde{R}^{({\rm II})}_{T'}=R_{++}-R_{--}\nonumber\\
\tilde{R}^{({\rm II})}_{LT}=2{\rm Im}(R_{0+}+R_{0-}) &
\tilde{R}^{({\rm II})}_{TT}=-2{\rm Im}R_{+-}  \nonumber\\
\tilde{R}^{({\rm II})}_{LT'}=2{\rm Re}(R_{0+}+R_{0-}) \, . 
\label{Rrela}
\end{eqnarray}
All of these quantities are written in the helicity 
basis, with $\lambda_\gamma$ the helicity of the
virtual photon, $\lambda_1$ and $\lambda_2$ the
helicities of particles 1 and 2  in the final-state,
and $\lambda_d$ the helicity of the initial
deuteron.   The matrix element of the 
helicity-basis current operator between helicity
states is represented by $\left<\lambda_1\lambda_2
\left|J_{\lambda_\gamma}(q)\right|
\lambda_d\right>$.   In cases where the deuteron target might be
polarized {\it only\/} in the $\hat y$ direction, and where
{\it only\/} the polarization of the outgoing
particle 1  might be measured, the spin density matrices for 
particle 1 in  the final state is given by
$\rho^{N({\rm X})}$ and that of the deuteron in the 
initial state by $\rho^d$, where 
\begin{eqnarray}
\rho^d_{\lambda\lambda'}={1\over3}\left\{\delta_{\lambda
\lambda'} +i\,p_y^d\,\left(-i\sqrt{2} {\cal
S}_y\right)_{\lambda\lambda'}\right\} = {1\over3}
\left(\begin{array}{ccc} 1 & -i\,p_y^d & 0\cr i\,p_y^d & 1 &
-i\,p_y^d\cr 0 & i\,p_y^d & 1 \end{array}\right)\nonumber\\
\rho^{N\,{\rm (I)}}_{\lambda\lambda'}={1\over2}
\biggl(1+p^N_y\,(\sigma_y)_{\lambda\lambda'}\biggr)
\qquad
\rho^{N\,({\rm II})}_{\lambda\lambda'}= {1\over2}
\biggl(p^N_x\,
(\sigma_x)_{\lambda\lambda'}+p^N_z\,
(\sigma_z)_{\lambda\lambda'}\biggr) \label{density}
\end{eqnarray}
where ${\cal S}_y$ is the $\hat y$ component of the
spin-one matrix, $p_y^d=\sqrt{3/2}\,{\rm Im}T_{11}$ is the
vector polarization of the deuteron target, $p_i$ is the
direction of the polarization of outgoing particle 1,
measured with respect to the $x''y''z''$ coordinate system
shown in Fig.~\ref{fig_cross}, and $\sigma_i$ are the Pauli
matrices.  Note that only those response functions of type I
(denoted by the superscript) are nonzero if all of the hadrons
are unpolarized; type II response functions require (for
the cases considered here)  measurement of the
polarization of the outgoing nucleon.  Further
details and additional cases can be found in
Ref.~\cite{dg89}. 

The familiar unpolarized inclusive cross section is easily 
obtained by integrating (\ref{crosssec}) in the c.m.~and summing
over electron polarizations.  The result is  
\begin{eqnarray}
{d^3\sigma\over d\Omega^{\prime}dE^{\prime}} &=& 
\sigma_M \biggl\{W_2(Q^2,\nu) + 2W_1(Q^2,\nu)\,
\tan^2\theta/2\biggl\}\, ,
\label{inclusive}
\end{eqnarray}
with
\begin{eqnarray}
\fl W_1(Q^2,\nu)= {|{\bf p}_{1_{c.m.}}|\over4m_d}\int_{-1}^1
d\cos\theta_{1_{c.m.}}\;
\tilde{R}^{({\rm I})}_T(Q^2,\nu,\theta_{1_{c.m.}})\nonumber\\
\fl W_2(Q^2,\nu)= {|{\bf p}_{1_{c.m.}}|Q^2\over2m_d\, {\bf q}_L^2}
\int_{-1}^1 d\cos\theta_{1_{c.m.}}\;\left\{
\tilde{R}^{({\rm I})}_L(Q^2,\nu,\theta_{1_{c.m.}})
+{1\over2}\tilde{R}^{({\rm 
I})}_T(Q^2,\nu,\theta_{1_{c.m.}}) \right\}
\, ,
\label{w1andw2}
\end{eqnarray}

\subsubsection{Photodisintegration}
\label{photodis1}

For real photons the longitudinal components are absent, and the 
cross section simplifies (there  is no electron scattering plane
and no electron kinematics).  The most general polarization of
the incoming photon, $\epsilon$, is therefore a superposition of
the circular polarization states $\epsilon_\pm$, which we write
as $\epsilon = a_+\epsilon_++a_-\epsilon_-$ with
$|a_+|^2 + |a_-|^2=1$.  The expansion coefficients $a_\pm$ can
therefore be written in terms of only three independent
parameters
\begin{eqnarray}
a_+=-\cos\beta\,e^{-i(\phi+\alpha)}\nonumber\\
a_-=\phantom{-}\sin\beta\,e^{\;i(\phi-\alpha)}
\end{eqnarray}
with $0\le\beta\le\pi/2$.  The coincidence cross section for a
polarized photon beam is then
\begin{eqnarray}
\fl{d^2\sigma\over d\Sigma_0} = 
{\textstyle{1\over2}}\tilde{R}^{({\rm I})}_T + 
{\textstyle{1\over2}}\left(|a_+|^2-|a_-|^2\right) 
\tilde{R}^{({\rm II})}_{T^{\prime}}  +{\rm
Re}(a_+a_-^*)\tilde{R}^{({\rm I})}_{TT} +  {\rm
Im}(a_+a_-^*)\tilde{R}^{({\rm II})}_{TT}\nonumber\\
= {\textstyle{1\over2}}\left\{\tilde{R}^{({\rm I})}_T + 
p_+ \tilde{R}^{({\rm II})}_{T^{\prime}} 
-p_\gamma \tilde{R}^{({\rm I})}_{TT}\cos2\phi-
p_\gamma \tilde{R}^{({\rm II})}_{TT}\sin2\phi\right\}
\, ,
\label{photonxsec}
\end{eqnarray} 
where $p_+=\cos2\beta$ and $p_\gamma=\sin2\beta$ are the
fractions of right circular and linear photon polarizations,
respectively, and $d\Sigma_0$ depends on the frame 
\begin{eqnarray}
d\Sigma_0 \vert_{c.m.}  =d\Omega_{1_{c.m.}} \left({|{\bf
p}_{1}|\over4\nu W}\right)_{c.m.} \nonumber\\  
d\Sigma_0 \vert_{LAB}  = d\Omega_{1_L} \left({ |{\bf
p}_{1}|\over4\nu m_d}\right)_L {\cal R}\, , \label{sigma0}
\end{eqnarray}
with ${\cal R}$ defined in Eq.~(\ref{defrecoil}).  We will
return to the cross section (\ref{photonxsec}) in
Sec.~\ref{photodis}.  

\subsubsection{Theoretical Issues} 

It is important to appreciate
that these formulae for the  cross section are {\it exact\/}
relativistic results (subject only to the one photon exchange
approximation).  All of our ignorance is confined to the
hadronic matrix elements of the current
\begin{equation}
J^{\lambda_\gamma}_{\lambda_1\lambda_2\lambda_d}(p_1p_2,q)=
\left<\lambda_1\lambda_2\left|J_{\lambda_\gamma}(q)\right|
\lambda_d\right>\, , \label{currents}
\end{equation}
and the structure functions (\ref{Rrela}) that are products
of these currents.  

In much of the older literature, particularly for studies
of the $(e,e'p)$ reaction from nuclei with mass number $A>2$
\cite{Ma88}, the cross section is written
\begin{equation}
{d^5\sigma\over d\Omega dE^{\prime}d\Sigma} = {\cal K}\,
\sigma_{ep}\,S(p,E_s)\, , \label{spectral}
\end{equation}
where ${\cal K}$ is a kinematic factor, $\sigma_{ep}$ is
cross section for scattering of an electron from an
``off-shell'' proton, and $S(p,E_s)$ is the proton spectal
function (which gives the probability of finding a proton
with momentum $p$ and separation energy
$E_s$ in the target nucleus).  The proton momentum
distribution is obtained by integrating the spectral
function over the separation energy 
\begin{equation}
n(p)=\int dE\,S(p,E) \, .
\end{equation} 
Some early experiements focused on ``measuring'' the
momentum distribution and the spectral function.  While
this picture has a nice physical interpretation [it is
motivated by the PWIA, Eq.~(\ref{PWIA})], and presenting
data this way is sometimes useful, particularly in the early
phases of the program,  it is important to realize that the
individual structure functions that enter the exact cross
section (\ref{crosssec}) are, in general, {\it independent\/}
functions which are {\it not\/} proportional to each other,
and that therefore Eq.~(\ref{spectral}) is {\it only an
approximation\/} to the cross section \cite{JVO00}.  Attempts
to refine the definitions of $\sigma_{ep}$ and $S(p,E_s)$ can
have limited value at best, and at worst can lead to many
unproductive debates about the precise definition of
the spectral function. 

Calculation of the hadronic current matrix elements
(\ref{currents}) is complicated by requirement that the
current be conserved, $q_\mu J^\mu=0$.  For elastic
scattering, where the initial and final states are
identical, invariance under time inversion usually
guarantees that even simple approximations to the dynamics
will satisfy this constraint.  But building in current
conservation for inelastic processes usually requires
consistent treatment of both final-state interactions and
interaction currents.  The failure of approximate
calculations (and the PWIA in particular) to satisfy current
conservation is often seen as a serious obstacle. Some {\it ad
hoc\/} prescription of the kind introduced by De Forest
\cite{deF83}, is needed.

We propose the simple prescription introduced recently in
the study of deep inelastic scattering \cite{BG98}.  Suppose
the {\it exact\/} current is composed of two parts
$J^\mu=J_1^\mu+J_2^\mu$.  In general, {\it neither\/} of these
two parts will satisfy current conservation {\it alone\/}; that
is $q_\mu J^\mu_i\ne0$ for each $i$.  However, since the exact
current satisfies current conservation,
$q_\mu(J^\mu_1+J^\mu_2)=0$.  We propose replacing each of the
individual terms in the current by 
\begin{equation}
J^\mu_i \to J^\mu_{Ti}=J^\mu_i-{q^\mu\, q_\nu J^\nu_i\over
q^2}\, . \label{redef}
\end{equation}
This procedure is covariant, guarantees that each
component conserves current (so that one can be calculated
without knowing the other), and that their sum is unchanged:
$J_1^\mu+J_2^\mu=J_{T1}^\mu+J_{T2}^\mu$.  Perhaps the
best argument can be found in Ref.~\cite{BG98} where it
was shown (for a very simple case) that the Born term
defined in this way dominates the final-state interaction
term in the deep inelastic limit, resolving a long
standing puzzle. Finally, note that
$J_{i\epsilon_\gamma}\equiv\epsilon_{\gamma\mu}J_i^\mu=
\epsilon_{\gamma\mu}J_{Ti}^\mu$ (where $\epsilon_\gamma$
are the virtual photon polarization vectors satisfying
$\epsilon_\gamma\cdot q=0$) so that the response tensor
(\ref{resptens}) is {\it unaffected\/} by the
redefinition (\ref{redef})! 

We now turn to a brief discussion of threshold
electrodisintegration.


\subsection{Threshold Electrodisintegration}

\subsubsection{Overview}

Threshold deuteron electrodisintegration measures the 
$d(e,e^{\prime}) pn$ reaction in kinematics in which the
proton and neutron, rather than remaining bound, are unbound
with a few MeV of relative kinetic energy in their center of
mass system.  If the final-state energy is low enough, the
final state will be dominated by transitions to the $^1S_0$
final state, and will be a pure $\Delta S = 1$, $\Delta I =
1$, $M1$  transition, similar to the $N\to\Delta(1232)$
transition.  This transition is a companion to the
$B$ structure function; both are magnetic transitions and
both are filters for exchange currents with only one isospin
($d\to d$ is $\Delta I=0$ and $d\to\ ^1S_0$ is $\Delta
I=1$).  To see the similarity, compare the top right panel of
Fig.~\ref{RABT20} with the threshold measurements shown in
Fig.~\ref{thresh}.  Both have a similar shape, and in both
cases the uncertainties in the theoretical predicitions are
large.  

The similarity of these two processes (elastic and
threshold inelastic) also holds for the theory.  These two 
processes can be used to separately determine the precise details
of the $I=0$ and $I=1$ exchange currents.   Once the exchange
currents are fixed, they can be used to predict the results of
$d(e,e'p)n$ over a wide kinematic region.  Any theoretical
approach that works for the form factors should also work equally
well for threshold electrodisintegration, yet very few of the
groups who have calculated form factors have also calculated the
threshold process. This may be due in part to the fact that the
$I=1$ interaction currents are larger than the $I=0$ interaction
currents, making the threshold electrodisintegration calculation
more difficult than the elastic calculation.  A more definitive
test of the various relativistic approaches discussed in the
previous sections will be possible once the elastic
calculations are extended to the threshold inelastic process.   

Previous threshold electrodisintegration experiments have
reported an {\it average\/} cross section that can be
obtained theoretically by integrating the relative $pn$
energy in the final state, $E_{pn}$, from 0 to 3, or (in some
cases) 0 to 10 MeV. The unbound $^1S_0$ final state dominates
at threshold (because the $^3S_1$-$^3D_1$ scattering state is
orthogonal to the deuteron state at threshold), but above
threshold there are contributions from the $^3S_1$-$^3D_1$
scattering state, and eventually from the $NN$ $L=1$
scattering states as well.  To
emphasize this magnetic transition, data have been taken at
large electron scattering angles, mostly 155$^\circ$ or
180$^\circ$.  

Threshold electrodisintegration provides strong evidence
(perhaps the best we have) for the existence of isovector
exchange currents \cite{hoekert73}.  The impulse
approximation calculation of the transition to the
$^1S_0$ final state has a zero arising from the negative
interference between the $^3S_1\to^1S_0$ and $^3D_1\to^1S_0$ 
pieces of the transition that lead to a minimum  at $Q^2$ near
0.5 GeV$^{2}$.  This minimum is not seen in the data, and
theoretical calculations of the $I=1$ exchange current
contribution fill in the minimum and explain the data. 

\subsubsection{Current status of theory and data}
\label{thresdandt}

%
\begin{figure}[t]
\begin{center}
\vspace*{0.1in} 
\mbox{
   \epsfxsize=4.0in
\epsffile{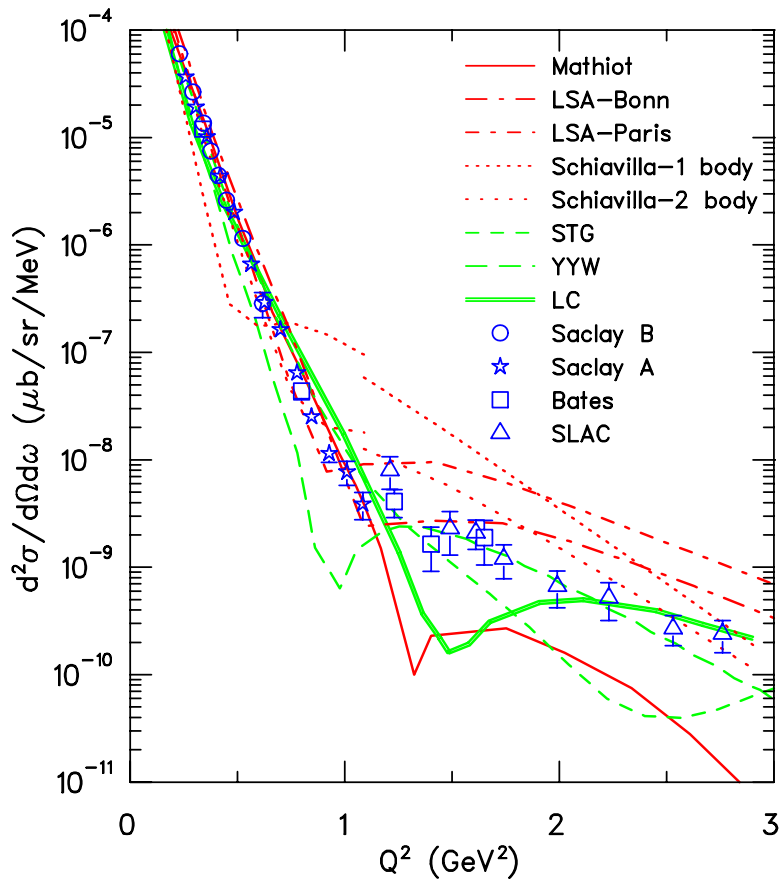}  
}
\end{center}
\caption{The cross section for threshold electrodisintegration 
of the deuteron. 
Data shown are from Saclay B \cite{bernheim81}, Saclay A 
\cite{auffret85b},  Bates \cite{lee91,schmitt97}, and SLAC
\cite{arnold90,frodyma93}. Theory calculations are from Mathiot
\cite{mathiot}, LSA \cite{arenhovelth},
Schiavilla \cite{CS98}, STG
\cite{smejkal}, YYW \cite{yamauchi}, and LC
\cite{lucheng}.}
\label{thresh}
\end{figure}

Experiments have measured $d(e,e^{\prime})pn$
threshold electrodisintegration
to $Q^2$ about 1.6 GeV$^2$ with better than 1 MeV ($\sigma$)
resolution, integrating up to $E_{pn}$ $=$ 3 MeV
\cite{ganichot72,rand67,simon79,bernheim81,auffret85b,lee91,schmitt97},
and to nearly 3 Gev$^2$ 
with 12 - 20 MeV (FWHM) resolution, and
integrating up to $E_{pn}$ $=$ 10 MeV
\cite{arnold90,frodyma93}.\footnote
{
One possible source of confusion in any close examination 
of threshold electrodisintegration is the use of two
different conventions for the cross section.
Early articles \cite{bernheim81,auffret85b} and the most
recent Bates article \cite{schmitt97} report the cross 
section as $d^2\sigma / d\Omega d\omega = d^2\sigma /
d\Omega dE^{\prime}$ where $\omega$ ($E^{\prime}$) is the
energy transfer  (scattered electron energy).
The SLAC articles \cite{arnold90,frodyma93} and 
first Bates article \cite{lee91}
use instead $d^2\sigma / d\Omega dE_{np}$, where $E_{np}$
is the total $np$ kinetic energy in their c.m. frame,
$E_{np} = W - m_d - 2.225 (MeV)$.
(A typographical error at one point in \cite{frodyma93} 
misidentifies
$E_{np}$ as the energy of a nucleon, rather than the two 
nucleons.) These cross sections are related by the
Jacobian $|dE_{np}/dE^{\prime}| = (m_d + 2 E \sin^2(\theta
/2) ) / W$ which numerically ranges from about 1.3 to 2.3
for the data that we present. The articles showing the
``$dE_{np}$'' cross sections appear to plot the Saclay
cross sections as published, as  ``$dE^{\prime}$'' cross
sections, rather than converted.
}
  
Figure \ref{thresh} shows the smooth rapid fall off of 
threshold electrodisintegration cross sections for $Q^2$ up
to about 1.2  GeV$^2$, and the change in slope for the
higher $Q^2$ SLAC and Bates data. There is good agreement
between the various measurements, including  those not
shown, considering the change in scattering angle
(155$^{\circ}$ for  Saclay, 160$^{\circ}$ for Bates,
180$^{\circ}$ for SLAC) and integration  region for
$E_{pn}$ (0-3 MeV for Saclay and Bates data shown, 0-10 MeV
for SLAC).  The figure also shows results from seven
theoretical calculations.
 
The oldest calculation shown in the figure is from Mathiot
\cite{mathiot}.  This model includes one and two body
currents based on $\pi$ and $\rho$ exchange, and also
contributions to the two-body current operator from the
exitation of the $\Delta$.  Mathiot confirms that the one body
current (the impulse approximation) produces a sharp minimum in
the cross section at about $Q^2=$ 0.5 GeV$^2$ (in agreement
with the recent Schiavilla-1 body curve shown in the figure
and in complete disagreement with the data) and that the $\pi$
exchange current fills this in, shifting the minimum to about 1
GeV$^2$.  The  $\rho$ exchange is also important, shifting the
mimimum to $Q^2\simeq$ 1.4 GeV$^2$.  Contributions from the
electroexication of a $\Delta$ are smaller, at least
below $Q^2\simeq 1$ GeV$^2$.  Figure \ref{thresh} shows that
this calculation breaks down above $Q^2\simeq1$ GeV$^2$,
probably because it does not include many of the contributions
included in modern calculations.  More recent calculations, also
based on hadronic degrees of freedom, are from Leidemann,
Schmidtt, and Arenh\"ovel [LSA] \cite{arenhovelth} and
Schiavilla.  (The  calculations by Schiavilla shown on the
figure are based on the work of Ref.~\cite{SR91}, but use the
more recent Argonne AV18 potential.  These curves were also
published in the review
\cite{CS98}.)  In both of these cases the details are largely
unpublished and the inelastic calculations are not as up-to-date
as the corresponding elastic calculations recently done by the
same groups, so it is premature to draw definite conclusions. 
The work of Smejkal, Truhl\'ik, and G\"oller [STG]
\cite{smejkal} obtains exchange current contributions from the
$\pi\rho a_1$ system using a chiral lagrangian, and does a good
job describing the data out to 2 GeV$^2$.  Predictions from two
quark cluster models, the early model of Yamauchi, Yamamoto, 
and Wakamatsu [YYW] \cite{yamauchi}, and the more recent model
of Lu and Cheng [LC] \cite{lucheng}, are also shown in
Fig.~\ref{thresh}.  These calculations both tend to have too
much structure in the region of the shoulder, but do show that
quark cluster models have the ability to describe the exchange
currents needed to account for the data.
  
The recent improvements in relativistic theory discussed in
Sec.~\ref{nucsec} will lead to a new generation of
calculations that will rely on threshold electrodisintegration
to provide details about the nature of the $I=1$ exchange
currents.  The most precise constraint on these currents
comes from the $d\to\, ^1S_0$ transition, and this part of the
transition is partly obscured by the poor energy resolution of
the existing high $Q^2$ measurements. A new and improved
experiment at JLab with higher resolution would allow the
threshold $d\to\, ^1S_0$ process to be better extracted, with a
better resulting determination of the isovector exchange
currents. It is also important to determine whether or not there
is a minimum near 1.2 GeV$^2$. This present indication of a
minimum might be an artefact of the end of the Saclay data
\cite{auffret85b} vs.\ the start of the SLAC data
\cite{arnold90,frodyma93},  along with systematic uncertainties
of these and the Bates measurements
\cite{lee91,schmitt97}.

A high statistics high resolution measurement at large $Q^2$ 
is feasible.
Measurements have been proposed \cite{pr00103}
with 1.5 MeV resolution at a scattering angle of 
160$^\circ$. The experiment would use Hall A with the HRS
spectrometer vacuum  coupled to the scattering chamber, and
a special cryotarget and collimators to enhance resolution
and reduce backgrounds. Measurements were proposed for 6
points from 1 to 3.7 GeV$^2$.

We note that SLAC NE4 simultaneously measured the threshold 
electrodisintegration along with the elastic structure 
function 
$B$. Because the large $Q^2$ threshold inelastic cross 
section is typically an  order of magnitude larger,
the $d(e,e^\prime)$ measurements see essentially only the 
inelastic processes, and $d(e,e^\prime d)$ is needed to 
determine the elastic scattering.
If it is possible to maintain
large solid angle for the elastic scattering, and 
good resolution for the threshold electrodisintegration, 
both data sets can be obtained simultaneously.





\section{Deuteron Photodisintegration} 
\label{photodis}

\subsection{ Introduction }

Deuteron photodisintegration was first investigated in the 
early 1930s, in order to understand the structure of the neutron.
After the discovery of the neutron by James Chadwick,
attention turned to its mass and structure.
Was the neutron a fundamental particle, like the proton and 
electron,  or was it a bound state of the electron and proton, 
different from the hydrogen atom?
If it was a bound state of the proton and electron, how
were the electrons confined into the small nuclear volume?
Conflicting experimental evidence on the neutron mass prevented 
resolution of the issue until 1934,
when Chadwick and Maurice Goldhaber used deuteron 
photodisintegration
\cite{cg34} to determine that the neutron mass was  
slightly heavier than that of the hydrogen atom.
Thus, the neutron, being heavier than the proton plus electron, 
was a fundamental particle, and there was no longer any basis for 
thinking electrons could be present in nuclei \cite{bethe99rmp}.

Subsequently, deuteron photodisintegration cross sections
have served as a standard test case for nuclear theory.
The effects, for example, of the $\Delta$ resonance in cross 
sections for beam energies near 300 MeV are pronounced, but until
recently large discrepancies between different experimental data
sets made precise tests of theories difficult \cite{oldreview}.
The 1960s-1970s saw the start of polarization measurements.
The earliest data were intermediate energy measurements
of the induced proton polarization \cite{feld60} and low
energy  measurements of the induced neutron polarization
\cite{john61}. Large induced polarizations were observed
\cite{liu68,kose69} soon afterward, particularly for energies
above the $\Delta$  resonance and for  center of mass angles
near 90$^{\circ}$. The combination of more extensive
confirming measurements
\cite{kamae77}  for $E_{\gamma}$ about 350 - 700 MeV, which could
not be reproduced theoretically, and interest in dibaryons led to
much excitement about deuteron photodisintegration in the late
1970s and early 1980s. There were many serious theoretical
efforts, numerological studies involving inclusion of dibaryon resonances,
and extensive experimental studies of
cross sections and polarization observables.
 
\begin{figure}[t]
\begin{center}
\vspace*{0.1in}
\mbox{
   \epsfxsize=4.0in
\epsffile{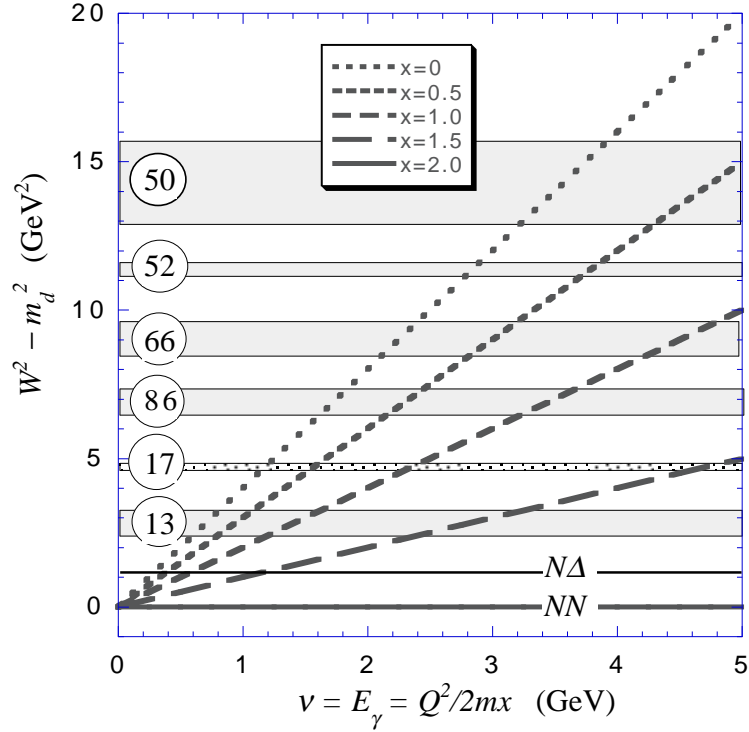}  
}
\end{center}
\caption{The variation of $W^2$ with the photon energy
$\nu$ for various values of $x$, as given by Eq.~(\ref{kineq1}).
The shaded regions show the approximate thresholds for the
production of bands of nucleon resonances, as discussed in the
text and shown in Table \ref{tab:channels}.  The numbers in the
small circles are the number of distinct channels in each band.}
\label{fig:dgpkin}
\end{figure}

\subsection{Relation between elastic scattering and
photodisintegration}

Since {\it both\/} the recent deuteron form factor 
measurements {\it and\/} the recent high energy deuteron
photodisintegration measurements have been made with 4 GeV
electron beams, it is sometimes assumed that the same
theory should work for both.  In this review we emphasize
that this need not be the case.  The kinematics of elastic
electron-deuteron scattering and deuteron
photodisintegration are very different, and the physics
being explored by these two measurements is also very
different.  The implications of this remarkable feature of
electronuclear physics, often not fully appreciated, will
be discussed briefly in this section. 

\begin{table}
\caption{\label{tab:thresh}
The 24 well established nucleon resonances listed in the 
{\it Particle Physics Booklet\/} \cite{groom00} fall into
the 8 bands listed below.  Masses of {\it neighboring\/}
resonances in each band are less than 150 MeV apart.  All
of these resonances can contribute to deuteron
photodisintegration for $W<4.5$ GeV.  All but $N_4$ and
$\Delta_4$ can contribute in all combinations, giving
$13+(13\times12)/2=91$ channels with two $I=1/2$ particles, 45
channels with two $I=3/2$ particles, and 117 channels with one
$I=1/2$ and one $I=3/2$ particle. The number of additional
channels contributed by $N_4$ and $\Delta_4$ is shown on
the table and totals 33 channels.  The total number of 
channels is 286. }
\begin{indented}
\item[]\begin{tabular}{@{}llll}
\br
 & $I=1/2$ &  & $I=3/2$ \\
\mr
$N_1$ & $N$ (939) & $\Delta_1$ & $P_{33}(1232)$\\
$N_2$ & $P_{11}(1440)$, $D_{13}(1520)$, $S_{11}(1535)$, $\quad$ &
$\Delta_2$ & $P_{33}(1600)$,  $S_{31}(1620)$, $D_{33}(1700)$ \\
& $S_{11}(1650)$, $D_{15}(1675)$, $F_{15}(1680)$,  & & \\
& $D_{13}(1700)$, $P_{11}(1710)$, $P_{13}(1720)$   & & \\
$N_3$ & $G_{17}(2190)$, $H_{19}(2220)$, $G_{19}(2250)$ & 
$\Delta_3$ & $F_{35}(1905)$, $P_{31}(1910)$, $P_{33}(1920)$, \\
& &   & $D_{35}(1930)$, $F_{37}(1950)$  \\
$N_4$ & $I_{1,11}(2600)\qquad 14$ channels   & $\Delta_4$  &
$I_{3,11}(2420)\qquad 19$ channels \\
\br 
\end{tabular}
\end{indented}
\end{table}

\begin{table}
\caption{\label{tab:channels}
The thresholds for the production of pairs of baryon 
resonances also fall into 8 bands.  {\it Neighboring\/}
thresholds within each band are less than 150 MeV apart. 
These bands are shown in Fig.~\ref{fig:dgpkin}. }
\begin{indented}
\item[]\begin{tabular}{@{}lclr}
\br
 Band & mass range & members & number of \\
& & & channels \\
\mr
$B_1=NN$      & 1878 & $N_1N_1$       & 1  \\
$B_2=N\Delta$ & 2171 & $N_1\Delta_1$  & 1  \\
$B_3$ & 2464 -- 2579  & $\Delta_1\Delta_1$, $N_2N_2$,
$N_1\Delta_2$  & 13 \\
$B_4$ & $\quad$ 2858 -- 2872 $\quad$ & $N_2\Delta_1$,
$N_1\Delta_3$,
$\Delta_1\Delta_2$  & 17 \\
$B_5$ & $\quad$ 3155 -- 3280 $\quad$ & $\Delta_1\Delta_3$,
$N_2N_2$, $N_2\Delta_2$, $\Delta_2\Delta_2$ & 86 \\
$B_6$ & $\quad$ 3452 -- 3652 $\quad$ & $N_3\Delta_1$,
$N_1\Delta_4$, $N_1N_4$, $N_2\Delta_3$, & \\
& & $\Delta_2\Delta_3$, $\Delta_1\Delta_4$ &  66 \\
$B_7$ & $\quad$ 3832 -- 3860 $\quad$ & $N_4\Delta_1$,
$\Delta_3\Delta_3$, $N_2N_3$, $N_3\Delta_2$, & 52 \\
$B_8$ & $\quad$ 4046 -- 4440 $\quad$ & $N_2\Delta_4$,
$\Delta_2\Delta_4$, $N_3\Delta_3$, $N_2N_4$, & \\
& & $\Delta_2N_4$, $\Delta_3\Delta_4$, $N_3N_3$ &  50 \\
\br 
\end{tabular}
\end{indented}
\end{table}

The kinematics of elastic scattering and 
photodisintegration are compared in Fig.\ref{fig:dgpkin},
which shows $W^2-m_d^2$ as a function of the photon (real
or virtual) energy
\begin{equation}
W^2-m_d^2=2m_d\nu\left(1-{m\, x\over m_d}\right) \, ,
\label{kineq1}
\end{equation}
where $x=Q^2/2m\nu$ and $\nu=E_\gamma$ for real photons.
The mass of the final excited state increases rapidly as 
$x$ decreases below its maximum allowed value of
$x=m_d/m\simeq2$.   For any energy $\nu$ or any $Q^2$,
elastic $ed$ scattering leaves the $pn$ system bound, with
no internal excitation energy added to the two nucleons. 
For quasifree scattering ($x=1$) the mass of the final $pn$
system grows with $\nu$, and as $x$ decreases below 1 the
mass grows more rapidly with
$\nu$.  As $x$ $\rightarrow$ 0, we approach the real photon
limit.  Real photons produce the maximum value of $W$ of 
any given beam energy.  With each 1 GeV of beam energy
$W-m_d$ increases by approximately 500 MeV, driving the
final state deep into the resonance region.  The well
established nucleon resonances, all of which can be excited
by 4 GeV photons, are listed in Table~\ref{tab:thresh}, and
the bands of thresholds at which these resonances are
excited, either singly or in pairs, are listed in
Table~\ref{tab:channels} and shown in
Fig.~\ref{fig:dgpkin}.   [The Fermi momentum of the struck
nucleon, and the widths of the resonances, will average
these thresholds over a wider kinematic region than shown.]
By $E_{\gamma}$ = 1.2 GeV, the final-state mass is already
reaching $W$ = 2 GeV, the nominal onset of deep inelastic
scattering (DIS) from a single nucleon if we assume one of
the nucleons in the deuteron remains at rest.  At
$E_{\gamma}$ = 4 GeV, the final-state mass is approximately
4.5 GeV, and at least 286 thresholds for the production of
pairs of baryon resonances have been crossed (and there are
probably more from unseen or weakly established
resonances).  A photon energy of 4 GeV corresponds to $np$
scattering with a laboratory kinetic energy of about 8
GeV! 
(See also the kinematic argument given 
by Holt \cite{holtkin}.)

It is clearly very difficult (if not impossible) to 
construct a theory of high energy photoproduction in which
all of the 24 established baryon resonances and their
corresponding 286 production thresholds are treated
microscopically.  By contrast, elastic electron deuteron
scattering requires a microscopic treatment of only {\it
one channel\/}.  All of  the 286 channels also contribute
to elastic scattering, of course, but in this case they are
not explicitly excited, and can probably be well described
by slowly varying short-range terms included in a meson
exchange (or potential) model.  In photodisintegration,
{\it each of these channels is excited explicitly\/}.  As
we shall see below, it is sufficiently difficult to
construct an adequate theory in the region of the $\Delta$
resonance, so it is difficult to  imagine this program
being extended to a realistic treatment of many resonances,
including intermediate states in which multiple mesons are
present. It would instead appear that an alternate
framework that {\it averages over the effects of many
hadronic states\/} is needed.  The alternatives are to use
a Glauber-like approach, or to borrow from our knowledge of
DIS and build models that rely on the underlying quark
degrees of freedom.  We will return to these issues in our
review of the theory in Sec.~\ref{phototheory} below.

\subsection{Cross section and polarization observables}

\begin{figure}[t]
\begin{center}
\vspace*{0.1in}
\mbox{
   \epsfxsize=4.5in
\epsffile{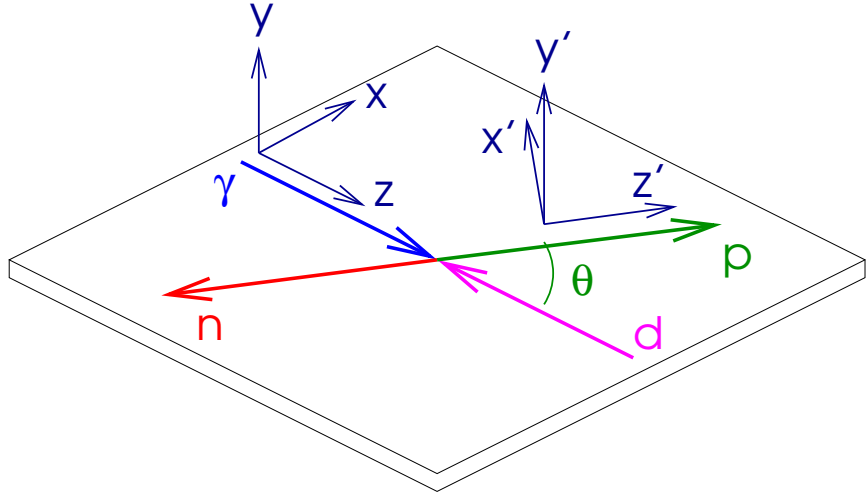}  
}
\end{center}
\caption{The cordinate system for deuteron photodisintegration.
The $\hat{y}$ $=$ $\hat{y}^{\prime}$ axis is given by
$\hat{y} = \hat{k} \times \hat{p} / |\hat{k} \times 
\hat{p}|$, and the $\hat{x}$, $\hat{x}^{\prime}$ axes are
chosen to make a right handed coordinate system.}
\label{dgpcoord}
\end{figure}

The photodisintegration cross section was previously given in
Sec.~\ref{photodis1}.  Including the polarization observables
defined by the density matrices of Eq.~(\ref{density}), and
using the notation of Ref.~\cite{dg89}, the structure functions
are 
\begin{eqnarray}
\tilde R_T^{({\rm I})}\equiv\tilde R^{({\rm I})}_T(p^N_y,{\rm
Im}T_{11},U) =R_T+p_y^N\,R_T(y)+p_y^d\,R_T({\rm
Im}T_{11})\,\nonumber\\
\tilde R_{TT}^{({\rm I})}\equiv\tilde R^{({\rm
I})}_{TT}(p^N_y,U) =R_{TT}+p^N_y\,R_{TT}(y)\nonumber\\
\tilde R_{T'}^{({\rm II})}\equiv \tilde R^{({\rm II})}_{T'}
(p^N_{x'},p^N_{z'}) =p^N_{x'}\,R_{T}(x')  +p^N_{z'}\,R_{T}(z')
\nonumber\\
\tilde R_{TT'}^{({\rm II})}\equiv \tilde R^{({\rm II})}_{TT'}
(p^N_{x'},p^N_{z'}) =p^N_{x'}\,R_{TT}(x') 
+p^N_{z'}\,R_{TT}(z')
\label{polex}
\end{eqnarray}
where the coordinate system, shown  in Fig.~\ref{dgpcoord} (a
simplified and relabeled version of that given in
Fig.~\ref{fig_cross}) is constructed from  the incident photon
direction $\hat{k}\equiv\hat{z}$ and  the outgoing proton
direction $\hat{p}\equiv\hat{z}^{\prime}$.   
Substituting the expansions (\ref{polex}) into the cross
section formula (\ref{photonxsec}), gives, in a notation
suggested by Ref.~\cite{bara86},
\begin{eqnarray}
{d\sigma\over d\Omega}= \left({d\sigma\over
d\Omega}\right)_0 \biggl[ 1 +p^N_y p_y +p^d_y \,T - p_\gamma
(\Sigma + p_yT_1)\cos2\phi\nonumber\\
\qquad\qquad\qquad +p_+(C_{x'}p_{x'} +
C_{z'}p_{z'}) +p_\gamma(O_{x'}p_{x'}
+O_{z'}p_{z'}) \sin2\phi \biggr] \label{gdcross}
\end{eqnarray}  
where $(d\sigma/d\Omega)_0$ is the differential cross section for
unpolarized photons, and explicit expressions for the asymmetry
parameters are given in Table~\ref{tab:polpara}.  Note that the
observables $C_x$ and $p_y$ are the real and imaginary parts
of the same combination of amplitudes, so that an experimental
measurement of both of these observables will fully determine
this linear combination of amplitudes.\footnote
{Some references, such as \cite{bara86}, use $P_y$ for the
induced polarization instead of $p_y$. 
}  
This is not true for any other pair of
observables shown in the table.  In the c.m.\ system
\begin{eqnarray}
\left({d\sigma\over d\Omega}\right)_0 =
{e^2\,p_1\over8\nu_0W}\,R_T={\alpha\, p_1\over 192\pi E_1^2
\nu_0}
\,f(\theta)
\end{eqnarray}
where
\begin{eqnarray}
f(\theta) = \sum_1^6 \left[ |F_{i +}|^2 + |F_{i -}|^2 \right] 
\, .
\end{eqnarray}
The $F$s are defined in Table \ref{tab:Fs} 

\fulltable{\label{tab:polpara}
Formulae for polarization observables.  Each structure function
in the second column is to be divided by $R_T$, and in the
third column by $f(\theta)$.  [The $F_{i\pm}$ are defined in
Table \ref{tab:Fs}.] }
{\small
\begin{tabular}{@{}lll}
\br
Observable  & Structure  & Helicity amplitude
combination  \\
& function & \\
\mr
$p_y$    & $R_T(y)$  & $2 {\rm Im} \sum_{i=1}^3 \left[
  F_{i+}^*\; F_{(i+3)-} +  F_{i-} \;F_{(i+3)+}^*\right]$  \\
$T$      & $R_T({\rm Im}T_{11})\quad$  &  $2 {\rm Im}
\sum_{i=1}^2 \sum_{j=0}^1\left[ F_{(i+3j)+}\;F^*_{(i+3j+1)+}
+F_{(i+3j)-}\;F^*_{(i+3j+1)-}\right]$  \\
$\Sigma$ &  $R_{TT}$ &  $2 {\rm Re} \sum_{i=1}^3    
  (-)^{i}\left[-F_{i+}\;F_{(4-i)-}^* + F_{(3+i)+}\;F_{(7-i)-}^*
  \right]$  \\
$T_1$    & $R_{TT}(y)$ &  $2{\rm Im}\sum_{i=1}^3 
(-)^{i} \left[- F_{i+}  F_{(7-i)+}^* +
F_{i-}  F_{(7-i)-}^*\right]$ \\
$C_{x'}$ & $R_{T}(x')$  & $2 {\rm Re} \sum_{i=1}^3\left[
F_{i+}^*\; F_{(i+3)-} +  F_{i-}\; F_{(i+3)+}^* \right]$   \\
$C_{z'}$ & $R_{T}(z')$  &  $\sum_{i=1}^6 \left\{ 
|F_{i+}|^2  - |F_{i-}|^2 \right\}$ \\
$O_{x'}$ & $R_{TT}(x')$    &  $2 {\rm Im} \sum_{i=1}^3(-)^{i+1}
\left[ F_{i+} F_{(7-i)+}^* +  F_{i-} F_{(7-i)-}^*
\right]$ \\
$O_{z'}$ & $R_{TT}(z')$  & $2 {\rm Im} \sum_{i=1}^3    
(-)^{i+1}\left[F_{i+}\;F_{(4-i)-}^* 
+ F_{(3+i)+}\; F_{(7-i)-}^* \right]$ \\
\br 
\end{tabular}}
\endfulltable 
%

\begin{table}[tbp]
\caption{\label{tab:Fs}
The relations between the helicity amplitudes used in
Refs.~\cite{dg89} and \cite{bara86}.}
\begin{indented}
\item[]\begin{tabular}{@{}lll}
\br
Amplitude  & Ref.~\cite{dg89}$\qquad$ & Ref.~\cite{bara86}\\
\mr
$\left<\pm{\textstyle{1\over2}}\pm {\textstyle{1\over2}}
\left|J\cdot\epsilon_+\right| 1\right>$ &
  $F_{1,2}$ &  $F_{1\pm}/2m$ \\
$\left<\pm{\textstyle{1\over2}}\pm {\textstyle{1\over2}}
\left|J\cdot\epsilon_+\right| 0\right>$ &
  $F_{3,4}$ &  $F_{2\pm}/2m$ \\
$\left<\pm{\textstyle{1\over2}}\pm {\textstyle{1\over2}}
\left|J\cdot\epsilon_+\right| -1\right>\qquad$ &
  $F_{5,6}$ &  $F_{3\pm}/2m$ \\
$\left<\pm{\textstyle{1\over2}}\mp {\textstyle{1\over2}}
\left|J\cdot\epsilon_+\right| 1\right>$ &
  $F_{7,8}$ &  $F_{4\pm}/2m$ \\
$\left<\pm{\textstyle{1\over2}}\mp {\textstyle{1\over2}}
\left|J\cdot\epsilon_+\right| 0\right>$ &
  $F_{9,10}$ &  $F_{5\pm}/2m$ \\
$\left<\pm{\textstyle{1\over2}}\mp {\textstyle{1\over2}}
\left|J\cdot\epsilon_+\right| -1\right>$ &
  $F_{11,12}$ &  $F_{6\pm}/2m$ \\
\br 
\end{tabular}
\end{indented}
\end{table}

The {\it single\/} polarization observables are the induced
proton polarization $p_y$, the linearly polarized photon
asymmetry $\Sigma$, and the vector polarized (along
$\hat{y}$) target asymmetry
$T$.   The quantities $p_y$ and $\Sigma$  are
defined by
\begin{eqnarray}
p_y={1\over p^N_y}\left({d\sigma/d\Omega_+ -
d\sigma/d\Omega_-\over d\sigma/d\Omega_+ + d\sigma/d\Omega_-}
\right) \qquad
\Sigma ={1\over p_\gamma}\left( {{d\sigma/d\Omega_{\perp} -
d\sigma/d\Omega_{||}}
\over{d\sigma/d\Omega_{\perp} + d\sigma/d\Omega_{||}}}
\right)
\end{eqnarray} 
where $\pm$ refers to $p_y=\pm1$, and in the expression
for $\Sigma$ the photon is polarized either in
plane along the $\hat{x}$ direction ($\phi$=0, denoted
$||$) or out of plane along the $\hat{y}$ direction
($\phi=\pi/2$, denoted $\perp$).  Note that the opposite 
phase convention for $\Sigma$ ($\sigma_{||}$ -
$\sigma_{\perp}$) is often used, and is
used in the figures below.  
The {\it double\/} polarization  transfer observables are
$T_1$ and $O_{x',z'}$ (for linearly polarized photons), and
$C_{x',z'}$ (for circularly polarized photons), with the 
subscripts on $O$ and $C$ giving the polarization direction
in the final state.


\subsection{ Theory } 
\label{phototheory}
 
\subsubsection{Models using meson-baryon degrees of freedom}
There have been many attempts to understand low energy deuteron
photodisintegration using a conventional
meson-baryon framework.  Since the first band of nucleon
resonances is not excited until about 400 MeV photon energy
(recall Fig.~\ref{fig:dgpkin}) it makes sense to
describe the process below 400 MeV using a model of coupled
$NN$, $N\Delta$ and $NN\pi$ channels.  Laget \cite{laget}
showed that the prominant shoulder in the total cross section at 
$E_\gamma$ = 300 MeV can be largely explained by the mechanism in
which a $\Delta$ is photoproduced at a nucleon followed by the
reabsorption of its decaying pion by the other nucleon.  He
also examined many other mechanisms, including rescattering up
to second order, but did not do a full calculation of the
final-state interaction. Later Leideman and Arenh\"ovel
\cite{la}  treated the $NN$, $N\Delta$ and $\Delta\Delta$ as
coupled channels and included final-state effects to all
orders.  Tanabe and Ohta \cite{to} followed with a more complete
treatment of the final state which is consistent with three-body
unitarity.  In a number of conference talks, Lee \cite{tshlee}
reported on coupled channel calculations using $N$, $\Delta$ and
the $P_{11}(1440)$ (Roper) resonances, which he extended to
$E_\gamma=2$ GeV.  His work suggests that final-state
interactions significantly enhance the cross section for photon
energies above 1 GeV. 

The recent calculations by Schwamb and Arenh\"ovel and
collaborators \cite{schwambarenhovel01} include the $NN$,
$N\Delta$ and $\pi d$ channels, and also contributions
from meson retardation, meson exchange currents, and the meson
dressing of the nucleon lines required by unitarity. 
All parameters are fixed from nucleon-nucleon scattering and
photoreactions such as $\Delta$ excitation from the nucleon, so
no new parameters are introduced into the calculations of the
deuteron photodisintegration process itself. They obtain a
reasonable description of $NN$ scattering up to lab
energies of 800 MeV, particularly for the important $^1D_2$
partial wave, and emphasize that the consistent inclusion of
retardation effects improves their results for
photodisintegration.    In another work it was shown
that inclusion of the $S_{11}$ and $D_{13}$ resonances \cite{saw}
seems to have only a small effect below about 400 MeV.  These
resonances enhance the total cross section by only 14\% at 680
MeV, although effects on double polarization observables can be
more significant. Hence there is some justification to limiting
the theory of low energy photodisintegration to the channels
considered in Ref.~\cite{schwambarenhovel01}.       
 
By comparison to the detailed and careful treatments developed
for lower energies, calculations specifically designed to
describe  higher energy photodisintegration are less complete. 
An unpublished Bonn calculation \cite{kang} includes pole
diagrams generated from
$\pi$, $\rho$, $\eta$, and $\omega$ exchange,  plus the 17
well-established nucleon and
$\Delta$ resonances with mass less than 2 GeV and $J$ $\leq$ 5/2
(listed in Table \ref{tab:thresh}).  The calculation uses
resonance parameters taken from the Particle Data Group. 
Nagornyi and collaborators
\cite{nagame,dienag} have introduced a covariant model based on
the sum of pole diagrams.  The latest version \cite{dienag} 
gives the photodisintegration amplitude as a sum of
contributions from only 4 Feynman diagrams: three pole diagrams
coming from the coupling of the photon to the three external
legs of covariant
$dnp$ vertex plus a contact interaction designed to maintain
gauge invariance.  This model divides the $dnp$ vertex into
``soft'' and ''hard'' parts, with the hard part designed to
reproduce the pQCD counting rules and its strength determined by
a fit to the data at 1 GeV.  The model  has no final-state
interactions or explicit nucleon resonance contributions.  
There is also a relativistic calculation of photodisintgration
in Born approximation using the  Bethe-Salpeter formalism 
\cite{kazakovshulga,kazakovshimovsky}.  It is found that the
cross section is a factor of 2 to 10 times too small, and
that relativistic effects are large. All of these
models are rather crude, and taken  together it is not
clear what one should conclude from them.  The calculations 
each appear to emphasize some aspects of what a comprehensive
meson-baryon theory of photodisintegration should entail. 
Perhaps we can say that conventional calculations that
neglect final-state interactions seriously underestimate the
cross section, but may be corrected in an approximate manner
by introducing diagrams with poles in the $s$ channel. 
S-channel pole diagrams can be regarded as a crude
approximation to the missing final-state interactions.  More
generally, using pole diagrams with form factors that have
the correct behavior at high momentum transfer may also
insure that  meson-baryon theories of deuteron
photodisintegration will also have the correct high energy
behavior \cite{carlc90}. 
  
\subsubsection{Models based on quark degrees of freedom}

\begin{figure}[t]
\begin{center}
\vspace*{0.1in}
\mbox{
   \epsfxsize=2.0in
\epsffile{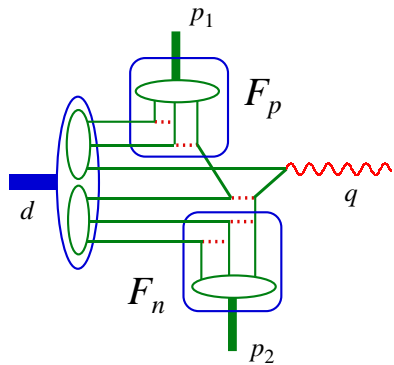}  
}
\end{center}
\caption{A ``typical'' pQCD diagram describing large angle
photodisintegration.  There are 5 hard gluons.  The RNA model
places the hard gluons in the rectangular boxes within
nucleon form factors.}
\label{fig:dg-pQCD}
\end{figure}

Perturbative QCD, discussed briefly in Sec.~\ref{pertQCD},
provides explicit, testable predictions for the cross section
and polarization observables.  For the case of elastic
scattering, the high energy (virtual) photon had to share its
mommentum equally with all of the constitutents, leading to
the typical diagram shown in Fig.~\ref{pQCDdiagrams}. 
Photo (or electro) disintegration differs in that the
momentum will not be distributed to {\it all\/} six quarks
unless momentum is transferred to {\it each\/} nucleon, and
this requires non-forward scattering, i.e.\ the angle
$\theta_{\rm cm}$ between the three-momentum of the outgoing
proton and the photon (in the c.m.\ system) must not be
0 or $\pi$.  More precisely, if $E_\gamma$ is the energy of the
photon in the lab system, the square of the momentum
transferred to each nucleon $t_i$ (where $i=$ 1,2) can be 
written
\begin{eqnarray}
\fl
t_1\equiv t=(q-p_1)^2=m^2-m_dE_\gamma\left(1-\sqrt{1-{4m^2\over
s}}
\cos\theta_{\rm cm}\right)\to -s\,\sin^2{\textstyle
{1\over2}}\theta_{\rm cm} \nonumber\\
\fl
t_2\equiv u=(q-p_2)^2=m^2-m_dE_\gamma\left(1+\sqrt{1-{4m^2\over
s}}
\cos\theta_{\rm cm}\right)\to -s\,\cos^2{\textstyle
{1\over2}}\theta_{\rm cm}\, , \label{t1t2}
\end{eqnarray}
where $s$, the square of the c.m.\ energy, is
\begin{equation}
s=(d+q)^2=m_d^2+2m_d E_\gamma \, ,
\end{equation}
and the limits in (\ref{t1t2}) are the result for
$E_\gamma>\!\!>m_d$.  If $\theta_{\rm cm}=90^\circ$
the momentum transferred to each nucleon is balanced and
the pQCD result is reached most rapidly.  A ``typical'' pQCD
diagram leading to large angle scattering is illustrated in
Fig.~\ref{fig:dg-pQCD}.  In general we cannot expect the
cross section to follow pQCD unless the minimum of the two
momentum transfers $t_1$ and $t_2$ is larger than some value
$t_{\rm min}$ at which pQCD holds.  

If the above conditions are satisfied, then the cross section
and polarization observables should satisfy results predicted
by pQCD: the constituent counting rules (CCR) and hadron
helicity conservation (HHC).  

The CCR \cite{bf,conrules} predicts the 
energy dependence of scattering cross sections at  fixed center
of mass angle
\begin{equation}
{ {d\sigma}\over{dt} } = { {1}\over{t^{n-2}} }  f(\theta_{\rm
cm})\, ,
\end{equation}
where $n$ is the total 
number of pointlike particles in the initial and final states
of the reaction.  For deuteron photodisintegration, $n=13$
(there is one photon and 6+6=12 quarks), and because of
Eq.~(\ref{t1t2}) $t$ may be replaced by $s$, as is commonly
done.  For elastic $ed$ scattering $n=14$ because there are
electrons in both the initial and final state, and $t=Q^2$, the
momentum transfered by the electron.  In this case $dt\sim
Q^2d\Omega$, and recalling Eq.~(\ref{mott}) we have
\begin{equation}
{1\over Q^2}{ {d\sigma}\over{d\Omega} } \sim {g(\theta)\over
Q^4} A(Q^2) \sim f(\theta)\left({1\over Q^2}\right)^{12}
\, ,
\end{equation}
recovering the pQCD prediction $A\sim Q^{-20}$.

The second rule from pQCD is that the total helicity of the
incoming and outgoing hadrons should be
conserved \cite{bl} 
%
\begin{equation}
\sum_i \lambda_i = \sum_f \lambda_f\, ,
\end{equation}
where $\lambda_{i}$ ($\lambda_{f}$) are the helicities of the
initial (final) hadrons.  In Sec.~\ref{pertQCD} we discussed the
implication of this rule for the deuteron form factors.  HHC
makes predictions for many spin observables, particularly vector
polarizations \cite{carlson}.  Here it predicts that the
amplitudes of Table~\ref{tab:Fs} will have the following
behavior at large $t$ 
\begin{eqnarray}
\begin{array}{ll}
F_{1+},\; F_{3-},\;F_{5\pm}\quad\qquad & {\rm leading}\cr
F_{2\pm},\; F_{4\pm},\; F_{6\pm} & {\rm suppressed by}\; t^{-1}
\cr 
F_{1-}, \; F_{3+} & {\rm suppressed by}\; t^{-2}
\end{array}
\end{eqnarray}
where we have used the fact that each helicity flip is 
suppressed by a power of $t$.  The implication of HHC for the
observables given in Table \ref{tab:polpara} is summarized in
Table~\ref{tab:dgppol}.  The limits at $\theta_{\rm cm}$ $=$
90$^{\circ}$ on some of the observables require assumptions
about relations between the helicity conserving variables
\cite{nagame,aadspc}.  

\begin{table}[tbp]
\caption{\label{tab:dgppol}
Implications of HHC for some polarization observables in 
$\gamma d \rightarrow p n$.  The $90^\circ$ limits require
additional assumptions.}
\begin{indented}
\item[]\begin{tabular}{@{}lll}
\br
Observable  & HHC limit & Approach\\
& &  to HHC \\
\mr
$d\sigma/d\Omega $ & $F_{1+}^2 + F_{3-}^2 + F_{5+}^2 + F_{5-}^2$ & $t^{-2}$ \\
$p_y$ &   0     &   $t^{-1}$ \\
$T$   &   0     &  $t^{-1}$  \\
$\Sigma$ & $2{\rm Re}\,(F_{1+}F_{3-}^* + F_{5+}F_{5-}^*) /
(d\sigma/d\Omega)\; \rightarrow -1$ at 90$^{\circ}$ \quad &
$t^{-2}$ \\
$T_1$ & 0  &    $t^{-1}$  \\
$C_{x'}$ & 0  &   $t^{-1}$   \\
$C_{z'}$ & $[F_{1+}^2 - F_{3-}^2 + F_{5+}^2 - F_{5-}^2] /
(d\sigma/d\Omega) \;\rightarrow 0$ at 90$^{\circ}$  
      & $t^{-2}$ \\
$O_{x'}$ &  0    &   $t^{-1}$ \\
$O_{z'}$ &  $-2{\rm Im}(F_{1+}F_{3-}^* + F_{5+}F_{5-}^*) /
(d\sigma/d\Omega)\; \rightarrow 0$ at 90$^{\circ}$  & $t^{-2}$
\\
\br 
\end{tabular}
\end{indented}
\end{table}
 
In an attempt to include some of the soft physics omitted from
pQCD, and to extend the region of applicability of pQCD down to
lower momentum transfers, Brodsky and Hiller introduced the
idea of reduced nuclear amplitudes (RNA)
\cite{brodskyhiller}. In this model the gluon exchanges that
contribute to identifiable subprocesses (such as nucleon form
factors) are collected together and their
contributions replaced by experimentally determined nucleon form
factors. It is hoped that the resulting expressions will
correctly include much of the missing soft physics,
and will therefore be valid to lower momentum transfers
than the original pQCD expressions from which they were
obtained.  When applied to deuteron photodisintegration, the
cross section is written
\begin{equation}
{{d\sigma}\over{dt}} = {m^2\over24\pi^2 (s-m_d^2)^2} 
\,\sum |J|^2 \to
{1\over (s-m_d^2)^2}\, F_p^2(\hat{t}_p) F_n^2(\hat{t}_n)
{{1}\over{p_T^2}}  f^2(\theta_{\rm cm}) \label{RNA}
\end{equation}
where the phase space factor of $1/(s-m^2_d)^2$ comes for 
a careful reduction of the phase space factors in
Eqs.~(\ref{resptens}), (\ref{sigma0}), and (\ref{t1t2})
\footnote{In some experimental fits the phase space 
factor used is $(s-m_d^2)^{3/2}\sqrt{s-4m^2}$ instead of
$(s-m_d^2)^2$.}, 
$f(\theta_{\rm cm})$ is the reduced nuclear amplitude, $F_p$ and
$F_n$ are the proton and neutron form factors with
$\hat{t}_p$ and $\hat{t}_n$ the average momentum transferred
to the proton and neutron
\begin{eqnarray}
\hat{t}_p=(p_1-d/2)^2 \qquad
\hat{t}_n=(p_2-d/2)^2\, ,
\end{eqnarray}
and the square of the transverse momentum is 
\begin{equation}
p_T^2=\left({s\over4}-m^2\right)\,\sin^2\theta_{\rm cm}\, .
\end{equation}
The power of the ``extra'' factor of
$p_T^{-2}$ is fixed once the phase space and nucleon form
factors have been taken into account. Note that this
model does not attempt to normalize
$f$.

The RNA form given in Eq.~(\ref{RNA}) is somewhat 
arbitrary, particularly in the specification of the
form of $p_T^2$ 
\footnote{In Ref.~\cite{brodskyhiller} it is
proposed that $p_T^2=tu/s$.  This choice of $p_T$ leads
to an increased energy dependence and worse agreement with the
data.}.  
Radyushkin \cite{radyushkinunpub} has argued that the elementary
process not accounted for by the nucleon form factors (i.e.\ the
absorption of a hard photon followed by exchange of a hard gluon
with another quark) should include nonperturbative
contributions.  The effect is to replace the $1/p_T^2$ factor in
Eq.~(\ref{RNA}) by a smooth function $f^2(s,t)$, which is assumed to
vary slowly in energy and angle.  In the fits described below,
$f^2$ will be taken to be a constant adjusted to fit the data,
implying that $d\sigma/ dt\sim s^{-10}$ instead of $s^{-11}$.
   
Alternatively, if the quark exchange mechanism shown in
Fig.~\ref{fig:dg-pQCD} is to be taken seriously, a more detailed
calculation is possible.  This is the motivation for the work of
Frankfurt, Miller, Sargsian, and Strikman
\cite{fmss,fmss2} where the quark exchange diagram (which
the authors refer to as a quark rescattering diagram) is
calculated in front-form dynamics using model wave functions
for the nucleons and the deuteron.  The matrix element is
written as a convolution of an elementary quark exchange
interaction with the initial and final nucleon wave functions.  
The final nucleons are free and the distribution of the
initial nucleons is given by the deuteron wave function.  Since 
the photon momentum is shared by the proton and neutron, there
is little sensitivity to the high momentum part of the deuteron
wave function.   The elementary interaction is a quark
exchange between the two nucleons, with the photon absorbed by
one quark which then gives up its momentum through a hard gluon
exchange with another quark.  The authors show that this can be
replaced approximately by the wide angle $np$ scattering cross
section (also dominated by quark exchange).  The final formula 
obtained from these arguments is
\begin{equation}
{d\sigma\over dt}\Bigg|_{\gamma d}= {4\alpha\pi^4m\over
9(s-m^2_d)}C\left({\hat{t}_p/ s}\right)
\,{d\sigma(s,\hat{t}_p)
\over dt}\Bigg|_{np}\;
\left|\int{d^2p_\perp\over (2\pi)^2} \Psi_d(p_z=0,p_\perp)
\right|^2\, ,
\end{equation}
where $C(-x)=x/(1-x)$ was used in the fits to the data
discussed below.  The authors propose using experimental
data for the $np$ cross section, but since data does not
exist for the actual kinematic conditions needed, it must
be extrapolated, and predictions for photodisintegration
are given as a band corresponding to the uncertainties
introduced by the extrapolations.  The authors believe
that their predictions should be valid for $E_{\gamma}$
$>$ 2.5 GeV, and nucleon momentum transfers $-t=-t_1$ and
$-u=-t_2$ $>$ 2 GeV$^2$.    

The high energy approaches described above all focus on the region
where both $\hat{t}_p$ and $\hat{t}_p$ are large (where
perturbative arguments can serve as the foundation for
the treatment).  Alternatively, we may ask what to expect when
one of these momentum transfers is small (but $s$ is still
large).  The authors in Refs.~\cite{qgs,qgs01} develop a model
(which they refer to as the ``quark-gluon string model'') based
on a Reggie generalization of the nucleon exchange Born term. 
Here the exchanged nucleon is replaced by a nucleon Reggie
trajectory that represents the sum of a tower of exchanged nucleon
resonances (or the exchange of three quarks dressed by an
arbitrary number of gluons).  The energy dependence of the
predicted cross section is 
\begin{equation}
{d\sigma\over dt}\to \left(s\over s_0\right)^{2\alpha_N(t)-2}  \,
,
\end{equation}
where $\alpha_N(t)$ is the nucleon Reggie trajectory, with
$\alpha_N(t)=-0.5+0.9t+0.25t^2/2$, where $t$ is in units
of GeV$^2$.  Recent work \cite{qgs01} emphasizes the importance of
the nonlinear term in the Regge trajectory.  The model is
intended to work for $E_\gamma>1$ GeV, with $-t$  less than about
1 GeV$^2$. 

We now turn to a review of the experimental data and to a
comparison between theory and experiment.


\subsection{Experimental Status}

\fulltable{\label{tab:dgppoldata}
Measurements of deuteron photodisintegration polarization 
observables.}
{\small
\begin{tabular}{@{}llllll}
\br
Laboratory  & Observable & $E$ (MeV)& $\theta_{\rm cm}$($^{\circ}$) & 
\# of Points & Reference \\
\mr
MIT               & $p_y $        & 250 & 49 & 1 & 
        \cite{feld60} \\
Livermore         & $p_y^n$       & 2.75 & 50 - 136 & 5 & 
        \cite{john61} \\
Z\"urich          & $p_y^n$       & 2.75 & 44, 94   & 2 & 
        \cite{bosch63} \\
Illinois          & $p_y^n$       & 12 - 23 & 148  & 4 & 
        \cite{frederick63} \\
RPI               & $p_y^n$       & 12 - 30 & 90   & 3 & 
        \cite{bertozzi63} \\
Purdue            & $p_y$       & 294 & 72 & 1 & 
        \cite{lpw63} \\
Stanford          & $\Sigma$    & 80 - 140 & 45, 90, 135 & 41 & 
        \cite{liu64,liu65} \\
Livermore         & $p_y^n$       & 2.75 & 32 - 152 & 7 & 
        \cite{jewell65} \\
Frascati          & $\Sigma$    & 235 - 404 & 90 & 8 & 
        \cite{barbiellini67} \\
Stanford          & $p_y$       & 172 - 436 & 39 - 126 & 19 & 
        \cite{liu68} \\
Bonn              & $p_y$       & 282 - 405 & 74 - 98 & 4 & 
        \cite{kose69} \\
Yale              & $p_y^n$      & 7 - 30 & 48, 94 & 20 & 
        \cite{nath72} \\
Yale              & $p_y^n$      & 7 - 13 & 90 & 3 & 
        \cite{drooks76} \\
Tokyo             & $p_y$       & 352 - 697 & 45 - 133 & 27 & 
        \cite{kamae77,kamae78,ikeda79,ikeda80} \\
Kharkov           & $\Sigma$    & 80 - 600 & 75 - 150 & 109 &
        \cite{gorbenko79,gorbenko80,gorbenkoerr,gorbenko82yf35,gorbenko82npa} \\
Frascati          & $\Sigma$    & 10 - 69 & 90 & 9 &
        \cite{delbianco80,delbianco81} \\
Kharkov           & $p_y$       & 375 - 700 & 43, 78, 90, 120 & 40 &
        \cite{brata80yf31,brata80yf32,brata80pzetf31} \\
Kharkov           & $p_y$       & 550 - 1125 & 90, 120 & 30 &
        \cite{brata81pzetf34,brata82pzetf36,brata86yf44} \\
Yerevan           & $\Sigma$    & 400 - 700 & 45, 55 & 5 &
        \cite{adamian81} \\
Tokyo             & $T$         & 324 - 672 & 72, 100, 130 & 24 &
        \cite{ishii82,ohashi87} \\
Kharkov           & $p_y,\Sigma, T_1$ & 300 - 600 & 75, 90, 120 & 22,20,20 &
        \cite{brata82pzetf35,brata86yf43,bara86,ganenko92} \\
Frascati          & $\Sigma$    & 20, 29, 39, 61 & 14 - 165 & 41 &
        \cite{depascale82,depascale85} \\
Bonn              & $\Sigma$    & 233 - 818  & 114, 135 & 103 &
        \cite{dahl82} \\
Argonne           & $p_y^n$     & 6 - 14     & 90       & 6   &
        \cite{holt83} \\
Kharkov           & $\Sigma$    & 40, 50, 60, 70 & 75, 90 & 8 &
        \cite{barannik83} \\
Bonn              & $T$         & 450, 550, 650 & 25 - 155 & 41 &
        \cite{althoff84,althoff89} \\
Yerevan           & $\Sigma$    & 395 - 795 & 45 - 95 & 30 &
        \cite{adamian84,agababyan85} \\
TRIUMPF           & $A_y^n$    & 180, 270 & 32 - 144 & 18 &
        \cite{cameron84,cameron86} \\
TRIUMPF           & $A_y^n$    & 370, 478 & 45 - 155 & 45 &
        \cite{hugi87} \\
Tomsk             & $\Sigma$    & 50 - 100 & 45, 60, 90 & 13 &
        \cite{vnukov86,vnukov88} \\
Wisconsin         & $A_y^n$    & 6, 13 & 94 & 2 &
        \cite{soderstrum87} \\
Novosibirsk       & $T_{21}$    &  33 - 125 & 50 & 4 &
        \cite{mostovoy87} \\
Yerevan           & $O_{x'}^n$    & 300, 400, 500 & 130 & 3 &
        \cite{adamian88} \\
Kharkov           & $p_y$       & 200 - 367 & 25 - 110 & 30 &
        \cite{zybalov90,zybalov91} \\
Yerevan           & $p_y,p_{xz}$ & 306 - 436 & 65, 75 & 2$\times$8 &
        \cite{avakyan90,avakyan90b} \\
Karlsruhe         & $A_y^n$     & 19 - 50   & 62, 98, 131 & 27 &
        \cite{fink91} \\
Yerevan           & $\Sigma$    & 284 - 999 & 45, 60, 75, 95 & 94 &
        \cite{adamian91} \\
BNL LEGS          & $\Sigma$    & 100 - 314 & 16 - 162 & 112 &
        \cite{blanpied91,blanpied95,blanpied99} \\
Novosibirsk       & $T_{20},T_{22}$    & 49 - 505 & 88 & 2$\times$9 &
        \cite{mishnev93} \\
IUCF              & $C_{nn},A_y,A_y^n$ & 183 & 48 - 125 & 3$\times$6 &
        \cite{pate93} \\
PSI               & $A_y^n$             & 68 & 69 - 144 & 5 &
        \cite{tuccillo94} \\
Mainz             & $\Sigma$    &  160 - 410 & 35 - 155 & 140 &
        \cite{wartenberg99} \\
Yerevan           & $\Sigma$    & 787 - 1566 & 90 & 6 &
        \cite{adamian00} \\
JLab              & $p_y,C_{x'},C_{z'}$ & 479 - 2411 & 90 & 
10,9,8 &        \cite{wije01} \\
\br
\end{tabular}}
\endfulltable


The world data set for cross sections (except for the most recent 
experiments) has been presented several times \cite{oldreview},
and we will not review the normalization problems of
older cross section data sets.
Table~\ref{tab:dgppoldata} presents an extensive
list of the published polarization data.
About 70 publications, starting in 1960,
present about 1200 data points for photodisintegration 
and the time reversed radiative capture reaction.
Table~\ref{tab:dgppoldata} generally lists photon lab energy
and proton c.m.\ angle (neutron c.m.\ angle for the $p_y^n$ data).
All of the radiative capture experiments have measured $A^n_y$,
and for these experiments we usually give the neutron beam kinetic energy
and outging photon c.m.\ angle; for the IUCF experiment we give their
reported proton c.m.\ angle.
Matching c.m.\ energies leads to $T_n = E_{\gamma}m_d/m_p + 
(m^2_d - (m_n+m_p)^2)/2m_p$ or $T_n
\approx 2 E_{\gamma}$ at high energies.
Comparison of these data indicates serious problems with
backgrounds and/or estimates of systematic uncertainties
in a number of cases, as will become clear in figures in
the sections below.

We review some of the lower energy data in the next subsection.  
High energy experiments, with photon energy above $\approx$1 GeV,
are covered in the following subsection.

\subsubsection{Low energy photodisintegration}

\begin{figure}  
\begin{center}
\mbox{\epsfxsize=4.0in \epsffile{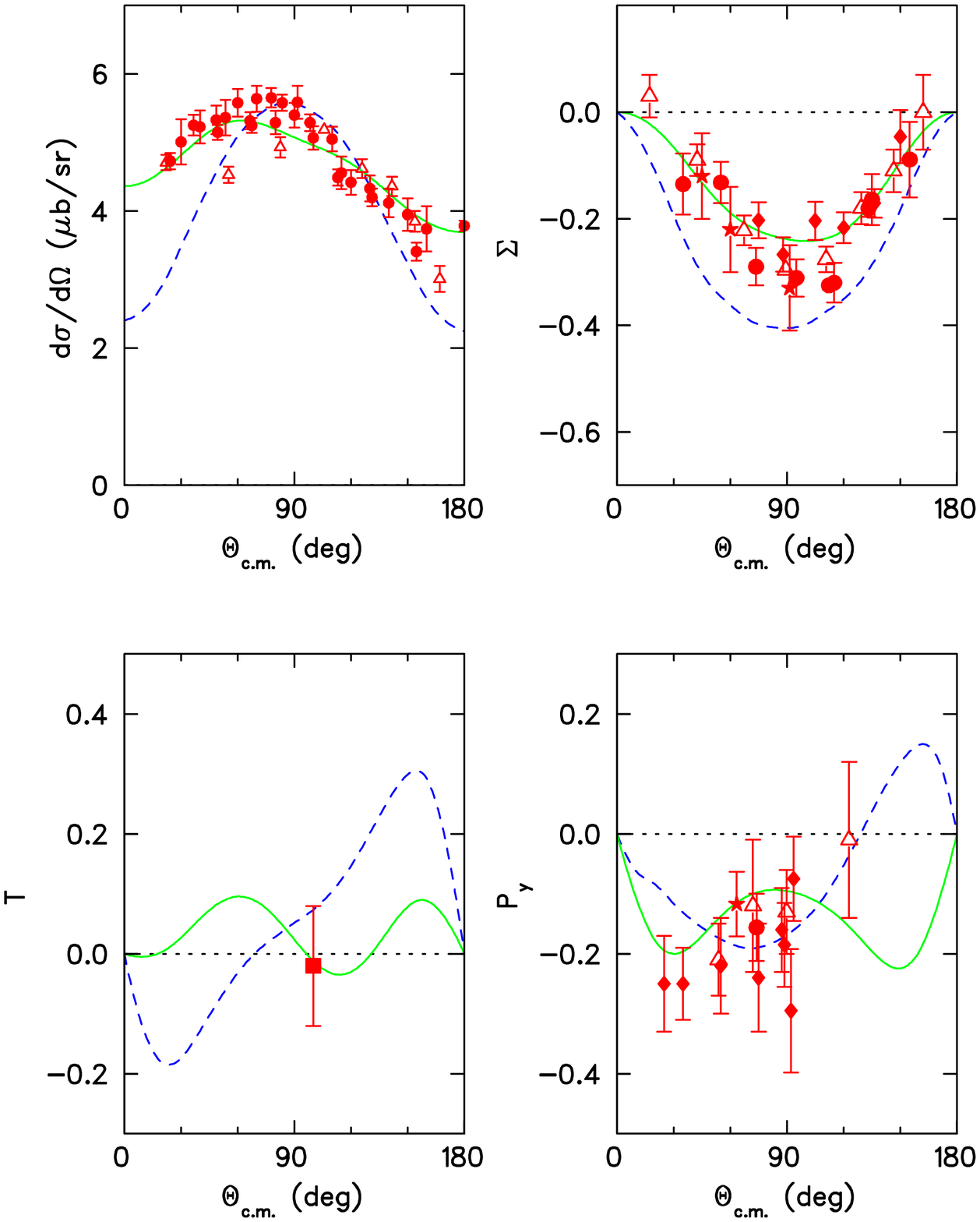} }   
\caption{\label{fig:dgpobs300} Four observables for 
$d(\gamma,p)n$ at 300 MeV. Calculations are from Schwamb and
Arenh\"ovel (solid line) \cite{schwambarenhovel01} and Kang,
Erbs, Pfeil and Rollnik (dashed line) \cite{kang}. The cross
section data are from LEGS (triangles) \cite{blanpied99} and
Mainz \cite{crawford96} plus Bonn
\cite{dougan76,althoff83,arends84} (circles). The $\Sigma$
data are from LEGS (triangles) \cite{blanpied99},  
Mainz (circles) \cite{wartenberg99},
Kharkov (diamonds) \cite{gorbenko82npa}, and Yerevan (stars)
\cite{adamian91}. The $T$ datum is from Tokyo
\cite{ohashi87}. The $p_y$ data are from 
Stanford (triangles) \cite{liu68}, Bonn (circle)
\cite{kose69}, Yerevan (star) \cite{avakyan90b}, and  Kharkov
(diamonds) \cite{ganenko92,zybalov91}. See
Table~\ref{tab:dgppoldata} for related references. }
\end{center}
\end{figure} 

\begin{figure} 
\begin{center}
\mbox{\epsfxsize=4.0in \epsffile{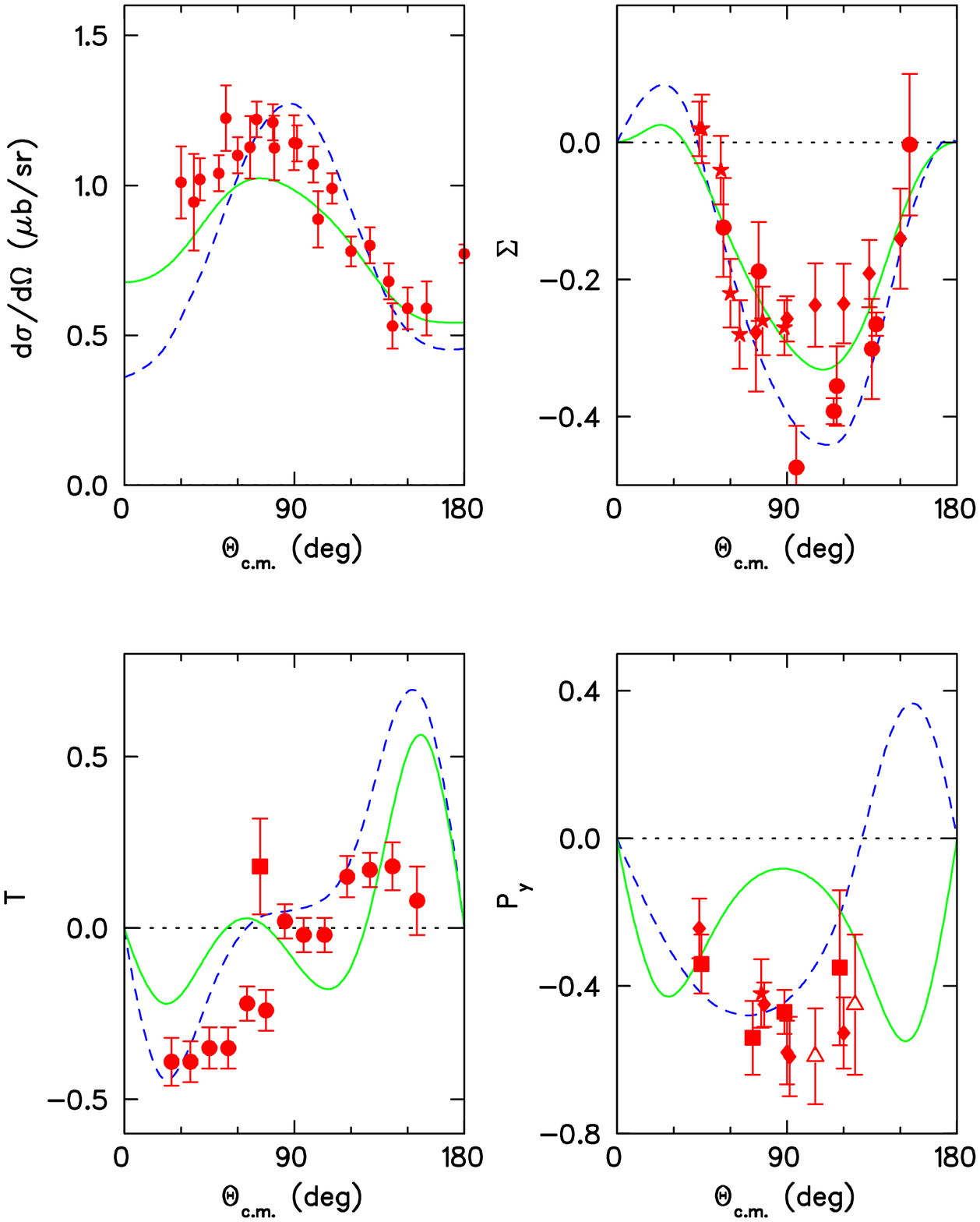} }
\caption{\label{fig:dgpobs410} 
Four observables for $d(\gamma,p)n$ at about 450 MeV (410 MeV
for $\Sigma$).   Calculations are from Schwamb and
Arenh\"ovel  (solid line) \cite{schwambarenhovel01} and Kang,
Erbs, Pfeil and Rollnik (dashed line) \cite{kang}. The cross
section data are from Mainz \cite{crawford96} plus  Bonn
\cite{dougan76,althoff83}. The $\Sigma$ data are from Mainz
(circles) \cite{wartenberg99}, Kharkov (diamonds)
\cite{gorbenko82npa}, and Yerevan (stars)
\cite{adamian91}. The $T$ data are from Tokyo (square)
\cite{ohashi87} and Bonn (circles)
\cite{althoff89}. The $p_y$ data are from Stanford (triangles)
\cite{liu68}, Tokyo (squares) \cite{ikeda80}, Yerevan (star)
\cite{avakyan90b}, and  Kharkov (diamonds)
\cite{brata80pzetf31,ganenko92}. See
Table~\ref{tab:dgppoldata} for related references. }
\end{center}
\end{figure}

In this section we review selected experiments
with beam energies from about pion production threshold
to several hundred MeV.
Tagged photon facilities, with their improved knowledge
of incident beam flux, have allowed significantly improved
cross section measurements in this region since the 1980s.
Of particular note are extensive recent data sets
from LEGS and Mainz. 
Figures~\ref{fig:dgpobs300} and \ref{fig:dgpobs410} show angular
distributions for the cross section, and $\Sigma$,
$T$, and $p_y$ at photon energies near 300 and 450 MeV,
respectively.

The LEGS tagged, backscattered, and linearly-polarized photon beam
was used to determined cross sections and  $\Sigma$
\cite{blanpied91,blanpied95,blanpied99}. Five independent measurements used three
detector systems, two targets, and two different laser
frequencies. Data were taken for $E_{\gamma}$ = 100 - 315 MeV
with eight laboratory angles from 15 - 155$^{\circ}$.
Cross section statistical uncertainties range from a few percent to
about 15\%, with systematic uncertainties of 5\%.
One observation in the LEGS data is of pion contamination
that may have been missed in earlier experiments, leading to
increased cross sections.

The MAMI Mainz experiment \cite{crawford96} used the Glasgow photon tagger
along with the large solid angle detector DAPHNE to determine
cross sections in the ranges $E_{\gamma}$ = 100 - 800 MeV and 
$\theta_{\rm cm}$ = 30 - 160$^{\circ}$.
Data were binned by 20 MeV in energy and 10$^{\circ}$ in angle.
Statistical uncertainties ranged from a few percent at lower energies
to about 25\% at the highest energies.
Systematic errors were also energy dependent, ranging from a few percent
to several percent.
The $\Sigma$ asymmetry \cite{wartenberg99} was obtained
by using a coherent bremsstrahlung radiator.

Agreement between the Mainz and LEGS cross section results is 
generally better than 10\%.
The $\Sigma$ asymmetries also agree well with each other and 
with earlier results from Yerevan \cite{adamian91}.
Measurements from Kharkov \cite{gorbenko79,gorbenko80,gorbenko82yf35,gorbenko82npa} 
generally agree, except for a tendency to be 
slightly smaller at many beam energies.

The vector polarized target asymmetry $T$ was measured at
Tokyo \cite{ishii82,ohashi87} and at Bonn \cite{althoff84,althoff89}.
The two measurements generally agree, with the data appearing to follow, 
very roughly, a $-\sin\theta$ dependence at each energy,
as do the calculations of \cite{kang}. 

The induced polarization $p_y$ has been measured at a number of
laboraties, with significant amounts of lower energy data from
Stanford \cite{liu68}, Tokyo
\cite{kamae77,kamae78,ikeda79,ikeda80}, and Kharkov. There were
several experiments at Kharkov, including  initial measurements
\cite{brata80yf31,brata80yf32,brata80pzetf31}, high-energy
measurements \cite{brata81pzetf34,brata82pzetf36,brata86yf44},
simultaneous measurements of $\Sigma$, $T_1$, and $p_y$
\cite{brata82pzetf35,brata86yf43,bara86,ganenko92}, and
lower-energy measurements \cite{zybalov90,zybalov91}.
The Kharkov data do not have a desirable level of consistency.
The simultaneous measurements of  $\Sigma$, $T_1$, and $p_y$
were taken as single-arm data, away from the photon endpoint,
and may suffer from backgrounds.
Polarizations below about 300 MeV, including the intermediate energy neutron
measurements \cite{cameron84,cameron86,hugi87}, 
appear to be well reproduced by theories.
There are numerous data, particularly at $\theta_{\rm cm}$ $=$ 
45$^{\circ}$, 78$^{\circ}$, 90$^{\circ}$, and 120$^{\circ}$, 
but there are few energies at which there are good angular
distributions.
The conclusion that polarizations are large, close to $-1$,
and peak at about 500 MeV near 90$^{\circ}$ is beyond dispute.

Figures~\ref{fig:dgpobs300} and \ref{fig:dgpobs410} show 
the good agreement of the recent Mainz \cite{schwambarenhovel01} 
and older Bonn \cite{kang} calculations with the cross section and $\Sigma$
asymmetry.  Theory seems to agree with
$d\sigma/d\Omega_{||}$ better than
$d\sigma/d\Omega_{\perp}$ \cite{blanpied95}.  The agreement
is better at the lower energy, and the newer Mainz
calculation is generally in better agreement. However, there
is difficulty, particularly at the higher energy, with $T$ and
$p_y$,  both imaginary parts of the interference of amplitudes. 
The large induced polarizations above the $\Delta$ resonance
have remained a puzzle for almost 30 years, and are still not
fully explained by the newest theories.  The Bonn calculation
was in sufficiently good  qualitative agreement with all
observables ($d\sigma/d\Omega$,
$\Sigma$, $T$, {\em and} $p_y$) for energies up several hundred
MeV for the authors to consider this $p_y$ puzzle solved. 
However, detailed examination of Fig.~\ref{fig:dgpobs410} shows
that neither the shape nor strength of the angular distribution
is accurately reproduced;
this will become clearer when we examine the energy
dependence of $p_y$ at $\theta_{\rm cm}$ = 90$^{\circ}$ below.


\subsubsection{High energy photodisintegration}

\begin{figure}
\begin{center}
\vspace*{0.2in}
\mbox{\epsfxsize=4.2in \epsffile{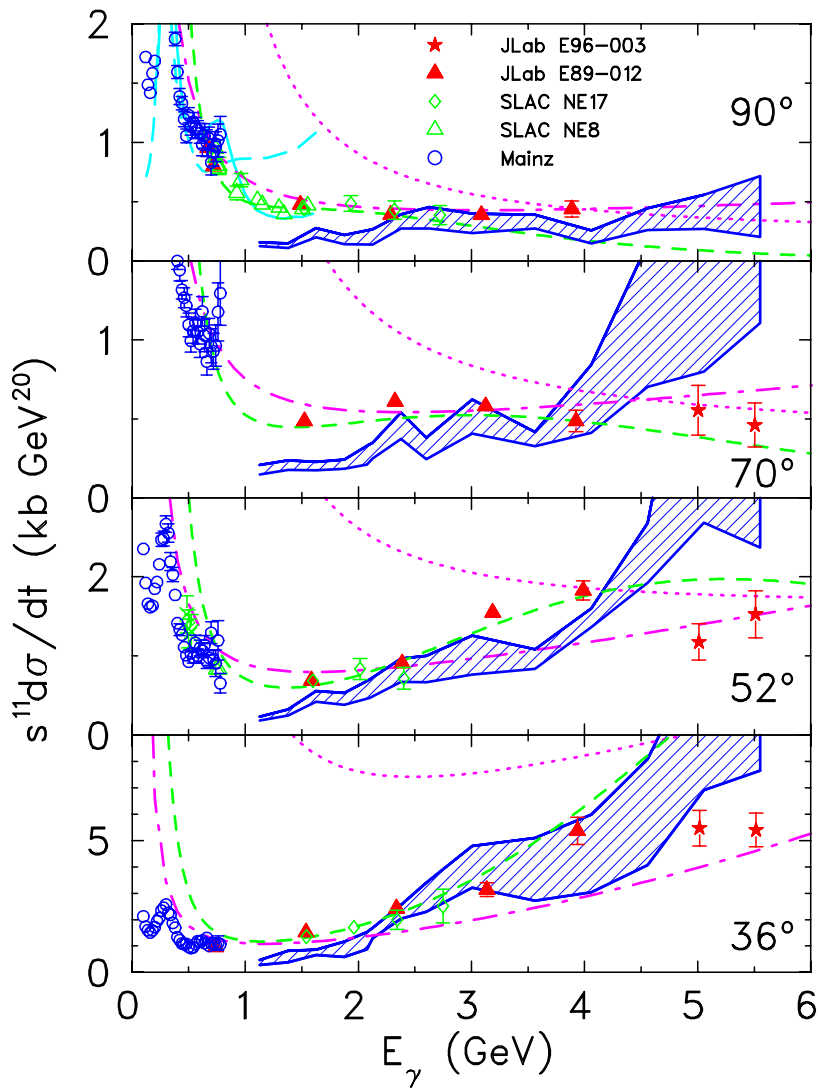} } 
\end{center}
\caption[]{\label{fig:dgpsigmas} Photodisintegration cross section
$s^{11}d\sigma/dt$ versus incident lab photon energy.
The calculations are from 
Kang, Erbs, Pfeil and Rollnik (solid line; 90$^\circ$ low
energy data only) \cite{kang}, Lee (long dashed line; 90$^\circ$
low energy data only) \cite{tshlee}, Raydushkin (dot-dashed line)
\cite{radyushkinunpub}, RNA of Brodsky and Hiller (dotted)
\cite{brodskyhiller}, quark qluon string model (short dashed) 
\cite{qgs,qgs01}, and Frankfurt, Miller,
Strikman and Sargsian (shaded region) \cite{fmss,fmss2}.}
\end{figure}

Figure~\ref{fig:dgpsigmas} shows the published
high energy photodisintegration data, from
experiments NE8 \cite{napolitano88,freedman93} and NE17 \cite{belz95}
at SLAC, and E89-012 \cite{bochna98} and E96-003 \cite{schulte01} 
at JLab.
These experiments determine cross sections for $\theta_{\rm cm}$ $\approx$ 
36$^{\circ}$, 52$^{\circ}$, 69$^{\circ}$,and 89$^{\circ}$ at energies
from about 0.7 to 5.5 GeV;
there are also some backward angle data up to 1.8 GeV from NE8.
These data overlap well;
the experiments, while all run by essentially the same collaboration,
used three spectrometers in two experimental halls at 
two laboratories.
There is also good overlap, variations of less than about 20\%,
with the highest energy Mainz
tagged photon data \cite{crawford96}, and with older untagged data
\cite{dougan76,arends84,myers61,ching66,baba83}.

Tagged photon measurements at low energies provide an accurate
measure of beam flux, and along with the measured proton angle and energy,
can determine a missing mass that allows background rejection.
At high energies, smaller cross sections cannot be determined with the
reduced flux of tagged photons.
The Bremstrahlung endpoint technique was used for all of the
SLAC and JLab measurements shown in Fig.~\ref{fig:dgpsigmas}.

In the endpoint technique, the measured proton momentum vector determines
the incident photon energy and neutron kinematics,
{\em assuming} the reaction is two body photodisintegration.
Low momentum protons are cut from the analysis to prevent contamination
from final states such as $p n \pi^0$, while
high momentum protons are cut
to eliminate the larger uncertainty in the photon flux close to
the photon endpoint. Backgrounds are determined by radiator out
and empty target measurements, and subtracted.
Events in the region beyond the endpoint can be used to check the subtraction.
The increase in time required for this subtraction makes it prohibitive
for the highest energy measurements;
for these the electrodisintegration background is calculated
\cite{tiatorwright} and subtracted.  To determine the cross
section, the incident photon flux is calculated  using the method
of Ref.~\cite{matthewsxx}. 
Thick radiator corrections are typically about 15\% for a radiator
thickness of 6\% of a radiation length.

The main feature of the cross section data above about 1 GeV is
the $s^{-11}$ ($s^{-10}$) fall off of the cross sections
$d\sigma/dt$  ($d\sigma/d\Omega$) at $\theta_{\rm cm}$ =
90$^{\circ}$ and 69$^{\circ}$, in  agreement with the CCR and
thus with perturbative QCD expectations. In contrast, the cross
sections at the forward angles 36$^{\circ}$ and  52$^{\circ}$
fall off more slowly, with approximate $s^{-9}$  scaling at lower
energies until the onset of the $s^{-11}$ behavior at about 4
and 3 GeV beam energy, respectively. At each angle, the onset of
the $s^{-11}$ behavior corresponds to $p_T$ $\approx$ 1 GeV. 

The RNA and the Radyushkin estimates in Fig.~\ref{fig:dgpsigmas}
were normalized to the datum at 89$^\circ$ and $E_\gamma=4$ GeV, 
fixing their one free parameter.  
The RNA is then almost a factor of 2 too large at 36$^\circ$, 
and also much too large at lower energies,
requiring that the soft physics missing from the RNA interfere
{\it destructively\/} with the leading terms.  
Suggested angular dependences of $f(\theta_{\rm cm}$) \cite{brodskyhiller}
would increase the RNA curve further at the forward angles,
worsening agreeement with the data.
In contrast, the
Radyushkin estimate gives a somewhat better account of both the
angular and energy dependence (even though it only goes
asympotically as $s^{-10}$), confirming that phase space and
nucleon form factors are all that is needed to account for much of
the kinematic variation of the cross section.
While the apparent onset of scaling at the forward angles suggests
this agreement is starting to break down, we conclude the
present data are insufficient to uniquely fix the asymptotic energy
dependence of the cross section.
 
The cross section data are also reasonably well
reproduced by the model of Frankfurt, Miller,
Strikman and Sargsian [FMSS] \cite{fmss,fmss2} and the quark
gluon string (QGS) model \cite{qgs,qgs01}.  The predictions of
FMSS are uncertain because (a) the high energy $NN$
scattering data has an uncertain energy dependence
reflecting the experimental errors, and (b) the extrapolation of
the $NN$ data required for the predictions introduces further
errors.  These two uncertainties combine to give the jagged region
shown in Fig.~\ref{fig:dgpsigmas}.  The QGS model describes the
forward angle data up to 4 GeV reasonably well, even for values of
$-t$ exceeding the nominal limits of the model. The newer work
\cite{qgs01} predicts that the angular distributions will become
increasingly symmetric at higher energies; 
older estimates \cite{qgs} had predicted an increasing {\em asymmetry\/}. 

\begin{figure}
\begin{center}
\mbox{\epsfxsize=3.0in \epsffile{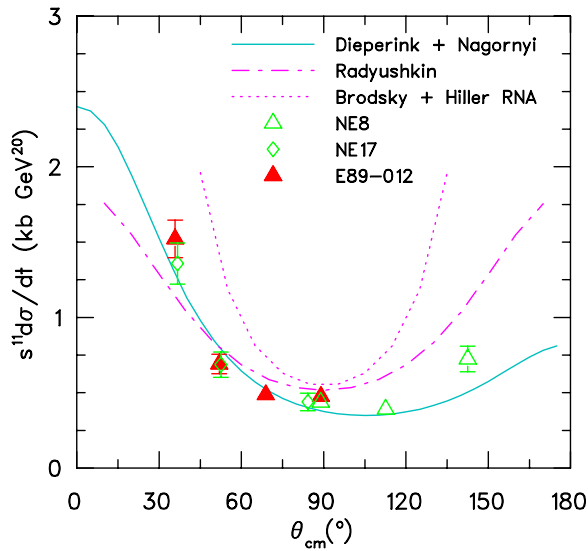} }  
\end{center}
\caption{\label{fig:ad} Deuteron photdisintegration angular 
distribution at $E_{\gamma}$ $\approx$ 1.6 GeV. Data are from
SLAC experiments NE8
\cite{napolitano88,freedman93} and NE17 \cite{belz95}, and JLab
experiment E89-012 \cite{bochna98}. Calculations are from 
Dieperink and Nagornyi \cite{dienag}, Radyushkin
\cite{radyushkinunpub}, and the RNA of Brodsky and
Hiller \cite{brodskyhiller}.  The Radyushkin
curve has been normalized to  the 4 GeV 90$^\circ$ datum, as in
Fig.~\ref{fig:dgpsigmas}, while the RNA curve has been reduced
60\% from that normalization to better agree with this data.} 
\end{figure}
 
Two experiments currently have unpublished data for cross sections
at photon energies up to 2.5 GeV.  
JLab Hall B E93-017
\cite{e93017} used the CLAS with tagged photons  to determine
nearly complete angular distributions. The preliminary data agree
well with earlier measurements, but are much more  comprehensive
than previous measurements in this energy range.  Hall A E99-008
\cite{e99008} has taken angular distributions at eight angles with
$\theta_{\rm cm}$ $=$ 30$^{\circ}$ -- 143$^{\circ}$,
at energies of 1.67, 1.95, and 2.50 GeV.  
Fig.~\ref{fig:ad} shows a sample angular distribution at
approximately 1.6 GeV.   


\subsubsection{Polarization observables in high energy
photodisintegration}

\begin{figure}
\begin{center}
\mbox{\epsfxsize=5.0in 
\epsffile{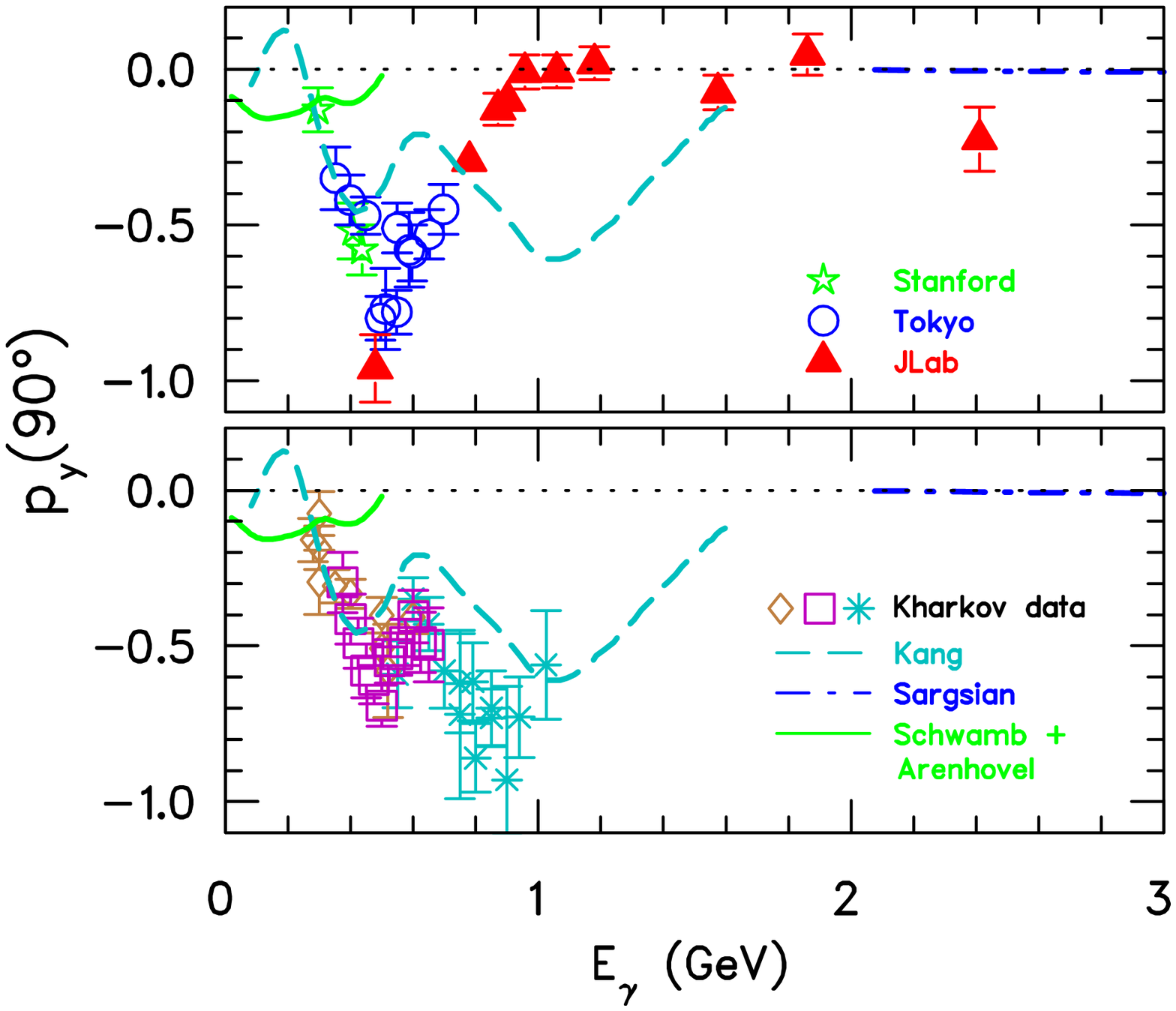} 
}  
\end{center}
\caption{\label{fig:89019py} Induced polarization $p_y$ for 
photodisintegration at $\theta_{\rm cm}$ = 90$^{\circ}$. 
Data sets from Stanford \cite{liu68}, Tokyo \cite{ikeda80},
and JLab \cite{wije01}, are in the top panel. 
Data sets from Kharkov (squares) \cite{brata80pzetf31}, 
(stars) \cite{brata86yf44}, and (diamonds) \cite{ganenko92} 
are in the lower panel.
See Table~\ref{tab:dgppoldata} for related references. 
The calculations are from 
Kang, Erbs, Pfeil and Rollnik (dash line) \cite{kang}, 
Sargsian (dash dot line) \cite{sargsianpc}
and from Schwamb and Arenh\"ovel (solid line) 
\cite{schwambarenhovel01}.}
\end{figure}

\begin{figure}
\begin{center}
\mbox{\epsfxsize=3.0in \epsffile{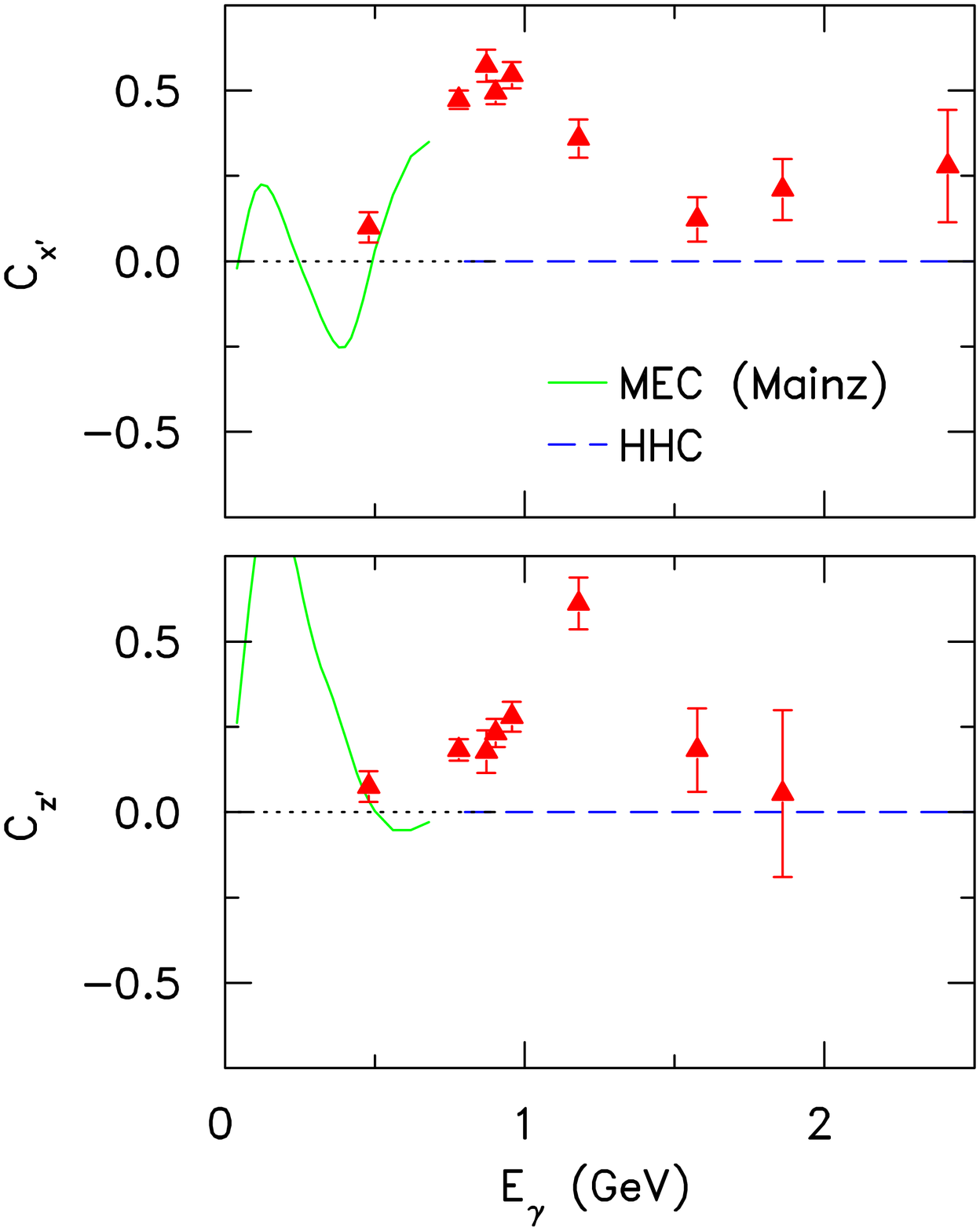} }
\end{center}
\vspace*{-0.2in}
\caption{\label{fig:89019cxcz} Polarization transfer for 
deuteron photodisintegration at $\theta_{\rm cm}$ $=$ 90$^\circ$.
The data are from Ref.\ \cite{wije01}. 
The calculation is from Schwamb and Arenh\"ovel 
\cite{schwambarenhovel01}.  HHC (with some additional 
assumptions for $C_{z'}$) predicts that both of these
amplitudes should vanish as $s\to\infty$.}
\end{figure}

\begin{figure}
\begin{center}
\mbox{\epsfxsize=3.5in \epsffile{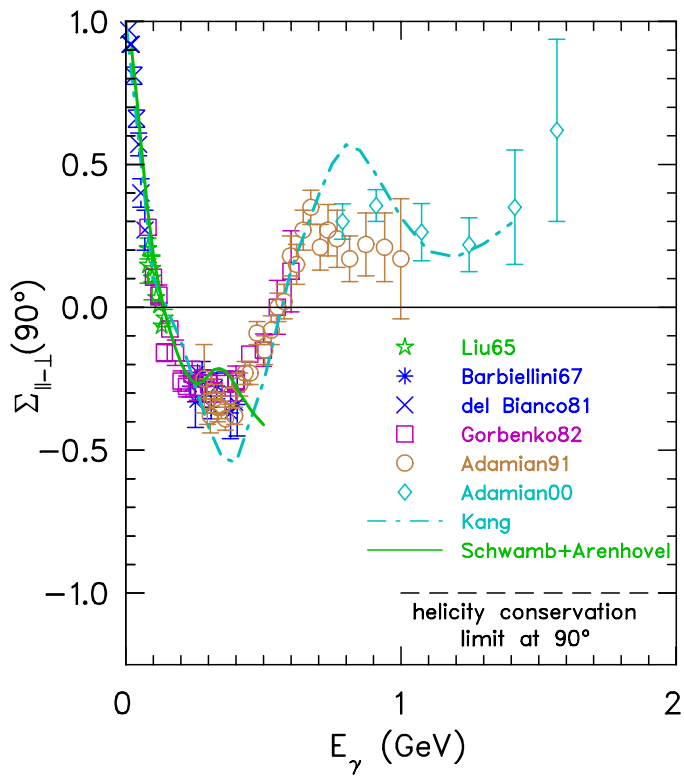} }  
\end{center}
\caption{\label{fig:sigfig} Polarized photon asymmetry
$\Sigma$  at $\theta_{\rm cm}$
$=$ 90$^\circ$. The calculations are from Kang, Erbs, Pfeil and
Rollnik (dot-dashed line) \cite{kang} and from Schwamb and
Arenh\"ovel (solid line) \cite{schwambarenhovel01}.
The data are from
Liu65 \cite{liu65},
Barbiellini67 \cite{barbiellini67},
del Bianco81 \cite{delbianco81},
Gorbenko82 \cite{gorbenko82npa},
Adamian91 \cite{adamian91},
and
Adamian00 \cite{adamian00}.
See Table~\ref{tab:dgppoldata} for related references.}
\end{figure}

There are only three sets of polarization data for
deuteron photodisintegration at energies near and above 1 GeV. 
The induced polarization $p_y$ was measured at Kharkov
\cite{brata81pzetf34,brata82pzetf36,brata86yf44},
the $\Sigma$ asymmetry was measured at Yerevan
\cite{adamian91,adamian00},
and $p_y$ and the polarization transfers $C_{x'}$ and $C_{z'}$
were measured at JLab \cite{wije01}.
Data for the energy dependence of these four observables at a fixed
$\theta_{\rm cm}=90^\circ$ are compared with theory in
Figs.~\ref{fig:89019py}, \ref{fig:89019cxcz}, and
\ref{fig:sigfig}.

The Kharkov measurements of $p_y$ at $\theta_{\rm cm}$ $=$
90$^{\circ}$ and 120$^{\circ}$ extend up to about 1.1 GeV (see
Fig.~\ref{fig:89019py}). These experiments were very difficult.
The small duty factor at Kharkov increases instantaneous
background  rates a factor of about 20,000 over those at JLab. 
These large backgrounds made it necessary to use multiple spark
chambers to track particle trajectories.  It was also difficult
to calibrate the polarimeter.  Calibrations of a  polarimeter are
best done by measuring its analyzing power using the known
$\vec{e} p\rightarrow e^{\prime} \vec{p}$ elastic scattering
reaction \cite{wije01,jones99,gayou01}, but Kharkov had no
polarized beam. 
The Kharkov measurements relied on a single elastic $ep$ point to check 
false asymmetries in their polarimeter, and used analyzing
powers from the literature. Finally, the polarimeter had a rear
trigger scintillator; any inefficiencies in the
scintillator would lead to false asymmetries. In contrast, the
recent JLab experiment \cite{wije01} had little background,
used $\vec{e} p \rightarrow e^\prime \vec{p}$ calibrations to
determine false asymmetries and analyzing powers at each
kinematic setting, and had no rear trigger scintillator.
Given the clear disagreement of the Kharkov data with the recent
JLab data, and noting that one of us (RG)
is a spokesperson of the JLab experiment, we conclude that the
highest energy set of Karkov data should not be trusted. 

The induced polarizations shown in Fig.~\ref{fig:89019py}
confirm our comments above concerning the Bonn calculation.
The large negative polarization near 500 MeV is not reproduced,
and the calculation is qualitatively incorrect at higher energies.
The imaginary part of the amplitude appears to be a problem in these
meson-baryon calculations; presumably this arises from an inadequate
treatment of resonances. 

Taken together, these recoil polarization data only weakly
confirm the predictions of HHC, which predicts that $p_y$ and
$C_{x'}$ should approach zero as $s\to\infty$, and that (with
additional assumptions 
\cite{nagame,aadspc} about relations between the helicity 
conserving amplitudes at $\theta_{\rm cm}$ = 90$^{\circ}$)
$\Sigma\to-1$ and $C_{z'}\to0$ as $s\to\infty$.  The highest 
energy polarization measurements of $p_y$ show that it is
consistent with vanishing at energies  above about 1 GeV, the
same energy at which the $s^{-11}$  cross section scaling
begins.  [In the Radyushkin model $p_y$ should be zero because the
amplitudes are all real if there is no gluon
exchange \cite{aaarpc}.] Similiarly, the polarization transfer
observables $C_{x'}$ and $C_{z'}$ both appear to peak near 1 GeV,
and decrease at higher energies.  However, $C_{x'}$ does not
appear to vanish sufficiently rapidly; the data might be
inconsistent with HHC.   So $p_y$ and $C_{z'}$ (and
perhaps $C_{x'}$) seem to have close to the correct behavior,
but $\Sigma$ does not.  The highest energy $\Sigma$ asymmetry 
measurements from Yerevan, with data up to about 1.6 GeV, give
the immediate impression that there is a minimum near 1.2 GeV and
that above this energy the asymmetry is tending to increase
towards 1, although the data are also statistically consistent
with a constant value of about 0.3. In either case the trend is
clearly not consistent with $\Sigma
\rightarrow -1$.  

In models based on meson-baryon degrees of freedom, the data
indicate that the combined effects of resonances plus final-state
interactions are small.  Calculations of $p_y$ in meson baryon
theories \cite{laget,la,to,saw,kang} indicate that $p_y$ at 
higher energies arises largely from resonance - background 
interference, with a small contribution from final-state
interactions. Calculations generally indicate that the $\Delta$
resonance generates a large polarization, though only perhaps
about 50\% of the magnitude seen in the experimental data at 
$\theta_{\rm cm}$ $=$ 90$^{\circ}$.
The Roper and $S_{11}$ have small effects, while
the $D_{13}$ has a large effect \cite{saw,kang}.
The $D_{33}$, included only by the Bonn group \cite{kang}, also
generates a large polarization.
(Of the 17 resonances included in the Bonn calculation
\cite{kang}, only those mentioned above had large effects on
$p_y$.)  As discussed above, it is hard to
imagine that a theoretically acceptable high energy 
model based on hadronic degrees of freedom will be constructed in
the near future.  Still, a modern relativistic calculation based
on hadronic degrees of freedom would nonetheless be desirable.

A calculation of $p_y$ has been done in the model of FMSS
\cite{sargsianpc}.  Since the helicity amplitudes in the
nucleon-nucleon scattering for this center of mass energy range
are not uniquely determined, some modelling was needed.
The calculation showns that $p_y$ is generally very small, going
from a negative value at low energies to positive values at
several GeV beam energy, and is consistent with the trend of the
experimental results. The polarization observable $C_x$ is also
expected to be small, and opposite in sign to $p_y$, while $C_z$
is expected to vanish.


\subsection{ Future prospects }
 
The rapid falloff of photodisintegration cross sections with
energy makes extension of the measurements difficult.
As beam energy increases from 4 to 5, 6, and 7 GeV, the
$s^{-11}$ dependence reduces cross sections by factors of 
6, 30, and 115, respectively.
Only a few experiments are possible without the proposed
12 GeV upgrade to JLab.
As indicated in Table~\ref{dexptab}, there are two approved 
experiments to continue the Hall A recoil polarization measurements
to additional angles, and to beam energies near 3 GeV.
Measurements of the $\Sigma$ asymmetry are
possible in JLab Hall B, over a range
of angles and energies well over 1 GeV.

In the longer term, the JLab 12 GeV upgrade offers additional
possibilities.  The luminosity increase planned for Hall B would
allow precise polarization measurements to continue above 2 GeV.
The proposed MAD spectrometer for Hall A \cite{MAD} would give about
a factor of 5 improvement in solid angle, and if it has the low
backgrounds characteristic of the current HRS spectrometers,
cross section measurements are possible up to 7 GeV and
polarization measurements are possible up to 4 GeV.


\subsection{Conclusions to Sec. 5}

Review of deuteron photodisintegration suggests the following:

\begin{itemize}

\item A microscopic meson-baryon theory of deuteron
photodisintegration must describe the $NN$
interaction at high energies, including pion production  and
the  contributions of hundreds of $N^*$ channels. It is
unlikely that such a theory will be constructed in the
foreseeable future.   The data might indicate
that the effect of many resonances is to increase the cross
section and decrease the polarization observables (by
averaging over may phases).  This suggests that it might be
possible to construct an effective theory based on hadronic
degrees of freedom.

\item For $p_T^2$ greater than about 1 GeV$^2$, cross sections
appear to follow the constituent counting rules, but it is
expected that an absolute pQCD calculation of the size of the
cross section would give a result much too small.
Similar observations may be made for other photoreactions,
and it remains to be seen how this behavior arises,
and if there is a general explanation for it.

\item The energy dependence of the photodisintegration cross 
sections has been shown to be potentially misleading
indicator of the success of pQCD. Models with asymptotic
behavior which differ from pQCD fit the data as well or
better than pQCD.  Further theoretical development and 
experimental tests of nonperturbative quark models would be
desirable.
 
\end{itemize}

\section{Overall Conclusions}

The new high energy measurements of the deuteron form
factors and the deuteron photodisintegration observables have
motivated much theoretical work.   In
particular:   

\begin{itemize}

\item Conventional meson theory works well in cases where
all of the {\it active\/} hadronic channels that can
contribute to a process are included in the calculations. 
This has been done for the deuteron form factors (where only
the $NN$ channel is active), but not for high energy
deuteron photodisintegration where 100's of $N^*N^*$ channels
are active.  At high energy any successful meson theory must
include relativistic effects.

\item New approaches, probably using quark degrees of freedom, 
are needed for high energy deuteron photodisintegration. 
While photodisintegration (as well as other reactions) seem
to follow the scaling laws, theoretical estimates using pQCD
give cross sections orders of magnitude too small.  The
data do not support hadronic helicity conservation.
Thus scaling is no longer seen as sufficient evidence for the
applicability of pQCD.

\end{itemize}

There is good evidence
that, in this energy region, the deuteron is undergoing a
transition from a region in which conventional hadronic
degrees of freedom describe the physics to a region in which
quark degrees of freedom are more appropriate. 



\ack

We thank H.\ Arenh\"ovel, B.\ L.\ G.\ Bakker,
J.\ Carbonell, W.\ H.\ Klink, D.\ Phillips, R.\ Schiavilla, and
G.\ Salm\`e and for supplying numerical values of
their deuteron form factor calculations, I.\ Sick for
values of the coulomb distortion corrections,
and M.\ Sargsian and M.\ Schwamb for supplying numerical values for
their photodisintegration calculations.
It is also a
pleasure to thank A.\ Afanasev, C.\ Carlson,  M. Gar\c{c}on, 
T.-S.\ .\ Lee, K. McCormick, G G Petratos, J.\ W.\ Van Orden,
A.\ Radyushkin, and G.\ Warren for several helpful
discussions, and S.\ Strauch for assisting with
photodisintegration figures.  This work was supported in
part by the US  Department of Energy. The Southeastern 
Universities Research Association (SURA) operates the Thomas
Jefferson National Accelerator Facility under DOE contract
DE-AC05-84ER40150.  RG acknowledges the support of the 
National Science Foundation from grant PHY-00-98642 to 
Rutgers University.


\end{document}